\documentclass[11pt,times]{elsarticle}
\usepackage{appendix}
\usepackage{lineno}

\journal{Geochimica et Cosmochimica Acta}

\usepackage{amssymb}
\usepackage[version=3]{mhchem}
\usepackage{hyperref}
\usepackage[capitalize]{cleveref}
\usepackage[figurename=Fig.]{caption}
\usepackage{booktabs}
\usepackage[referable]{threeparttablex}
\usepackage{tabularx}
\usepackage[T1]{fontenc}
\usepackage{textcomp}
\usepackage{wasysym}
\usepackage{pdflscape}
\usepackage{setspace}

\graphicspath{{fig/}}

\usepackage{natbib}
\bibpunct[]{(}{)}{;}{a}{,}{,}

\hypersetup{
	pdftitle={},
	pdfauthor={Takashi Yoshizaki},%
	pdfsubject={},%
	pdfkeywords={}
}

\biboptions{authoryear}

\begin{document}
	\begin{frontmatter}
				
		\title{Nebular history of an ultrarefractory phase bearing CAI from a reduced type CV chondrite}
		%
		%
		%
		%
		
		\author[tu]{Takashi Yoshizaki\corref{cor1}}
		\ead{takashiy@tohoku.ac.jp}
		\author[tu]{Daisuke Nakashima}
		\author[tu]{Tomoki Nakamura}
		\author[ko]{Changkun Park}
		\author[hu]{Naoya Sakamoto}
		\author[tu]{Hatsumi Ishida}
		\author[ku]{Shoichi Itoh}
		
		\cortext[cor1]{Corresponding author.}
		
		\address[tu]{Department of Earth Science, Graduate School of Science, Tohoku University, Aoba, Sendai, Miyagi 980-8578, Japan}
		\address[ko]{Division of Earth-System Sciences, Korea Polar Research Institute, Incheon 21990, South Korea}
		\address[hu]{Creative Research Institution Sousei, Hokkaido University, Kita, Sapporo 001-0021, Japan}
		\address[ku]{Department of Earth and Planetary Sciences, Graduate School of Science, Kyoto University, Sakyo, Kyoto 606-8502, Japan}

		\begin{abstract}
			\label{abst-1}


			Ultrarefractory (UR) phases in calcium-aluminum-rich inclusions (CAIs) could have formed at higher temperature compared to common CAI minerals and thus they potentially provide constraints on very high temperature processes in the solar nebula. We report a detailed characterization of the mineralogy, petrology and oxygen isotopic composition of an UR phase davisite (\ce{CaScAlSiO6}) bearing CAI from a reduced type CV chondrite. The CAI is an irregular-shaped, compound inclusion that consists of five units that are composed of melilite + spinel + Al,Ti-rich pyroxene $ \pm $ perovskite with various modal abundances of minerals and lithologies, and surrounded by the Wark-Lovering (WL) rim. Davisite occurs only in one lithological unit that consists of three chemically and isotopically distinct parts: i) \ce{^{16}O}-poor ($ -20$\textperthousand ~$ \leq \delta$\ce{^{18}O} $ \leq  0$\textperthousand) regions with reversely-zoned melilite and davisite; ii) \ce{^{16}O}-rich ($-50 $\textperthousand ~$ \leq \delta $\ce{^{18}O} $ \leq -40$\textperthousand) regions consisting of unzoned, gehlenitic melilite, Al,Ti-rich diopside and spinel; and iii) spinel framboids composed of \ce{^{16}O}-rich spinel and \ce{^{16}O}-poor melilite. Absence of secondary iron- and/or alkali-rich phases, occurrence of low-iron, manganese-enriched (LIME) olivine, and random distribution of the oxygen isotopic heterogeneity indicate that primitive chemical and isotopic compositions are preserved in the inclusion. The occurrence of chemical and isotopic heterogeneities with sharp boundaries in the CAI indicates formation of the inclusion by an aggregation of mineral assemblages formed and processed separately at different time and/or space in the solar nebula. Although isotope exchange between \ce{^{16}O}-rich solids and  \ce{^{16}O}-poor gases prior to the final agglomeration of the CAI cannot be ruled out, we suggest that modification of chemical and isotopic composition of porous CAI precursors or aggregation of isotopically distinct mineral assemblages are alternative scenarios for the origin of oxygen isotopic heterogeneity in CAIs. In either case, coexistence of spatially and/or temporally distinct \ce{^{16}O}-rich and \ce{^{16}O}-poor gaseous reservoirs at the earliest stage of the solar system formation is required. The grain-scale oxygen isotopic disequilibrium in the CAI indicate that post-formation heating of the inclusion (i.e., the WL rim formation event) was short (e.g., $ \lesssim  10^3 $ hours at 1400 K; $ \lesssim  10^5 $ hours at 1100 K), which can be achieved by rapid outward transport of the CAI. High \ce{Ti^{3+}}/\ce{Ti^{tot}} ratios of pyroxene from CAI interior and the rim and LIME composition of the olivine rim document that the entire CAI formation process took place under highly reducing conditions. 
			
		\end{abstract}
		
		
	\end{frontmatter}

	\section{Introduction}
	\label{sec:intro_chap1}
	
	
	Calcium-aluminum-rich inclusions (CAIs) are the oldest dated solar system solids \citep{,connelly2012absolute} and the key objects for understanding of physicochemical processes in the earliest stage of the solar system evolution \citep[e.g.,][ and references therein]{,macpherson2014calcium}. CAIs mostly occur as $ \mu $m- to cm-sized inclusions in chondritic meteorites, especially in CV carbonaceous chondrites \citep[e.g.,][]{,hezel2008modal}, and consist of various kinds of refractory minerals (e.g., melilite, spinel, Al,Ti-rich diopside) that are predicted by thermodynamic models to be among the first solids to condense from a cooling gas of the solar composition \citep[e.g.,][]{,grossman1972condensation}. They are considered to have formed during high-temperature processes including condensation, evaporation, aggregation and melting that occurred near the proto-Sun \citep[e.g.,][]{,mckeegan2000incorporation,sossi2017early}. Shortly after their formation, CAIs were transported to the asteroid- or even comet-forming region \citep[e.g.,][]{,zolensky2006mineralogy} via X-winds \citep[e.g.,][]{,shu1996toward}, outwards turbulent flow \citep[e.g.,][]{,ciesla2007outward}, or disk winds \citep{,van2016isotopic}.
	
	Bulk CAIs sometimes show the group II Rare Earth element (REE) patterns in which the most refractory (ultrarefractory~\textendash~UR) and the most volatile REEs are depleted compared to moderately refractory ones \citep[e.g.,][]{,tanaka1973rare}. Such a pattern can only be produced by fractional condensation in which earlier condensate were removed from a gas before the condensation is completed \citep{,boynton1975fractionation}. Therefore, the earlier condensates should be enriched in the UR elements. In some rare cases, UR phases such as davisite \citep[\ce{CaScAlSiO6};][]{,ma2009davisite} and Zr and/or Y-rich perovskite form UR CAIs, whose bulk chemical composition show a large enrichment ($ > 1000 ~\times $ CI) in the most refractory elements \citep[e.g.,][]{,davis1985trace,simon1996unique,el2002efremovka}.
	 UR phases also occur in a very minor population of CAIs which mostly consist of non-UR phases \citep[e.g.,][]{,lin2003fassaites} and amoeboid olivine aggregates (AOAs) \citep{,noonan1977zr,ma2012panguite,komatsu2018first}. Since evaporation experiments have failed to reproduce the REE patterns observed in UR CAIs \citep{,davis1995volatility,davis1995isotopic} and the natural UR CAIs show no clear isotopic evidence for mass fractionation during evaporation \citep{,simon1996unique}, these inclusions are unlikely to be evaporation residues. In contrast, thermodynamic calculations demonstrated that the UR REE pattern could be produced by condensation \citep[e.g.,][]{,davis1979condensation,simon1996unique,davis2018titanium}. As the UR phases are expected to have formed at higher temperatures compared to common CAI minerals, they can potentially provide constraints on physicochemical processes occurred in the beginning of the solar system formation \citep{,krot2018mineralogy}. However, due to their rare occurrences and small grain sizes, the detailed formation history of the UR phases is poorly understood.
	
	Oxygen isotopes in CAI minerals show mass-independent fractionation \citep[e.g.,][ and references therein]{,clayton1977distribution,yurimoto2008oxygen}. On an oxygen three-isotope diagram, the individual CAI minerals distribute along a slope $ \sim $1 line, which is known as the carbonaceous chondrite anhydrous minerals (CCAM) line \citep{,clayton1977distribution}, except for minerals in Fractionation and Unidentified Nuclear effects (FUN) CAIs \citep[e.g.,][]{,krot2014calcium}. Most CAIs in unmetamorphosed (petrologic types $ \leq $ 3.0) carbonaceous chondrites have uniformly \ce{^{16}O}-rich compositions ($-50$\textperthousand ~$ \leq $ $ \delta $\ce{^{18}O} $ \leq $ $-40$\textperthousand), suggesting that the majority of CAIs are originated in a solar-like, \ce{^{16}O}-rich gaseous reservoir \citep[][ and references therein]{,mckeegan2011oxygen,yurimoto2008oxygen}. On the other hand, mineral-scale oxygen isotopic heterogeneity is commonly observed in CAIs from CV and metamorphosed (petrologic types $ > $ 3.1) chondrites \citep[e.g.,][ and references therein]{,yurimoto2008oxygen}. In most of these CAIs, spinel, hibonite, Al,Ti-rich diopside and forsterite are \ce{^{16}O}-rich ($-50$\textperthousand ~$ \leq $ $ \delta $\ce{^{18}O} $ \leq $ $-40$\textperthousand) whereas melilite, anorthite and secondary alkali- and/or iron-rich phases are \ce{^{16}O}-poor ($-30$\textperthousand ~$ \leq $ $ \delta $\ce{^{18}O} $ \leq $ $10$\textperthousand). To date, several models have been proposed for a possible origin of the oxygen isotopic heterogeneity: (i) modification of isotope composition during fluid-assisted thermal metamorphism on the chondrite parent body \citep[e.g.,][]{,wasson200116,aleon2005fine}; (ii) gas-melt isotope exchange during partial or disequilibrium melting in the solar nebula \citep[e.g.,][]{,yurimoto1998oxygen,aleon2002calcium,kawasaki2017crystal,aleon2018closed}; (iii) isotope exchange by solid-state diffusion during nebular reheating events \citep[e.g.,][]{,clayton1977distribution,itoh2003contemporaneous,simon2011oxygen,simon2016oxygen}; and (iv) changes in the isotopic composition of a nebular gas during CAI condensation \citep[e.g.,][]{,katayama2012oxygen,kawasaki2012oxygen,kawasaki2016chronological,park2012oxygen}.
	
	The UR minerals sometimes show \ce{^{16}O}-depletion \citep{,ivanova2012compound,ivanova2017oxygen,zhang2015mineralogical,aleon2018o}, whose origin remains controversial. \citet{,ivanova2012compound} reported that UR oxide, Zr,Sc-rich pyroxene and Y-rich perovskite in an UR inclusion 3N-24 in a forsterite-bearing CAI from NWA 3118 ($ \mathrm{CV_{ox}} $) are \ce{^{16}O}-poor, whereas spinel and Al,Ti-rich diopside are \ce{^{16}O}-rich. They also showed that in an UR inclusion 33E-1 in a fluffy type A CAI from Efremovka, UR oxide, davisite and Y-rich perovskite are depleted in \ce{^{16}O}, and Al,Ti-rich diopside and spinel are enriched in \ce{^{16}O}. The authors suggested that these inclusions originated in a \ce{^{16}O}-rich gaseous reservoir and subsequently experienced isotope exchange in a \ce{^{16}O}-poor gaseous reservoir, and they preferred post-crystallization isotope exchange in such UR phases due to their faster O self diffusion compared to spinel and diopside. However,  the detailed mechanism and timing of the isotopic modification are unclear. \citet{,zhang2015mineralogical} reported that an UR inclusion A0031 from Sayh al Uhaymir 290 (CH), which is composed of various UR phases such as panguite \citep[\ce{(Ti^{4+}{,}Sc{,}Al{,}Mg{,}Zr{,}Ca)_{1.8}O3};][]{,ma2012panguite}, davisite and Sc-rich anosovite (\ce{(Ti^{4+}{,}Ti^{3+}{,}Mg{,}Sc{,}Al)3O5}) associated with spinel, anorthite, and perovskite, is uniformly depleted in \ce{^{16}O}. They concluded that the inclusion condensed from a \ce{^{16}O}-depleted nebular gas. \citet{,ivanova2017oxygen} revealed that spinel and rubinite \citep[\ce{Ca3Ti^{3+}2Si3O12};][]{,ma2017discovery} are \ce{^{16}O}-rich; Zr,Sc-bearing grossmanite  \citep[\ce{CaTi^{3+}AlSiO6};][]{,ma2009grossmanite} and melilite are \ce{^{16}O}-poor; and perovskite is intermediate in an UR CAI 40E-1 in a compound CAI. \citet{,aleon2018o} showed that in an UR compact type A CAI E101.1 from Efremovka ($ \mathrm{CV_{red}} $), Zr,Sc-rich diopside and nepheline are similarly depleted in \ce{^{16}O}, Zr,Sc-poor diopside is \ce{^{16}O}-rich, and UR perovskite and melilite have variable oxygen isotopic compositions.  The authors suggested crystallization of the Zr,Sc-rich diopside by interaction between \ce{^{16}O}-poor melt and the UR perovskite, and relict origin for the \ce{^{16}O}-rich diopside. They proposed that the CAI originated in a \ce{^{16}O}-rich gas and experienced isotopic exchange with a \ce{^{16}O}-poor gaseous reservoir during partial melting. Thus, the origin of oxygen isotopic heterogeneity observed in UR phase bearing CAIs can be variable among different inclusions but is not fully understood. Further isotopic studies on the UR phase bearing inclusions can provide us insights into the early solar system processes that provided isotopic heterogeneities in these highly refractory objects.
	
	Here we report petrology, mineral chemistry and oxygen isotopic composition of an UR phase bearing CAI R3C-01-U1 from the reduced type CV chondrite Roberts Massif (RBT) 04143. In this inclusion, Ti,V-rich davisite coexists with common CAI minerals such as melilite, Al,Ti-rich diopside, and spinel, and they are isotopically heterogeneous. Based on the occurrence, mineral chemistry, and oxygen isotope compositions of Ti,V-rich davisite and associated phases including spinel framboids \citep{,el1979spinel}, we discuss origin and thermal history of the CAI, and physicochemical conditions of the solar nebula.

	\section{Analytical Methods} \label{sec:methods_1}
	
	\subsection{Scanning electron microscope analyses}
	\label{sec:methods_1_SEM}
	
	Petrology and mineralogy of the CAI R3C-01-U1 in a polished thick section of the CV chondrite RBT 04143 were initially characterized with backscattered electron (BSE) imaging using Hitachi S-3400N scanning electron microscope (SEM) and JEOL JSM-7001F field-emission scanning electron microscope (FE-SEM) at Tohoku University (TU). Semi-quantitative analyses were performed using Oxford INCA energy-dispersive spectrometers (EDS) that were installed on both SEM, at an accelerating voltage of 15 kV and a beam current of 1.0--1.4 nA.
	
	\subsection{Electron probe microanalyzer analysis}
	\label{sec:methods_1_EPMA}
	
	Quantitative X-ray microanalyses of melilite, spinel, perovskite and pyroxene were performed at TU and Korea Polar Research Institute (KOPRI) using JEOL JXA-8530F field-emission electron probe microanalyzers (FE-EPMA) equipped with wavelength-dispersive X-ray spectrometers (WDS). All electron probe analyses are available as Supplementary Information. A  defocused electron beam of 3 $ \mu $m in diameter accelerated at 15 kV with a beam current of 10 nA was used to quantify 14 elements (Si, Ti, Al, Cr, Fe, Mn, Mg, Ca, Na, V, Zr, Hf, Sc and Y) in melilite, spinel and perovskite. The peak counting times were 10 s for Na; 20 s for Si, Al, Mg and Ca; and 40 s for Ti, Cr, Fe, Mn, V, Sc and Y. Silicon, Ti, Al, Cr, Fe, Mn, Mg, Ca, Na, V and Sc were measured using $ K\alpha $ and Zr, Hf and Y were quantified using the $ L\alpha $. Natural and synthetic crystalline minerals were used as standards. Matrix corrections were applied using the atomic number (Z), absorption (A), and fluorescence (F) (ZAF) correction method. The detection limits were 0.01 wt\% for \ce{Cr2O3}, FeO, MnO, MgO, CaO, \ce{Na2O} and \ce{Sc2O3}; 0.02 wt\% for \ce{TiO2}, \ce{Al2O3} and \ce{V2O3}; 0.03 wt\% for \ce{SiO2}, \ce{ZrO2} and \ce{Y2O3}; and 0.05 wt\% for \ce{HfO2}. Analyses with totals below 98 wt\% and above 102 wt\% were excluded from consideration. A focused electron beam accelerated at 10 kV with a beam current of 20 nA was used to quantify 13 elements (Si, Ti, Al, Cr, Fe, Mn, Mg, Ca, Na, V, Zr, Sc and Y) in pyroxene. The peak counting times were 10 s for Na and 20 s for other elements. Other conditions were similar to analyses of melilite, spinel and perovskite. Detection limits were 0.01 wt\% for MgO, CaO and \ce{Na2O}; 0.02 wt\% for \ce{Al2O3}, \ce{Cr2O3}, MnO, \ce{V2O3}, \ce{ZrO2}, \ce{Sc2O3}, and \ce{Y2O3}; and 0.03 wt\% for \ce{SiO2}, \ce{TiO2} and FeO. The amounts of \ce{Ti^{3+}} in pyroxene with \ce{TiO2} $ \geq $ 4 wt\% were calculated by assuming the pyroxene is ideally stoichiometric (4 cations per 6 oxygen atoms) and titanium is the only element with various valance states \citep{,beckett1986origin}. Quantitative chemical composition of olivine was measured using the JEOL JXA-8800 electron microprobe equipped with WDS detectors at TU. Analyses were conducted at an accelerating voltage of 15 kV and a beam current of 10 nA with a counting time 20 s for each element. 11 elements (Si, Ti, Al, Cr, Fe, Mn, Mg, Ca, Na, K and Ni) were quantitatively determined. Other conditions were similar to the aforementioned analyses. Detection limits are 0.02 wt\% for \ce{Al2O3}, MgO, CaO, \ce{Na2O} and \ce{K2O}; 0.03 wt\% for \ce{SiO2}, \ce{Cr2O3}, FeO and MnO; and 0.04 wt\% for \ce{TiO2} and NiO.
	
	X-ray elemental maps of Si, Ti, Al, Fe, Mg, Ca, Na, V and Sc were obtained using FE-EPMA at TU and KOPRI. For both instruments, a focused electron beam was used at 15 kV accelerating voltage, 20--50 nA beam current and 20--100 ms per pixel dwell time. Obtained X-ray elemental maps were processed using ImageJ, Python and R.
	
	The bulk chemical composition of R3C-01-U1 was obtained using average chemical compositions of each phase and their modal abundances and densities. The modal abundance of each mineral was calculated using high-resolution X-ray elemental maps processed using the ImageJ software and Python codes. A quantitative map of \aa{}kermanite content in melilite was obtained using spot EPMA data within a mapped area as referential composition.
	
	\subsection{Electron back-scatter diffraction analysis}
	\label{sec:methods_1_EBSD}
	
	Electron back-scatter diffraction (EBSD) analyses were performed to obtain crystal orientation maps of melilite using Oxford AZtec EBSD system on the FE-SEM at TU. Prior to the measurements, the sample surface was polished using 1 $ \mu $m alumina lapping film followed by a chemomechanical polish by a combination of colloidal silica and a Buehler VibroMet 2 Vibratory Polisher for 3 hours. The polished sample was then coated with a thin layer of carbon. Analyses were conducted at an acceleration voltage of 15 kV and beam current of 3.4 nA under a high vacuum. The grain boundaries of melilite crystals were identified based on crystallographic orientation maps, BSE images and X-ray elemental maps.
	
	\subsection{Oxygen isotope analysis}
	\label{sec:methods_1_isotoporgaphy}
	
	Spot oxygen isotope analyses were performed with the Cameca ims-1270 ion microprobe at Hokkaido University. A primary \ce{Cs+} ion with a beam current of 0.1 nA was used to achieve a count rate of $ \sim5\times10^7 $ cps for \ce{^{16}O-}. The \ce{Cs+} ions were accelerated with 20 kV to sputter the $ \sim 5$--$7$ $ \mu $m spot (in diameter) of the sample surface coated by gold. An electron flood gun was used for a charge compensation. Negative secondary ions were detected using a Faraday cup (for \ce{^{16}O-}) and an electron multiplier (for \ce{^{17}O-} and \ce{^{18}O-}). The mass resolution was set to $ \mathrm{M/\Delta M} \sim 7000$ to resolve \ce{^{16}O^{1}H-} and \ce{^{17}O-}. Each mass peak in a cycle was measured for 1 s, 2 s and 1 s for \ce{^{16}O-}, \ce{^{17}O-} and \ce{^{18}O-}, respectively, in the magnet peak switching mode. San Carlos olivine standard was used to correct instrumental mass fractionation. Results are reported using the conventional delta-notation relative to the Standard Mean Ocean Water (SMOW): 
	\begin{eqnarray}
	\delta \ce{^{17,18}O} = \left(  \frac{\ce{^{17,18}O}_{\mathrm{Sample}}}{\ce{^{17,18}O}_{\mathrm{SMOW}}} -1 \right) \times 1000 ~(\permil).
	\end{eqnarray}
	Analytical uncertainties are estimated based on an internal error of each measurement and reproductivity of the standard measurements and are reported in 2$ \sigma $. The sample was investigated using the SEM to make sure that there was no contamination due to beam overlapping onto neighboring phases and the sampling area was free of cracks.
	
	Quantitative oxygen isotope imaging (isotopograph) of melilite, spinel, perovskite and pyroxene was obtained using the isotope microscope system consisting of the Cameca ims-1270 ion microprobe and a high-efficiency Stacked CMOS-type Active Pixel Sensor (SCAPS) ion imager at Hokkaido University \citep{,takayanagi1999stacked,yurimoto2003high,sakamoto2008discovery}. Analytical conditions were generally similar to those in \citet{,park2012oxygen} and \citet{,zhang2015mineralogical}, otherwise described below. A \ce{Cs+} primary beam was set to $ \sim 65 ~\mu $m $ \times  ~55 ~\mu $m and intensity of $ \sim 3 $ nA. Two measurement cycles were performed to obtain secondary ion images of \ce{^{16}O-} and \ce{^{18}O-}. Each cycle consists of 1 frame for \ce{^{16}O-} and 40 frames for \ce{^{18}O-} with each frame of 5 s. Before the oxygen isotope mapping, \ce{^{27}Al^{16}O-} was also mapped for 1 frame to identify the locations of minerals in the imaged area. The $ \delta $\ce{^{18}O} isotopograph was obtained by calculating the secondary ion ratios of \ce{^{18}O-}/\ce{^{16}O-} for each pixel and then calibrated with a spinel grain in the imaging area, whose oxygen isotopic composition was determined by the spot analysis. The obtained $ \delta $\ce{^{18}O} isotopograph was smoothed with a moving-average of 3 $ \times $ 3 pixels to reduce statistical error to be $ \pm ~ 6$\textperthousand ~($ 1\sigma $) in the processed image. Spatial resolution of isotopography is calculated to be $ \sim $ 1 $ \mu $m, which corresponds $ \sim 4.3$ pixels (see Appendix B for more details).
		
	\section{Results} \label{sec:results-1}
	
	\subsection{Mineralogy and petrology}
	\label{sec:results-1_min}
	
	\subsubsection{Overview of mineralogy and petrology of R3C-01}
	\label{sec:results-1_min_R3C-01}
	
	The UR phase bearing CAI R3C-01-U1 (1.0 $ \times $ 0.6 mm) is one of five lithological units (\cref{tab:R3C-01_units}) that form an irregularly-shaped, compound CAI R3C-01 (1.5 $ \times $ 0.8 mm; \cref{fig:BSE_R3C-01}). Petrology and mineralogy of R3C-01-U1 are distinct from the common CAIs and the other four units of R3C-01: for instance, davisite and high concentration of spinel framboids are observed only in this unit (\cref{fig:BSE_R3C-01,fig:RGBP_R3C-01,fig:Sc-Px_map,fig:BSE_R3C-01_details}). The detailed mineralogy and petrology of R3C-01-U1 is described later in this section. 
	
	Unit 2 of R3C-01 consists of melilite, irregular-shaped clusters of spinel grains, tiny perovskite and Al,Ti-rich diopside (\cref{fig:BSE_R3C-01_details}f). Al,Ti-rich diopside commonly surrounds perovskite or it is associated with spinel but no davisite is observed in this unit. There are some spinel clusters in the Unit 2 which show framboid-like texture, but their abundance is much lower than that in R3C-01-U1. The Unit 2 is separated from R3C-01-U1 by the Wark-Lovering (WL) rim. Unit 3 of R3C-01 is a less-irregular domain with melilite-rich core and spinel-rich mantle (\cref{fig:BSE_R3C-01_details}g). The grain size of melilite is larger in the core ($ \sim 50 ~\mu$m in diameter) than in the mantle ($ \leq 20 ~\mu$m). The core melilite show reverse zoning (\AA{}k$ _{5} $ and \AA{}k$ _{30} $ in the core and rim, respectively). The core contains some large cracks and pores, and the mantle has numerous tiny pores. No davisite grain occurs in this unit. Spinel occurs as euhedral small ($ \sim 10 ~\mu$m) grains, some of which are forming small ($ \sim 20$--$50 ~\mu$m) clusters. Perovskite occurs as irregular $\sim$ subrounded grains embedded in the outer core of this unit, and its \ce{V2O3} content is higher than that in R3C-01-U1 (0.7--1.3 wt\% and 0.1--0.8 wt\%, respectively), whereas \ce{ZrO2} concentration is lower in the Unit 3 (0.05--0.15 wt\% and 0.15--0.35 wt\%, respectively; \cref{tab:R3C-01_units}).  Irregular Al,Ti-rich pyroxene occurs in the mantle of this subinclusion. The continuous occurrence of spinel-rich mantle around the core distinguishes the Unit 3 from associated Unit 2 and Unit 4. Unit 4 consists of unzoned melilite, fine-grained spinel and Al,Ti-rich diopside intergrown with spinel (\cref{fig:BSE_R3C-01_details}h). Melilite in this subinclusion show lower \AA{}k content (\AA{}k$ _{5-15} $) relative to melilite in the other units (\cref{tab:R3C-01_units}). Spinel has lower \ce{V2O3} content (0.1--0.2 wt\%)) than in other units (\cref{tab:R3C-01_units}).  Perovskite is rare and davisite is absent in this domain. Unit 5 consists of compact melilite, thick layers of spinel, perovskite and Al,Ti-rich diopside surrounding perovskite or intergrown with spinel (\cref{fig:BSE_R3C-01_details}g). Based on these mineralogical and petrological observations, it is likely that R3C-01 is a compound of at least five melilite-rich CAIs. Note that the number of lithological units in the CAI is a minimum since we only see a 2D section of the inclusion.
	
	Tiny grains of refractory metal nuggets (RMNs) are observed in each lithological unit. Secondary iron- and/or alkali-rich phases such as nepheline, sodalite and hydrous silicates are not observed in any units. The outer margin of the inclusion is surrounded by the multi-layered WL rim \citep{,wark1977marker} composed of spinel, Al,Ti-rich/poor diopside and olivine + Fe,Ni-metal layers (\cref{fig:BSE_R3C-01,fig:BSE_R3C-01_details}d). Based on the constant thickness of the olivine layer, we consider it as a part of the WL rim rather than a forsterite-rich accretionary rim, which is commonly observed around CV CAIs with a highly variable thickness that is controlled by the topography of the host inclusion \citep[e.g.,][]{,krot2001forsterite}. High Ca content in the olivine (up to 0.8 wt\%) is also distinct from those in common forsterite-rich accretionary rim \citep[mostly $ < 0.4$ wt\%; e.g.,][]{,krot2001forsterite}.
		
	\subsubsection{Mineralogy and petrology of R3C-01-U1}
	\label{sec:results-1_min_R3C-01-U1}
	
\begin{figure}[p]
		\centering
		\includegraphics[width=1\linewidth]{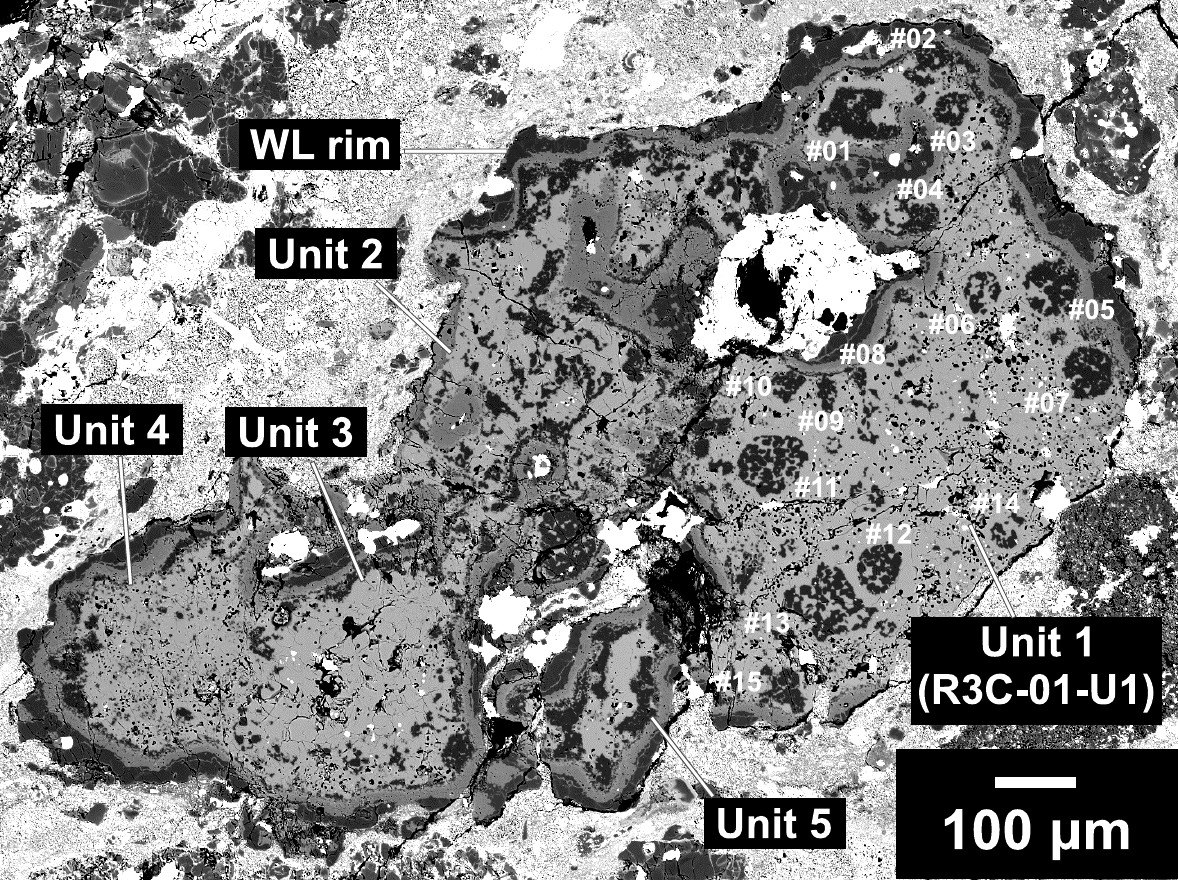}
		\caption[BSE image of R3C-01]
		{BSE image of the compound CAI R3C-01 from RBT 04143. Spinel framboids in R3C-01-U1 (\#01--15) are indicated by labels. Enlarged image of a part of the CAI is shown in \cref{fig:RGBP_R3C-01,fig:RGBP_R3C-01,fig:Sc-Px_map,fig:isotopograph}. Some holes and cracks in the thick section are filled with a gold, which had been coated for previous SIMS study and could not be removed by polishing. Abbreviations: WL rim\textendash Wark-Lovering rim.
		}
		\label{fig:BSE_R3C-01}
	\end{figure}

\begin{figure}[p]
		\centering
		\includegraphics[width=1\linewidth]{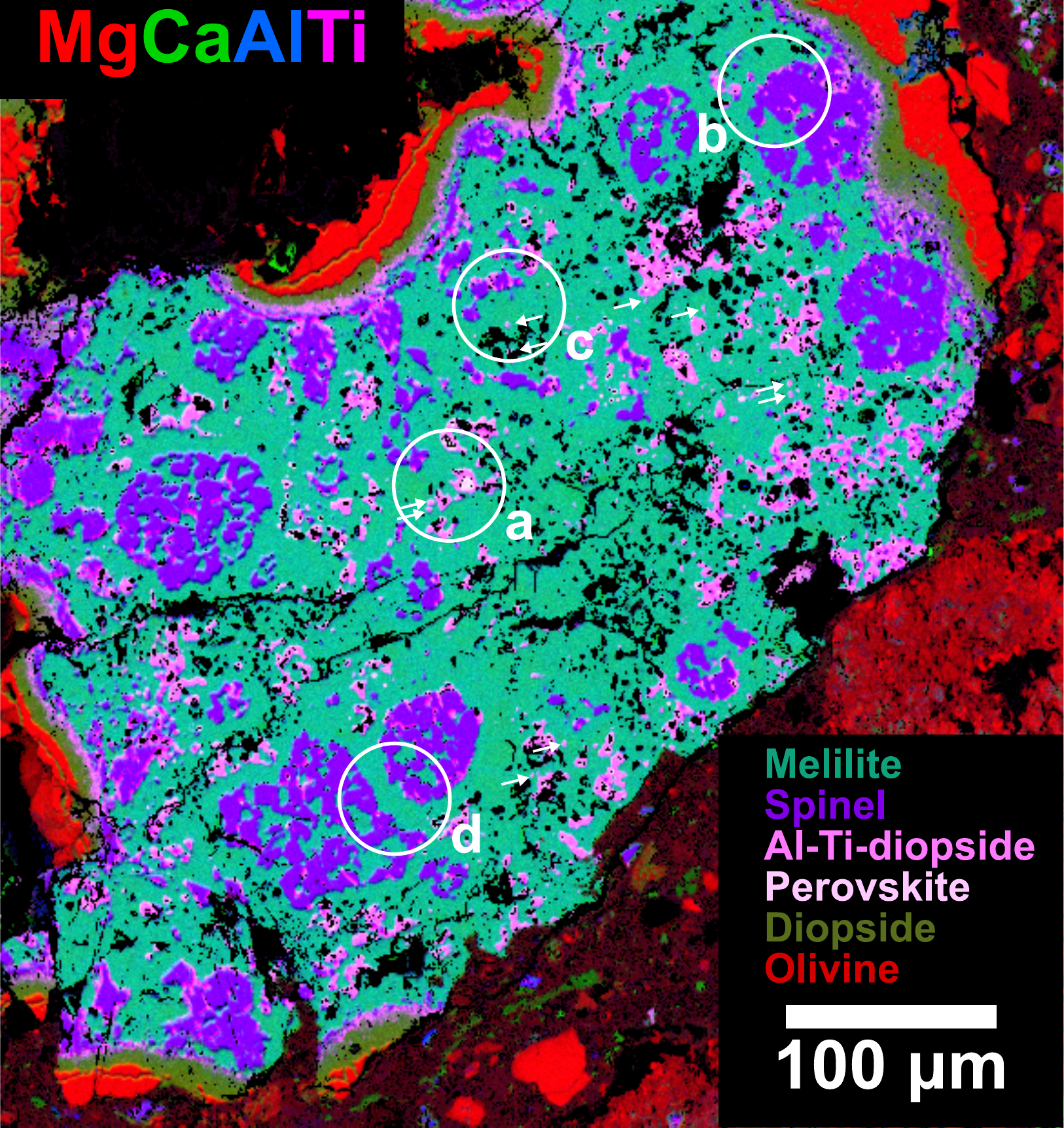}
		\caption[Combined X-ray map of R3C-01-U1]
		{Combined elemental map in Mg (red), Ca (green), Al (blue) and Ti (pink) $ K\alpha $ X-rays image of R3C-01-U1. The analysis areas for isotopography (\cref{fig:isotopograph}) are also shown by white circles with labels. Arrows show occurrences of Ti,V-rich davisite.
		}
		\label{fig:RGBP_R3C-01}
	\end{figure}

	R3C-01-U1, the largest part of the compound CAI R3C-01, mainly consists of fine-grained ($ < $ 50 $ \mathrm{\mu m} $) melilite and spinel, with minor perovskite, Al,Ti-rich pyroxene (\cref{fig:BSE_R3C-01,fig:RGBP_R3C-01,fig:Sc-Px_map,fig:isotopograph,fig:BSE_R3C-01_details}a--e), and RMNs. The CAI does not show any obvious core-mantle structure (\cref{fig:BSE_R3C-01,fig:RGBP_R3C-01}). Representative microprobe analyses of minerals in R3C-01-U1 are listed in \cref{tab:R3C-01_px_EPMA,tab:R3C-01_dav_EPMA,tab:R3C-01_mel_EPMA,tab:R3C-01_sp_EPMA,tab:R3C-01_pv_EPMA,tab:R3C-01_ol_EPMA}.  
	
	The X-ray elemental map and EBSD crystal orientation map show that R3C-01-U1 consists of small ($ < $ 50 $ \mathrm{\mu m} $) grains of reversely-zoned, relatively \AA{}k-rich melilite (typically \AA{}k$ _{15-20} $ in the core and \AA{}k$ _5 $ at the rim) and unzoned \AA{}k-poor melilite (typically \AA{}k$ _5 $), with random crystal orientation and 120$^\circ$ triple junctions at some grain boundaries (\cref{fig:isotopograph}).
	
	Al,Ti-rich pyroxene in the interior R3C-01-U1 shows various occurrences (\cref{fig:Sc-Px_map,fig:BSE_R3C-01_details}a--e). Most Al,Ti-rich pyroxene in R3C-01-U1, which occurs as irregularly-shaped grains of $ < $ 5 $ \mathrm{\mu m} $ in diameter attached to perovskite and/or spinel, is a member of Al,Ti-rich diopside with 
	20.9--25.2 wt\% \ce{Al2O3}; 
	10.2--17.0 wt\% \ce{TiO^{tot}_2} (= \ce{Ti^{3+}_2O3} + \ce{Ti^{4+}O2}); 
	0.1--0.5 wt\% \ce{Sc2O3}; 
	0.2--1.4 wt\% \ce{V2O3}; 
	and 0--0.3 wt\% \ce{ZrO2} (\cref{fig:Px-comp,fig:Sc-Px_comp,tab:R3C-01_px_EPMA}). 
	On the other hand, euhedral pyroxene grains ($ < $ 5 $ \mathrm{\mu m} $; mostly 2--3 $ \mathrm{\mu m} $) isolated in reversely-zoned melilite are Ti,V-rich davisite with 
	20.8--22.8 wt\% \ce{Al2O3}; 
	10.8--13.6 wt\% \ce{TiO^{tot}_2};
	6.2--8.9 wt\% \ce{Sc2O3}; 
	2.2--3.7 wt\% \ce{V2O3}; 
	and 0.5--0.7 wt\% \ce{ZrO2} (\cref{fig:Px-comp,fig:Sc-Px_comp,tab:R3C-01_dav_EPMA}). 
	The calculated \ce{Ti^{3+}}/\ce{Ti^{tot}} ratios mostly range from 0.3 to 0.8 for Al,Ti-rich diopside and from 0.6 to 0.9 for Ti,V-rich davisite (\cref{fig:Px-comp,fig:Sc-Px_comp,tab:R3C-01_px_EPMA,tab:R3C-01_dav_EPMA}). 
	Note that quantitative analyses of Al,Ti-rich pyroxene attached to spinel described above are possibly contaminated by beam overlapping onto spinel due to tiny grain size of pyroxene, as seen in their elevated Al and Mg contents and lower  \ce{Ti^{3+}}/\ce{Ti^{tot}} ratio (open symbols in \cref{fig:Px-comp}). Thus, its \ce{Ti^{3+}}/\ce{Ti^{tot}} ratio should be considered as the lower limit for its true ratio \citep{,dyl2011valence}. Importantly, measurements of pyroxene which is not attached to spinel (filled symbols or asterisks in \cref{fig:Px-comp}) do not show clear contamination by other phases.
	
\begin{figure}[p]
		\centering
		\includegraphics[width=1\linewidth]{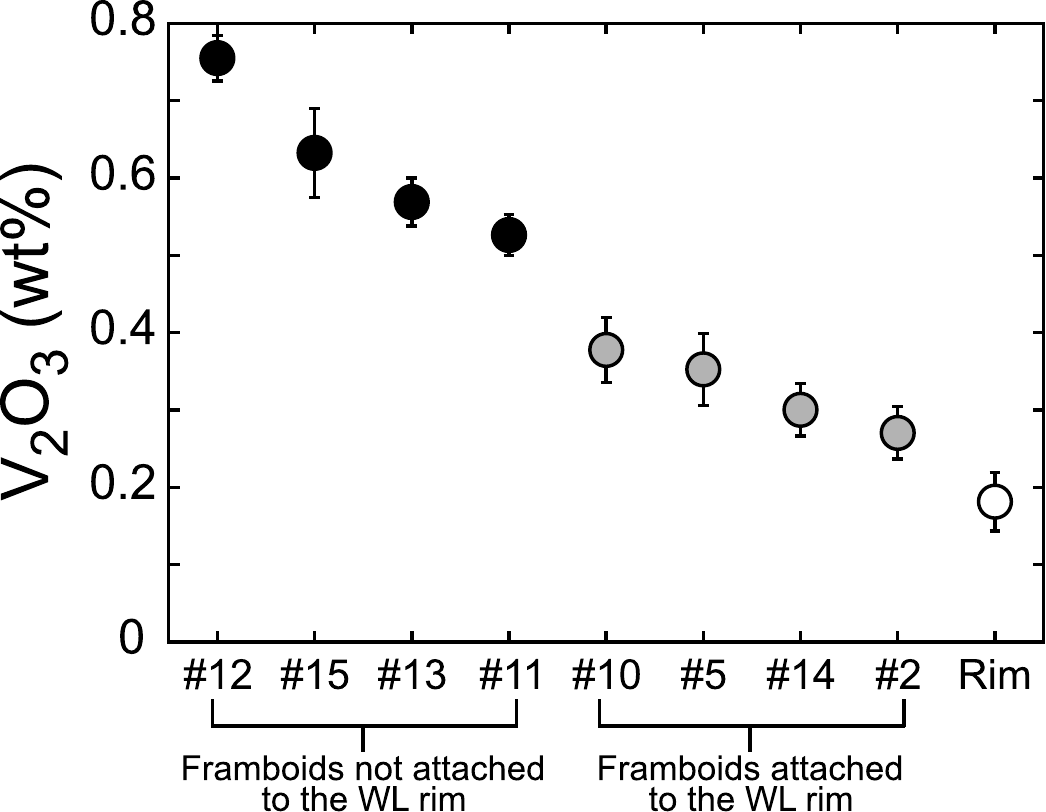}
		\caption[\ce{V2O3} contents of spinel in framboids and the WL rim in R3C-01-U1]
		{\ce{V2O3} contents in spinel in framboids and WL rim in R3C-01-U1. 
			Error bars represent 1$ \sigma $ standard deviation ($ n = 5 $ for each framboids and 10 for the rim, respectively). 
		}
		\label{fig:V2O3}
	\end{figure}

	Spinel in R3C-01-U1 occurs as euhedral grains ($ < $ 20 $ \mathrm{\mu m} $) and its chemical composition is nearly pure \ce{MgAl2O4} with FeO and \ce{Cr2O3} $ \leq $ 0.2 wt\% (\cref{tab:R3C-01_sp_EPMA}). Most spinel grains form framboidal aggregates (\cref{fig:BSE_R3C-01,fig:RGBP_R3C-01,fig:BSE_R3C-01_details}a,b,e). Fifteen framboids (\#01--15) are identified in R3C-01-U1. Palisade bodies, spheroidal shells of spinel enclosing typical CAI minerals that commonly coexist with framboids \citep{,simon1997situ}, are apparently absent in this inclusion. All framboids are surrounded by relatively coarse (> 10 $ \mathrm{\mu m} $), reversely-zoned melilite (typically \AA{}k$ _{15-20} $ in the core and \AA{}k$ _5 $ at the rim) and unzoned melilite (typically \AA{}k$ _5 $) grains and enclose tiny melilite ($ < $ 10 $ \mathrm{\mu m} $) (\cref{fig:isotopograph}b,d). In most framboids, spinel occurs with small amounts of Al,Ti-rich diopside ($ < $ 5 $ \mathrm{\mu m} $), whereas tiny perovskite is also associated with spinel and Al,Ti-rich diopside in one framboid (\#13; \cref{fig:isotopograph}d). Vanadium oxide content of spinel in framboids ranges between 0.2 to 0.8 wt\%, but the value is nearly constant within each framboids (\cref{fig:V2O3,tab:R3C-01_sp_EPMA}). Framboidal spinel attached to the rim show lower \ce{V2O3} contents compared to those from the inner part of the CAI, and spinel in the WL rim shows the lowest value. The V content in spinel is not correlated well with abundances of other elements (e.g., Ti, Cr). 
	
	Perovskite in R3C-01-U1 is smaller than 10 $ \mathrm{\mu m} $ and mostly surrounded by Al,Ti-rich diopside (Figs. \ref{fig:isotopograph}a and \ref{fig:BSE_R3C-01_details}b,c,e). Minor element concentration in perovskite in the inclusion (e.g., $ < $ 0.3 wt\% \ce{ZrO2} and \ce{Y2O3} and $ < $ 0.8 wt\% \ce{V2O3}; \cref{tab:R3C-01_pv_EPMA}) is in the range of those in typical non-UR CAIs \citep[e.g.,][]{,brearley1998chondritic}.
	
\begin{figure}[p]
		\centering
		\includegraphics[width=0.9\linewidth]{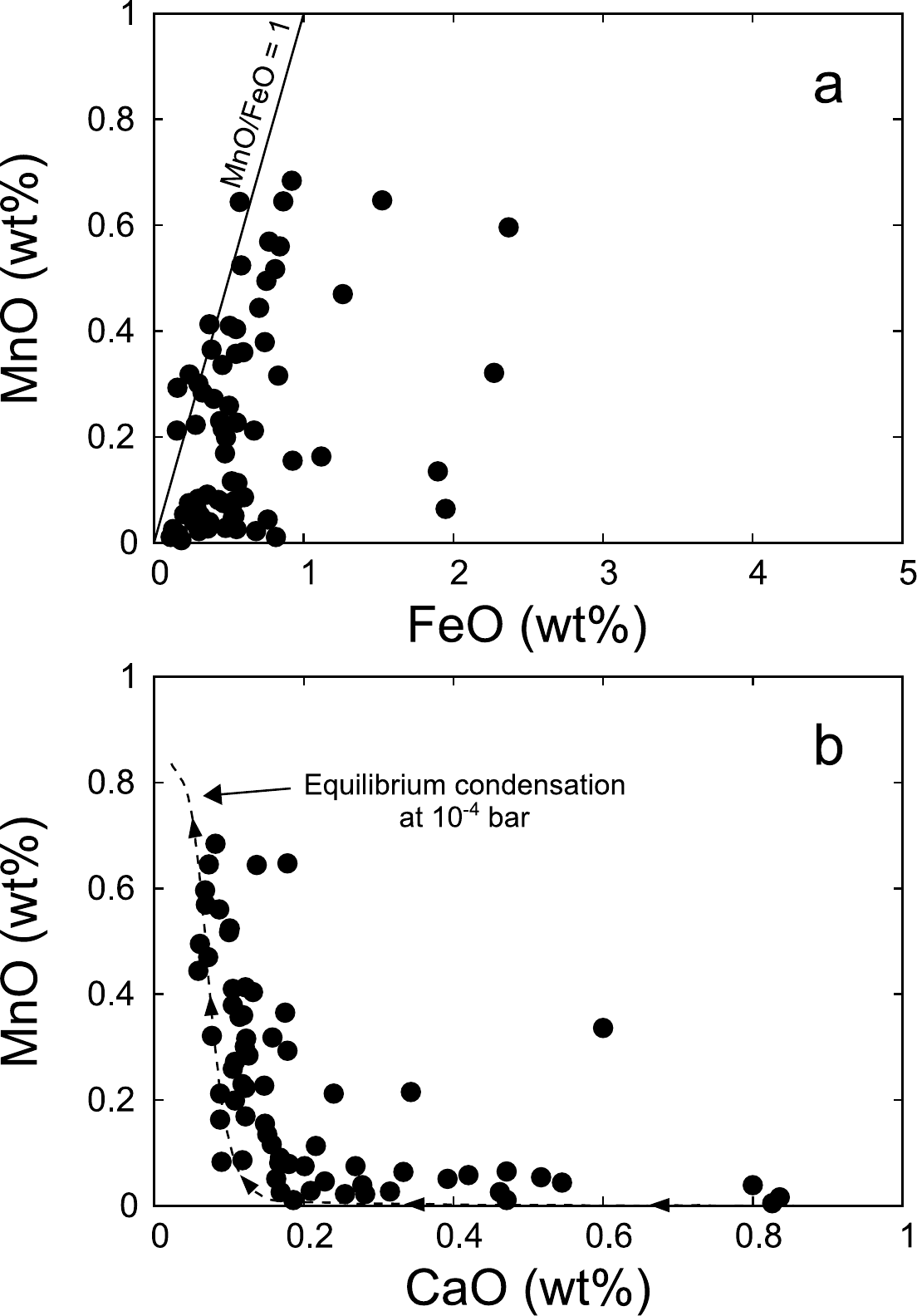}
		\caption{Plot of (a) MnO vs FeO and (b) MnO vs CaO for olivine in the WL rim on R3C-01-U1. A broken line in (b) shows a model composition of forsterite formed by equilibrium condensation in a cooling gas of solar composition composition at $ 10^{-4} $ bar traced from \citet{,sugiura2009nebular}
		}
		\label{fig:MnO-FeO}
	\end{figure}

	The innermost layer of the WL rim surrounding R3C-01-U1 and other units is semi-continuous, ribbon-like spinel layer with 0.1--0.2 wt\% \ce{V2O3} (\cref{fig:V2O3}) with thickness of $ < $ 10 $ \mathrm{\mu m} $ and intergrown with tiny Al,Ti-rich diopside grains (\cref{fig:BSE_R3C-01_details}d). The second layer consists of intergrowths of melilite and spinel, which is surrounded by diopside layer thereafter. The diopside layer is $\sim$10 $ \mathrm{\mu m} $ in thickness and compositionally zoned outward from Al,Ti-rich to Al,Ti-poor (\cref{fig:Px-comp,tab:R3C-01_px_EPMA}). Similar to Al,Ti-rich pyroxene in the CAI interior, the rim pyroxene shows high \ce{Ti^{3+}}/\ce{Ti^{tot}} ratio (0.4--1.0). Anorthite (An$ _{\sim 100}$) rarely occurs along cracks as thin layers ($ < $ 10 $ \mathrm{\mu m} $) between melilite in inner CAI and the WL rim. The outermost layer of the WL rim is compact-textured olivine rim (\cref{fig:BSE_R3C-01,fig:RGBP_R3C-01}), with highly forsteritic composition (mostly Fa$ _{0.5-1} $) (\cref{fig:Fa,tab:R3C-01_ol_EPMA}). Some olivine grains from the outermost part of the rim have the low-iron, manganese-enriched (LIME) compositions with MnO/FeO $\sim$ 1 and FeO $ < $ 1 wt\% (\cref{fig:MnO-FeO}a and \cref{tab:R3C-01_ol_EPMA}). CaO contents in olivine decreases from the inner to the outer rim whereas MnO contents gradually increases (\cref{fig:MnO-FeO}b). The continuous changes in CaO and MnO contents in olivine match a model composition of forsterite condensed from a cooling nebular gas of solar composition at $ 10^{-4} $ bar \citep{,sugiura2009nebular}.
	
	
	In the diagram of \ce{Al2O3}-\ce{Ca2SiO4}-\ce{Mg2SiO4}, bulk compositions of R3C-01 and R3C-01-U1 are similar to that of type A CAIs (\cref{fig:bulk_tplot}). The CI-normalized bulk composition of R3C-01-U1 is distinguished from those of previously studied UR inclusions in abundances of highly refractory elements such as Zr, Hf, Sc, Y, and Ti (\cref{fig:bulk_sdiagram}).

	\subsection{Oxygen isotopes}
	\label{sec:results-1_O-isotope}
	
	On the oxygen three-isotope diagram, spinel in two framboids (\#01 and \#05) from R3C-01-U1 and the WL rim minerals (Al-diopside and olivine) plot on the CCAM line (\cref{fig:SIMS_spot,tab:SIMS_spot}), showing no evidence for a mass-dependent fractionation of oxygen isotopes as observed in FUN CAIs \citep[e.g.,][]{,krot2014calcium}. All of these minerals are enriched in \ce{^{16}O}, with $ \delta $\ce{^{18}O} of $\sim$ $-$47\textperthousand ~for spinel, $\sim$ $-$40\textperthousand ~for Al-diopside and olivine in the WL rim.
	
	We obtained 4 isotopographs (areas (a)-(d)) from R3C-01-U1 (\cref{fig:RGBP_R3C-01,fig:isotopograph}). These areas include Ti,Vi-rich davisite, melilite, Al,Ti-rich diopside, spinel and perovskite. As O-isotopes of spinel in the CAI lay on the slope-1 line  (\cref{fig:SIMS_spot}), low- and high-$ \delta $\ce{^{18}O} areas in isotopographs can be regarded as \ce{^{16}O}-rich and -poor regions, respectively. The obtained isotopographs of R3C-01-U1 minerals show that this inclusion is mostly composed of three mineralogically and isotopically distinct regions: 
	i) \ce{^{16}O}-poor ($ -20$\textperthousand ~$ \leq \delta$\ce{^{18}O} $ \leq  0$\textperthousand) regions with reversely-zoned melilite and Ti,V-rich davisite; ii) \ce{^{16}O}-rich ($-50 $\textperthousand ~$ \leq \delta $\ce{^{18}O} $ \leq -40$\textperthousand) regions consisting of unzoned, gehlenitic melilite, Al,Ti-rich diopside and spinel; and iii) spinel framboids composed of \ce{^{16}O}-rich spinel and \ce{^{16}O}-poor melilite. These three chemically and isotopically distinct regions show clear boundaries and seem to be distributed randomly in this inclusion.
	
	In the area (a), some reversely-zoned melilite, spinel and Al,Ti-rich diopside are enriched in \ce{^{16}O} ($ \delta $\ce{^{18}O} $\sim$ $-$40\textperthousand) (\cref{fig:isotopograph}a). On the other hand, Ti,V-rich davisite and most reversely-zoned melilite are depleted in \ce{^{16}O} ($ \delta $\ce{^{18}O} $\sim$ 0\textperthousand ~for Ti,V-rich davisite; $ \delta $\ce{^{18}O} $\sim$ $-$10 to 0\textperthousand ~for melilite). A perovskite grain rimmed by \ce{^{16}O}-rich Al,Ti-rich diopside is depleted in \ce{^{16}O} ($ \delta $\ce{^{18}O} $\sim$ $-$20\textperthousand). The $ \delta $\ce{^{18}O} values change sharply at the grain boundaries between \ce{^{16}O}-rich regions and \ce{^{16}O} -poor regions.
	
	In the area (b), reversely-zoned melilite grains surrounding a spinel framboid are uniformly depleted in \ce{^{16}O} ($ \delta $\ce{^{18}O} $\sim$ $-$10\textperthousand) (\cref{fig:isotopograph}b). Tiny melilite grains enclosed in the spinel framboid are also depleted in \ce{^{16}O}. On the other hand, all spinel grains and Al,Ti-rich diopside grains enclosed in melilite or surrounding the spinel framboid are uniformly enriched in \ce{^{16}O} ($ \delta $\ce{^{18}O} $\sim$ $-$47\textperthousand).
	
	In the area (c), unzoned \AA{}k-poor melilite in the upper-left region of the area, spinel and Al,Ti-rich diopside are uniformly \ce{^{16}O}-rich ($ \delta $\ce{^{18}O} $\sim$ $-$45\textperthousand) (\cref{fig:isotopograph}c). In contrast, Ti,V-rich davisite and relatively coarse ($\sim$ 30 $ \mathrm{\mu m} $) reversely-zoned melilite are uniformly depleted in \ce{^{16}O} ($ \delta $\ce{^{18}O} $\sim$ $-$5\textperthousand). The $ \delta $\ce{^{18}O} values change clearly at the grain boundaries.
	
	In the area (d), spinel grains from two framboids and unzoned \AA{}k-poor melilite between these framboids are enriched in \ce{^{16}O} ($ \delta $\ce{^{18}O} $\sim$ $-$45\textperthousand) (\cref{fig:isotopograph}d). In contrast, reversely-zoned melilite between these framboids are depleted in \ce{^{16}O} ($ \delta $\ce{^{18}O} $\sim$ $-$15\textperthousand). Line profiles in the two melilite grains revealed a possible narrow zoning of oxygen isotopic composition at their grain boundaries (\cref{fig:SCAPS_line}). Tiny reversely-zoned or unzoned melilite enclosed in a spinel framboid (\#12) show moderate depletion in \ce{^{16}O} ($ \delta $\ce{^{18}O} $\sim$ $-$25\textperthousand).

	\section{Discussion} \label{sec:discussion-1}
	
	
	
	\subsection{Origin of oxygen isotopic heterogeneity in R3C-01-U1}
	\label{sec:discussion-1_O-isotope_heterogeneity}
	
	Here we discuss five possible scenarios for the origin of grain-scale O-isotopic heterogeneity observed in R3C-01-U1, based on combination of the oxygen isotope mapping with micron-scale petrology and mineralogy of the inclusion.
	
	\subsubsection{Parent-body processes}
	\label{sec:discussion-1_O-isotope_heterogeneity_p-body}
	
	R3C-01-U1 consists of three chemically and isotopically distinct regions: 
	i) \ce{^{16}O}-poor ($ -20$\textperthousand ~$ \leq \delta$\ce{^{18}O} $ \leq  0$\textperthousand) regions with reversely-zoned melilite and Ti,V-rich davisite; ii) \ce{^{16}O}-rich ($-50 $\textperthousand ~$ \leq \delta $\ce{^{18}O} $ \leq -40$\textperthousand) regions consisting of unzoned, gehlenitic melilite, Al,Ti-rich diopside and spinel; and iii) spinel framboids composed of \ce{^{16}O}-rich spinel and \ce{^{16}O}-poor melilite (\cref{fig:RGBP_R3C-01,fig:Sc-Px_map,fig:isotopograph}). 	It has been recognized that most CV3 chondrites have experienced various degrees of thermal metamorphism assisted by \ce{^{16}O}-poor fluid on their parent bodies \citep[e.g.,][]{,brearley2013metasomatism,krot2018evidence}. During the parent-body processes, primary anhydrous CAI minerals such as melilite, anorthite and forsteritic olivine are replaced by alkali- and/or iron-rich, \ce{^{16}O}-poor phases such as nepheline, sodalite, fayalitic olivine and phillosilicates \citep[e.g.,][]{,hashimoto1985sem,krot1995mineralogical,brearley2013metasomatism}. In addition, alteration by \ce{^{16}O}-poor thermal fluid could also result in \ce{^{16}O}-depletion in melilite compared to other phases in CAIs \citep[e.g.,][]{,fagan2004oxygenalt,fagan2004oxygen,itoh2004petrography,krot2018evidence}, because oxygen diffuses much rapidly in melilite among the common CAI phases \citep{,yurimoto1989diffusion,ryerson1994determination}.
	
\begin{figure}[p]
		\centering
		\includegraphics[width=1\linewidth]{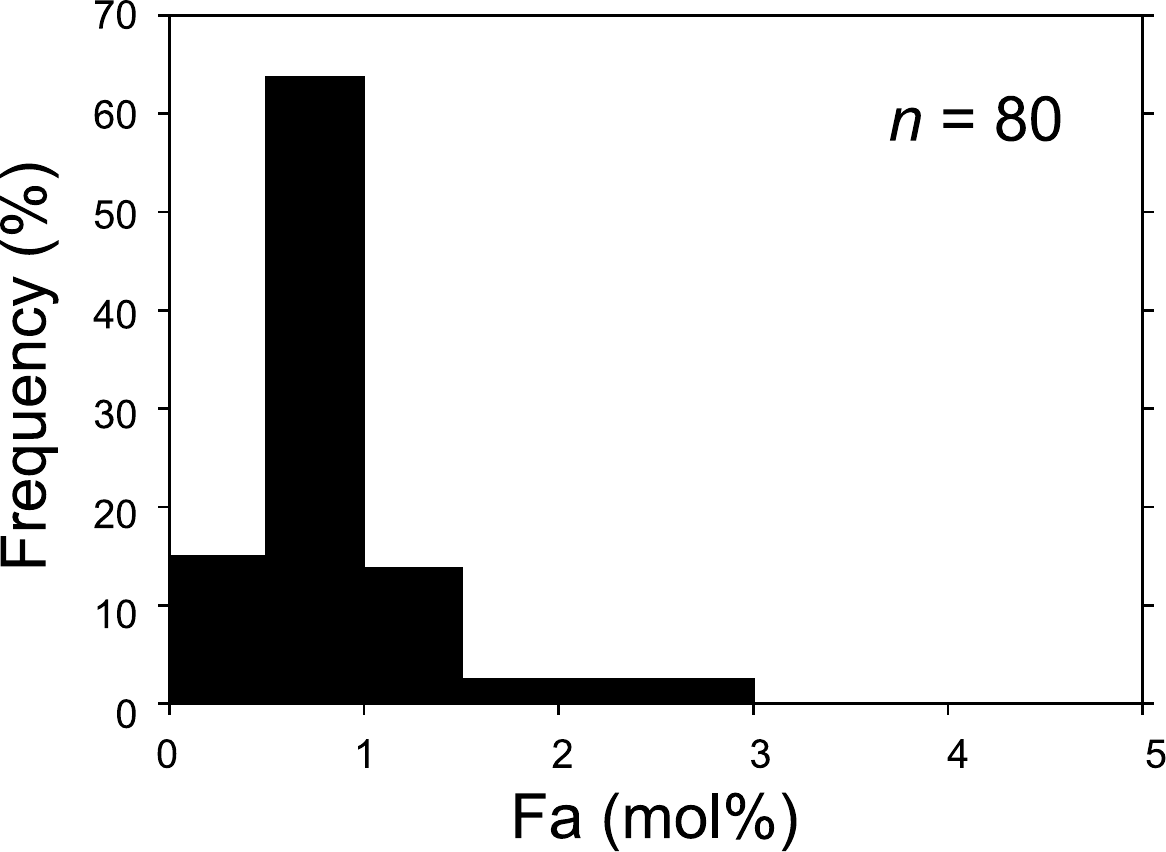}
		\caption{Histogram of Fa contents of olivine in the WL rim on R3C-01-U1.
		}
		\label{fig:Fa}
	\end{figure}

	R3C-01 lacks clear evidence for a fluid-driven thermal metamorphism in the CV parent body. Presence of intact melilite and anorthite, the most fragile phases that are easily replaced by secondary minerals during parental body processes \citep{,hashimoto1985sem,krot1995mineralogical}, and absence of iron- and/or alkali-rich phases in R3C-01 documents that this inclusion largely escaped fluid-assisted thermal metamorphism on the CV parent body. The fayalite content of olivine rim surrounding R3C-01 (mostly Fa$ _{0-1} $: \cref{fig:Fa}) is similar to that in AOAs from unmetamorphosed carbonaceous chondrites \citep[e.g.,][]{,krot2004amoeboid}. The presence of LIME olivine in the rim (\cref{fig:MnO-FeO}) is also inconsistent with thermal-fluid metamorphism that would result in increase of FeO content and decrease of MnO/FeO ratio of silicates \citep[e.g.,][]{,komatsu2015lime}. Unlike an Allende CAI that shows clear evidence for parent-body alteration \citep{,park2012oxygen}, the \ce{^{16}O}-poor region in R3C-01-U1 is not spatially related to cracks or pores and the $ \delta $\ce{^{18}O} value clearly changes at grain boundaries between primary CAI minerals (\cref{fig:isotopograph}). The coexistence of \ce{^{16}O}-rich and \ce{^{16}O}-poor melilite grains with similar \AA{}k content at their grain boundaries (\cref{fig:isotopograph}c,d) rules out the asteroidal modification scenario, because the oxygen isotopic exchange should occur similarly in the chemically comparable melilites if their isotopic compositions were originally similar.	The occurrence of uniformly \ce{^{16}O}-rich WL rim minerals also strongly suggest that there was no subsequent parent-body processing driven by a \ce{^{16}O}-poor thermal fluid  \citep[e.g.,][]{,matzel2013oxygen,bodenan2014oxygen,ushikubo2016long,krot2017high}. Thus, it is suggested that the oxygen isotopic heterogeneity in R3C-01-U1 did not result from fluid-rock interaction on CV parent body. Considering that RBT 04143 is a brecciated chondrite \citep{,ishida2012diverse,lunning2016cv}, it is inferred that the clast in which the CAI occurs escaped the parent body processing. R3C-01 is likely to be one of the most pristine CV CAIs which preserve primitive mineralogical, petrological and isotopic features, as CAIs from pristine chondrites do \citep[e.g.,][]{,itoh2004petrography,krot2008multiple,makide2009oxygen,bodenan2014oxygen,ushikubo2016long,simon2018condensate}.
	
	\subsubsection{Progressive isotopic evolution of a nebular gas}
	\label{sec:discussion-1_O-isotope_heterogeneity_changes}
	
	Significant variation in oxygen isotopic compositions within single grains is not observed in R3C-01-U1 (\cref{fig:isotopograph}), demonstrating that the oxygen isotopic heterogeneity in this inclusion does not reflect gradual changes of isotopic compositions of a nebular gas  \citep{,katayama2012oxygen,kawasaki2012oxygen,kawasaki2016chronological,park2012oxygen}.
	
	\subsubsection{Partial melting in the solar nebula}
	\label{sec:discussion-1_O-isotope_heterogeneity_partial_melting}
	
	The irregular shape without any apparent core-mantle structure and fine-grained nature of R3C-01-U1 (\cref{fig:BSE_R3C-01,fig:RGBP_R3C-01}) is distinct from typical igneous CAIs \citep[e.g.,][ and references therein]{,brearley1998chondritic}, indicating an igneous origin of this sub-inclusion in unlikely. In R3C-01-U1, reversely-zoned melilite and Ti,V-rich davisite enclosed in the melilite are depleted in \ce{^{16}O} (\cref{fig:isotopograph}a,c). \citet{,macpherson1984fluffy} suggested that formation of reversely zoned melilite crystals in unmelted CAIs cannot be explained by crystallization from a melt, but by direct condensation from a gas with decreasing pressure. If this is the case for reversely-zoned melilite in R3C-01-U1, they are expected to record oxygen isotopic compositions of the nebular gas in which the melilite condensed. On the other hand, \citet{,grossman2002formation} suggested that reversely-zoned melilite could be crystallized from an incomplete melt accompanied by a melt evaporation. Although we cannot exclude the melt evaporation scenario for the origin of reversely-zoned melilite in R3C-01-U1, it is less likely given that significant Mg isotopic fractionation has never been observed in reversely-zoned melilite in the melilite-rich CAIs like R3C-01-U1 \citep[e.g.,][]{,kawasaki2016chronological} and this scenario requires evaporation and crystallization of melilite grains without grain coarsening before aggregation of R3C-01-U1. In addition, uniform oxygen isotopic compositions of individual melilite grains with sharp changes in $ \delta $\ce{^{18}O} values at their grain boundaries (\cref{fig:isotopograph}) suggest that \ce{^{16}O}-depletion in R3C-01-U1 did not result from partial melting that could produce abrupt oxygen isotope variation within a single crystal \citep{,yurimoto1998oxygen}. Furthermore, the lack of Ti-V correlation in spinel also indicates that the inclusion has never experienced partial melting or subsolidus equilibration \citep{,connolly2003petrogenesis,connolly2003type}. The compact texture of minerals with triple junctions in R3C-01 (\cref{fig:BSE_R3C-01,fig:RGBP_R3C-01}) can be achieved with a subsolidus heating of less than few hours \citep{,karato1989grain,nichols1991grain,faul2006grain,whattam2008granoblastic,komatsu2009high} and does not require partial melting. Thus, oxygen isotope exchange during nebular partial melting processes is unlikely for the origin of oxygen isotopic heterogeneity in R3C-01-U1.
	
	On the other hand, it is possible that other lithological units in R3C-01 (e.g., the Unit 3 with a core-mantle structure; \cref{fig:BSE_R3C-01}) have experienced partial melting before aggregation of R3C-01 so that the irregular shape of R3C-01 is originated in a coagulation of igneous and non-igneous CAIs \citep{,rubin2012new}. Although it is beyond the scope of the present work to test this possibility, further isotopic study of this inclusion will provide us insights into formation process of irregularly-shaped CAIs.
	
	\subsubsection{Solid-state isotopic exchange during nebular heating}
	\label{sec:discussion-1_O-isotope_heterogeneity_solid-state}
	
	The oxygen isotopic composition of melilite in R3C-01-U1 does not correlate with the distance from the outer margin of the inclusion (\cref{fig:RGBP_R3C-01,fig:isotopograph}). The sharp isotopic changes at grain boundaries of melilite and the random distribution of \ce{^{16}O}-rich and \ce{^{16}O}-poor regions are apparently inconsistent with gas-solid isotopic exchange from the outer edge of the inclusion during nebular heating processes \citep{,simon2011oxygen,simon2016oxygen}.
	
	Importantly, the grain boundary diffusion is much faster than the volume diffusion \citep[e.g.,][]{,zhang2010diffusion}. Small grain sizes in R3C-01-U1 indicate that the grain boundary diffusion could have been more effective in the CAI. Furthermore, since compact-textured CAIs are likely the result of annealing of once porous aggregates of mineral grains, surface diffusion also should not be ignored as the effective diffusion mechanism in the pre-annealed CAIs. 
	
	Here we assume that R3C-01-U1 was originally porous aggregates of fine-grained minerals. If such an object experienced solid-state isotope exchange with an external gaseous reservoir, oxygen atoms are transported by diffusion in the gas between crystals, followed by the volume diffusion in each mineral crystal. In this case, diffusion profiles correlate with a distance from grain boundaries, rather than with a distance from the outer edge of the object. Thus, difference in diffusivity between phases can result in variation in degree of chemical and isotopic modification of the minerals. The diffusion coefficient of oxygen ($ D\mathrm{^{Ox}} $) in \aa{}kermanite is higher than that of gehlenite \citep{,yurimoto1989diffusion}. Since the melting temperature ($ T_{\mathrm{m}} $) of \aa{}kermanite is lower than that of gehlenite \citep{,osborn1941ternary}, \citet{,yurimoto1989diffusion} suggested that $ D\mathrm{^{Ox}_{Mel}} $ and $ T_{\mathrm{m}} $ follow the relationship $ \log D  \propto  -T_{\mathrm{m}}/T $, which has been demonstrated in general solid solution systems of alloys and semiconductors \citep[Flynn's rule;][]{,flynn1972point}.
	\citet{,nagasawa2001diffusion} demonstrated that $ D $ value of  interdiffusion of Al + Al vs Mg + Si can be the same order with that of oxygen self-diffusion in melilite at $ > 1200 $ K, and the $ D\mathrm{^{AlAl-MgSi}_{Mel}} $ value also follows the Flynn's rule. Therefore, if unzoned \AA{}k-rich melilite coexisted with uniformly \AA{}k-poor one and they experienced gas-solid interaction with another gaseous reservoir, selective modification of chemical and isotopic composition of the \AA{}k-rich one is possible. When both original \AA{}k-rich and \AA{}k-poor melilite grains were originally \ce{^{16}O}-rich and they reacted with Al-rich, \ce{^{16}O}-poor gaseous reservoir, occurrence of  reversely-zoned, \ce{^{16}O}-poor melilite adjacent to \AA{}k-poor, \ce{^{16}O}-rich melilite can be achieved. In this case, the reversely-zoned melilite should have been originally more  \AA{}k-rich than their present core . Alternatively, it is possible that the precursor of the reversely-zoned, \ce{^{16}O}-poor melilite was originally \ce{^{16}O}-rich and it has experienced solid state isotopic (and chemical?) modification with the ambient gas before aggregation of R3C-01-U1. Thus, subsolidus processes between crystals and \ce{^{16}O}-poor gaseous reservoir at several stages of the CAI formation could have resulted in the oxygen isotopic heterogeneity. In contrast, the uniform \ce{^{16}O}-enrichment in minerals of the WL rim (\cref{fig:SIMS_spot}) that seems to have completely surrounded R3C-01 before its fragmentation indicates that the solid-state isotopic and chemical modification might predate formation of the WL rim.
		
	A key issue with the gas-solid exchange model for the origin of \ce{^{16}O}-depletion ($ \pm $ reverse zoning) in the reversely-zoned, \AA{}k-rich melilite is the occurrence of \ce{^{16}O}-depleted davisite (i.e., Sc-rich pyroxene) in a center of some of these melilite grains (\cref{fig:Sc-Px_map,fig:isotopograph}). Diffusion coefficients of oxygen in diopside is much lower than in melilite \citep{,ryerson1994determination} although a diffusion coefficient of oxygen in davisite has not yet been determined. If pyroxene also follows the Flynn's rule, davisite \citep[$ T_{\mathrm{m}} \sim 1534$ \textcelsius~ at 1 atm;][]{,ohashi1978studies} can have much lower oxygen diffusivity than diopside \citep[$ T_{\mathrm{m}} \sim 1390$ \textcelsius;][]{,bowen1914ternary}. In this case, solid-state isotopic modification of davisite discussed above is less likely for the origin of \ce{^{16}O}-depletion of this phase and associated melilite.	Experimental determination of diffusivity of oxygen in davisite is a key to understand the origin of \ce{^{16}O}-depletion in the CAI.
	
	\subsubsection{Aggregation of isotopically distinct mineral assemblages}
	\label{sec:discussion-1_O-isotope_heterogeneity_aggregation}
	
	
	Alternatively, the heterogeneous chemical and oxygen isotopic composition in R3C-01-U1 can be explained by another scenario: aggregation of crystals formed separately in physicochemically and isotopically distinct gaseous reservoir. Mineralogy and chemical and oxygen isotopic composition of spinel framboids suggest a separate origin of spinel framboids in R3C-01-U1 (see \cref{sec:discussion-1_fram}), which is another evidence for the accumulation of mineral assemblages that formed separately. Even in the gas-solid isotopic (and chemical) modification model, aggregation of at least three mineralogically distinct mineral assemblages (\AA{}k-rich melilite + davisite, \AA{}k-poor melilite + Al,Ti-rich diopside + spinel, and spinel framboids) is required. Therefore, it is suggested that R3C-01-U1 consists of mineral assemblages that formed separately in different gaseous reservoir under variable physicochemical conditions. We suggest that this scenario will be supported if the oxygen diffusivity in pyroxene follows the Flynn's rule.
	
	\subsubsection{Implications for the oxygen isotopic reservoirs in the early solar system}
	
	In either scenario (gas-solid exchange and aggregation of isotopically distinct materials; \cref{sec:discussion-1_O-isotope_heterogeneity_solid-state,sec:discussion-1_O-isotope_heterogeneity_aggregation}), it is highly likely that oxygen isotopic composition of minerals in the interior of R3C-01-U1 was established before the WL rim formation since the rim minerals are uniformly enriched in \ce{^{16}O} (\cref{fig:SIMS_spot}). Thus, it is suggested that there were temporally and/or spatially distinct \ce{^{16}O}-rich and \ce{^{16}O}-poor gaseous reservoirs when the UR phase bearing CAI formed. This suggestion is consistent with the coexistence of an \ce{^{16}O}-rich gaseous reservoir whose oxygen isotopic composition is close to that of the Sun \citep[$ \delta $\ce{^{18}O} $\sim$ $ \delta $\ce{^{17}O} $\sim -60$\textperthousand:][]{,mckeegan2011oxygen} and a planetary-like, \ce{^{16}O}-poor gaseous reservoir \citep[$ \delta $\ce{^{18}O} $\sim$ $ \delta $\ce{^{17}O} $\sim$ 0\textperthousand: e.g.,][]{,clayton1993oxygen} in the earliest stage of the solar system formation  \citep[e.g.,][]{,krot2002existence,krot2017calcium,aleon2018o},	which is considered to have formed during photochemical self-shielding effects in CO occurred in the protosolar molecular cloud \citep{,yurimoto2004molecular,lee2008oxygen}, the inner protoplanetary disk \citep{,clayton2002solar}, or the outer protoplanetary disk \citep{,lyons2005co}.
	
	\subsection{Thermal processing of CAIs in the solar nebula}
	\label{sec:discussion-1_thermal_history}
	
	The occurrence of high-temperature, multilayered WL rim and compact texture of minerals in R3C-01 indicate that the inclusion has experienced heating event(s) after its aggregation. Decreasing refractoriness of mineralogy outwards the WL rim on R3C-01 (spinel + melilite + Al,Ti-rich diopside  \textrightarrow ~Al,Ti-rich diopside \textrightarrow ~Al,Ti-poor diopside  \textrightarrow ~Ca-rich, Mn-poor olivine $ \pm $ Fe,Ni-metal \textrightarrow  ~Ca-poor, Mn-rich olivine $ \pm $ Fe,Ni-metal) (\cref{fig:RGBP_R3C-01,fig:Px-comp,fig:MnO-FeO}) is generally consistent with the predictions of thermodynamic calculations of condensation from a cooling nebular gas \citep[e.g.,][]{,petaev2005meteoritic,ebel2006condensation}. Thus, the WL rim surrounding R3C-01 likely have formed by a condensation from a cooling nebular gas \citep[e.g.,][]{,wark1977marker,bolser2016microstructural,krot2017high}.
	
	On the other hand, the oxygen isotopic heterogeneity in the CAI interior (\cref{fig:isotopograph}) indicates that the subsequent heating of the CAI (i.e., the WL rim formation event) was short enough to preserve the grain-scale isotopic disequilibrium. Although most of \ce{^{16}O}-rich and \ce{^{16}O}-poor regions show clear isotopic boundaries, some melilite grains possibly have experienced micron-scale oxygen isotope exchange (\cref{fig:SCAPS_line}).	Thus, the duration of subsequent heating of the CAI can be estimated assuming that the observed isotopic zoning in melilite has resulted from this event. Here we consider a homogeneously \ce{^{16}O}-rich ($ \delta $\ce{^{18}O} $ \sim $ $ -45 $\textperthousand) melilite occurred adjacent to an isotopically unzoned \ce{^{16}O}-poor ($ \delta $\ce{^{18}O} $ \sim $ $ -15 $\textperthousand) melilite grain before the WL rim formation. We assume that the temperature of the CAI interior is uniformly similar to that of the ambient gas due to high thermal diffusivity of minerals \citep[e.g., $ > 10^{-6}$ $ \mathrm{m^2/s} $ at $ > 1000$ K;][]{,clauser1995thermal}. The duration of subsequent heating of the inclusion required to achieve the $ \mu $m-scale isotopic zoning (\cref{fig:SCAPS_line}) can be calculated by solving the diffusion equation
	\begin{eqnarray}
	\frac{\partial }{\partial t}c(x,t) = D(T) \frac{\partial^2}{\partial x^2}c(x,t),
	\end{eqnarray}
	
	\noindent where $ c $ is concentration of species $ i $ of interest, $ x $ is distance from the grain boundary, $ t $ is time, $ D(T) $ is the diffusion coefficient of oxygen in melilite
	\begin{equation}
	D(T) = D_0 \mathrm{e}^{-\frac{Q}{RT}},
	\end{equation}
	\noindent where $ D_0 $ is the pre-exponential factor, $ Q $ is the activation energy, $ R $ is the universal gas constant, and $ T $ is temperature. 
	Since both \ce{^{16}O}-rich and \ce{^{16}O}-poor melilite grains in R3C-01-U1 are highly gehlenitic at their rim (\AA{}k$ _5 $: \cref{fig:isotopograph}), we use the diffusion coefficient for gehlenitic (\AA{}k$ _0 $) melilite \citep{,yurimoto1989diffusion} in the calculation. Temperature conditions of the WL rim formation event can be constrained based on mineralogy of the rim: in the WL rim sequence, spinel is the highest temperature condensate \citep[1397 K at $ 10^{-4} $ bar;][]{,lodders2003solar} and the LIME olivine is the lowest \citep[$ \sim $ 1100 K at $ 10^{-4} $ bar;][]{,ebel2012thermochemical}. By fitting the model curve to the observed isotopic zoning, the duration of the WL rim formation event is estimated to be $ \sim $ 400 hours at 1400 K (\cref{fig:SCAPS_line}) and $ \sim  3\times 10^4$  hours at 1100 K. Since it is possible that some of the micron-scale isotopic zoning was achieved by nebular heating before the WL rim formation, these estimates are upper limits of heating duration of R3C-01 during the WL rim formation.
	
	
	Our estimate of timescale of the WL rim formation is also in harmony with the \ce{^{26}Al}-\ce{^{26}Mg} systematics of CAIs and their rims, which demonstrated that a time interval between initial CAI crystallization and the WL rim formation was 0--2 Myr  \citep{,taylor200426al,taylor2005high,cosarinsky2005magnesium,simon2005short,mane2015resolved,mane2016formation,matzel2015aluminum,kawasaki2016chronological,aleon2018o}, although \citet{,aleon2018o} suggested that the secondary processes and Mg self-diffusion could have resulted in longer duration estimates. Furthermore, the compact texture of minerals in R3C-01 (\cref{fig:BSE_R3C-01,fig:RGBP_R3C-01}) can be achieved with a subsolidus heating of less than few hours \citep{,karato1989grain,nichols1991grain,faul2006grain,whattam2008granoblastic,komatsu2009high}, supporting the short timescale of heating during the WL rim formation. 
	
	The short duration of post-formation heating of CAIs suggests that the high-temperature materials were rapidly removed from hot inner region of the solar nebula by outward transport processes \citep[e.g.,][]{,ciesla2007outward,van2016isotopic}. This rapid transport process might have transported CAIs beyond several AU before proto-Jupiter formation, which created a pressure bump and hindered inward drift of CAIs to the inner solar system \citep{,kruijer2017age,desch2017effect,scott2018isotopic}, and thus caused a refractory element enrichment in carbonaceous meteorites compared to non-carbonaceous ones \citep[e.g.,][]{,rubin2011origin,rubin2018differences}. 
	

	
		\subsection{Redox conditions of the solar nebula}
	\label{sec:discussion-1_redox}

	The presence of \ce{Ti^{3+}} in Ti-bearing phases in refractory inclusions (e.g., diopside, hibonite, spinel and forsterite) reflects highly reducing condition in the solar nebula \citep[$ \log f_{\ce{O2}}  \sim  \mathrm{IW} - 6.8 $;][]{,grossman2008redox}. \ce{Ti^{3+}}/\ce{Ti^{tot}} ratios in \ce{^{16}O}-rich, Al,Ti-rich diopside and \ce{^{16}O}-poor, Ti,V-rich davisite (0.3--0.8 and 0.6--0.9, respectively; \cref{fig:Px-comp,tab:R3C-01_px_EPMA,tab:R3C-01_dav_EPMA}) are similar to those of Ti-rich pyroxene in the interior of other CAIs investigated so far \citep[e.g.,][]{,simon1991fassaite,simon2005short,simon2007valence,simon2006comparative,simon2011refractory,grossman2008redox,dyl2011valence,ma2012panguite,ma2013kangite,zhang2015mineralogical}, suggesting solar-like, highly reducing settings for both \ce{^{16}O}-rich and \ce{^{16}O}-poor gaseous reservoirs in which these pyroxenes formed.
	
	We also observed clear evidence for the presence of \ce{Ti^{3+}} in  Al,Ti-rich diopside in the WL rim on R3C-01 (\ce{Ti^{3+}}/\ce{Ti^{tot}} $ \sim $ 0.4--1.0; \cref{fig:Px-comp,tab:R3C-01_px_EPMA}) There is no clear variation of \ce{Ti^{3+}}/\ce{Ti^{tot}} ratios between pyroxenes from interior and WL rim, as reported by J.~\citet{,simon2005short} and \citet{,dyl2011valence}, suggesting that the redox condition of a nebular gas remained highly reducing until the pyroxene layer formation. Occurrence of the LIME olivine at the outer edge of the WL rim (\cref{fig:MnO-FeO}) also supports highly reducing conditions during the WL rim formation, since the LIME olivine can condense only in a very reducing gas of solar composition \citep[e.g., $ \log f_{\ce{O2}} < \mathrm{IW}-4.17 \pm 0.06$ at total pressure of $ 10^{-4} $ bar;][]{,komatsu2015lime}. These results suggest that $ f_{\ce{O2}} $ of the nebular gas remains at solar-like redox condition \citep[$ \log f_{\ce{O2}} \sim  \mathrm{IW}-6.8$; e.g.,][]{,grossman2008redox} during the entire formation process of R3C-01. 
	
	CAIs likely have experienced outward transport after their formation near the proto-Sun \citep{,brownlee2006comet,mckeegan2006isotopic,zolensky2006mineralogy,simon2008refractory}, although the mechanism is still controversial. The X-wind model \citep[e.g.,][]{,shu1996toward} is one of the possible scenarios in which objects formed close to the Sun is transported by a bipolar outflow. However, as pointed out by \citet{,krot2009origin} and \citet{,desch2010critical}, in the X-wind model, high-temperature materials (e.g., CAIs) form in a dust-rich, hydrogen-depleted nebular setting, which is inconsistent with highly reducing condition during the entire R3C-01 formation inferred from our observation. Therefore, the X-wind model is unlikely and we support turbulent flow \citep[e.g.,][]{,ciesla2007outward} or disk winds \citep{,van2016isotopic} as the possible transportation mechanism of R3C-01.

	\subsection{Possible scenarios for formation of Ti,V-rich davisite}
	\label{sec:discussion-1_dav}
	
	Here we discuss two possible origin of \ce{^{16}O}-poor, Ti,V-rich davisite isolated in \ce{^{16}O}-poor, reversely-zoned melilite in R3C-01-U1 (\cref{fig:Sc-Px_map,fig:isotopograph}): condensation from a gas and relict of pre-existing CAIs. Melt origin of Ti,V-rich davisite can be ruled out since their V/Sc and Zr/Sc ratios are different from clinopyroxene/melt ratios \citep{,simon1991fassaite,hart1993experimental} and partial melting of R3C-01-U1 is unlikely (see \cref{sec:discussion-1_O-isotope_heterogeneity_partial_melting}).
	
	\subsubsection{Condensation origin}
	\label{sec:discussion-1_dav_condensation}
	
	As discussed in \cref{sec:discussion-1_O-isotope_heterogeneity_partial_melting}, the reversely-zoned melilite is likely to be a nebular condensate \citep{,macpherson1984fluffy}. Occurrence of davisite indicates its formation prior to the condensation of melilite. Although condensation temperature of davisite has never been constrained due to lack of knowledge of its thermodynamic properties, high concentration of very refractory elements such as Ti and Sc in davisite might have stabilized davisite at temperature higher than condensation temperature of diopside \citep[1347 K at $ 10^{-4} $ bar;][]{,lodders2003solar} and melilite (1529 K). The high condensation temperature of davisite is also inferred from common occurrence of davisite with UR phases in the UR CAIs in which lower temperature condensate such as melilite and diopside are not abundant \citep[see Table 1 of \citet{,ivanova2012compound}; e.g.,][]{,davis1984scandalously,simon1996unique,ma2009davisite,zhang2015mineralogical}.
	
	
	In the \ce{^{16}O}-poor region of R3C-01-U1, phases that are expected to condense at higher (e.g., perovskite, hibonite and UR element rich phases) and lower temperature (e.g., Ti-poor pyroxene) than melilite are absent (\cref{fig:isotopograph}). Spinel that possibly condenses in the similar temperature range with melilite and Al,Ti-rich pyroxene under equilibrium conditions is also absent. These results could be attributed to condensation from a gas of fractionated, non-solar composition and/or disequilibrium condensation due to isolation of earlier condensates \citep[e.g.,][]{,petaev1998condensation}.
	
	Ti,V-rich davisite in R3C-01-U1 shows higher V/Sc and lower Zr/Sc values compared to davisite in the UR inclusions previously studied (\cref{fig:Sc-Px_comp}). The 50\% condensation temperature of elements in a gas of solar composition at $ 10^{-4} $ bar are 1741 K for Zr, 1659 K for Sc, and 1429 K for V \citep{,lodders2003solar}, suggesting that the Zr/Sc ratio of a condensate decreases with increasing V/Sc value as a nebular gas temperature decreases. Therefore, the condensation temperature of Ti,V-rich davisite in R3C-01-U1 might be lower than davisite in the previously-reported UR inclusions. The relatively lower condensation temperature of Ti,V-rich davisite  in R3C-01-U1 and absence of any other high-temperature condensates in the inclusion suggest that higher temperature condensates had been isolated from a gas before the condensation of Ti,V-rich davisite. 
	
	\subsubsection{Relict origin}
	\label{sec:discussion-1_dav_relict}
	
	\citet{,lin2003fassaites} observed Sc- or V-rich Al,Ti-diopside grains isolated in melilite in compact type A CAIs and suggested that these grains are relicts of UR pyroxene grains from pre-existing UR inclusions \citep[e.g.,][]{,el2002efremovka}. In davisite-bearing UR CAIs reported so far, the mineral occurs with other refractory phases that are more refractory than Al,Ti-rich diopside and melilite, such as Zr- and/or Y-rich perovskite \citep{,ulyanov1982efremovka,davis1984scandalously,simon1996unique,el2002efremovka,ivanova2012compound}, hibonite \citep{,ulyanov1982efremovka,davis1984scandalously}, and/or other Zr-, Sc-, and/or Y-rich minerals \citep{,ma2009davisite,ma2011thortveitite,ma2012panguite,ma2013kangite,ma2017discovery,ivanova2012compound,ivanova2017oxygen,ma2012discovery,krot2015forsterite,zhang2015mineralogical,daly2017nebula,komatsu2018first}. Therefore, it is expected that davisite is associated with these highly refractory phases if the mineral is relict that originated in pre-existing UR inclusions. However, such highly refractory phases are completely absent in R3C-01-U1, suggesting that a relict origin of Ti,V-rich davisite in the CAI is unlikely.

	\subsection{Origin of spinel framboids}
	\label{sec:discussion-1_fram}
	
	In some cases, spinel occurs as "palisade bodies" and "framboids" in CAIs, whose origins remain controversial \citep[e.g.,][]{,el1979spinel,wark1982nature,simon1997situ,kim2002oxygen}. Palisade bodies are spheroidal shells of spinel enclosing typical CAI minerals such as melilite, spinel, Al,Ti-rich diopside and anorthite. Based on mineralogical investigations of natural occurrences and experimental works, \citet{,simon1997situ} suggested that they have formed \textit{in situ} in their host inclusions as a result of crystallization of spinel around the vapor-melt interfaces of vesicles. In contrast, \citet{,kim2002oxygen} showed that oxygen isotopic composition of palisade bodies and associated phases cannot be reconciled with this scenario and suggested an external origin of these bodies. Framboids are tightly packed spheroidal clusters of spinel enclosed within phases such as melilite, Al,Ti-rich diopside and anorthite. \citet{,el1979spinel} suggested that framboids are direct condensates from the nebular gas crystallized around pre-existing phases like pyroxene, melilite and anorthite. On the other hand, \citet{,simon1997situ} interpreted framboids as polar sections of palisade bodies and concluded that there is no genetic distinction between them. A furnace experimental study by \citet{,wark1982nature} showed that framboids likely formed by near-solidus annealing rather than liquid crystallization.
	
	Concentration of framboids only in one lithological unit (R3C-01-U1) of the compound CAI R3C-01 and absence of palisade bodies (\cref{fig:BSE_R3C-01}) suggest that formation of framboids predated the aggregation of R3C-01, and there is no genetic relationship between framboids and palisade bodies. Systematic variations in \ce{V2O3} contents of spinel (\cref{fig:V2O3}) and associated Ca,Ti-rich phases (perovskite $ \pm $ Al,Ti-rich pyroxene) among framboids cannot be explained by co-crystallization of all framboids in a single process. Therefore, \textit{in situ} formation scenario of framboids during annealing \citep{,wark1982nature} or melting \citep{,simon1997situ} that took place after aggregation of R3C-01 are unlikely for the origin of framboids in R3C-01-U1. The lack of Ti-V correlation in spinel is also inconsistent with the \textit{in situ} formation scenario \citep{,connolly2003petrogenesis,connolly2003type}. 
	
	The mineralogical and chemical variations among spinel framboids very likely reflect difference in formation conditions and/or chemical composition of precursor of each framboid. Distinct occurrence, chemistry and oxygen isotopic composition of melilite enclosed in and surrounding framboids (\cref{fig:isotopograph}b,d) also support separate and complex origin of framboids. Therefore, it is suggested that formation of spinel framboids took place separately prior to the formation of their host CAI R3C-01-U1. This scenario is consistent with the formation of R3C-01-U1 by aggregation of mineral assemblages formed separately in the solar nebula (see \cref{sec:discussion-1_O-isotope_heterogeneity}).
	
	The fine grain size of melilite enclosed in framboids, absence of glasses and textural and mineralogical zoning are distinct from known terrestrial and extraterrestrial igneous objects such as igneous CAIs and chondrules \citep[e.g.,][]{,wark1982nature}.  Although the spheroidal shape could have been achieved by partial melting before incorporation into R3C-01-U1, this process would have resulted in grain coarsening. Therefore, we prefer non-igneous, condensation origin for the framboids \citep{,el1979spinel}.  The oxygen isotopic disequilibrium between spinel and melilite inside framboids might have resulted from gas-solid or gas-melt interaction between originally \ce{^{16}O}-rich minerals in proto-framboids and \ce{^{16}O}-poor gaseous reservoir, or aggregation of isotopically distinct minerals. Systematically lower \ce{V2O3} content observed in framboids attached to the WL rim and the spinel layer in the rim sequence implies that the V-poor spinel formed later than V-rich spinel. Although V becomes less refractory at oxidizing conditions \citep{,kornacki1986abundance}, multiple observations indicate a highly reducing condition during the whole formation process of R3C-01-U1 (\cref{sec:discussion-1_redox}). We suggest that vanadium can increase the condensation temperature of spinel. 
	
	The origin of spheroidal shape of framboids is beyond the scope of this paper, but some irregularly-shaped framboids (e.g., framboid \#01; \cref{fig:BSE_R3C-01}) might be affected by deformation of R3C-01-U1. Since the thickness of the WL rim remains to be constant even regions near the irregularly-shaped framboids, it is suggested that the deformation event took place in the nebula before the WL rim formation. 

	\section{Conclusions}
	\label{sec:conclusions-1}
	
	The petrological, mineralogical and isotopic characteristics of  R3C-01 suggest the following formation history:
	\begin{enumerate}
		\item Three chemically distinct mineral assemblages (reversely-zoned or unzoned, \AA{}k-rich melilite + Ti,V-rich davisite; 
		unzoned, gehlenitic melilite + Al,Ti-rich diopside + spinel; and spinel framboids) formed and processed separately in the solar nebula under various physicochemical conditions. 
		They can originate in isotopically distinct gaseous reservoirs 
		(solar-like,  \ce{^{16}O}-rich reservoir and planetary-like, \ce{^{16}O}-poor one), or formed in the \ce{^{16}O}-rich reservoir and subsequently experienced solid-state modification of oxygen isotopic composition during subsolidus heating in the solar nebula before and/or during the aggregation of R3C-01-U1.
		\item Assemblages of \ce{^{16}O}-poor and \ce{^{16}O}-rich minerals aggregated to form R3C-01-U1. 
		\item R3C-01-U1 and four (or more) melilite-rich CAIs that formed separately aggregated to form the compound CAI R3C-01.
		\item 	A multi-layered Wark-Lovering (WL) rim condensed onto  R3C-01 from a cooling \ce{^{16}O}-rich nebular gas. The duration of the WL rim formation process was $ < 10^3 $ hours at 1400 K and $ < 10^5$ hours at 1100 K.
		\item All of the formation processes mentioned above occur under highly reducing conditions in which \ce{Ti^{3+}}-rich pyroxene and low-iron, manganese-enriched (LIME) olivine can be stable.
		\item R3C-01 was transported outward by the turbulent flow or the disk winds to the region of CV chondrite formation and incorporated into the RBT 04143 parent body.
	\end{enumerate}
	
	In this study, two scenarios for the origin of oxygen isotopic heterogeneity commonly observed in a variety of CAIs are proposed: i) solid-state isotopic exchange between \ce{^{16}O}-rich porous CAI precursors and \ce{^{16}O}-poor gaseous reservoir via grain boundary and/or surface diffusion followed by volume diffusion during subsolidus heating in the solar nebula, and ii) agglomeration of isotopically distinct mineral assemblages. These scenarios can be tested by experimental determination of oxygen diffusivity in davisite, which has not yet been determined. In either scenario, it is strongly indicated that spatially and/or temporally distinct \ce{^{16}O}-rich and \ce{^{16}O}-poor gaseous reservoirs existed in the period of UR phases and UR CAI formation.
	
	\section*{Acknowledgments}
	
	We thank Y. Ito, I. Narita and J. Muto for technical assistance. We are grateful to NASA/JSC for loan of RBT 04143. This work was supported by Grant-in-Aid for JSPS Research Fellow (No.~JP18J20708), GP-EES Research Grant and DIARE Research Grant to TY, and Korea Polar Research Institute project PE19230 to CP. SIMS analyses were carried out at the Isotope Imaging Laboratory, Creative Research Institution Sousei, Hokkaido University, Japan, which is supported by the Ministry of Education, Culture, Sports, Science and Technology (MEXT), Japan. TY gratefully acknowledges the Barringer Crater Company for travel grants to attend the Annual Meeting of the Meteoritical Society in 2017 and 2018 to present parts of this work.  We thank Associate Editor S. Russell and three anonymous reviewers for constructive and detailed comments that have helped improve this manuscript.
		
%
	
	\clearpage
	
	\begin{landscape}
		\begin{figure}[p]
			\centering
			\includegraphics[width=1\linewidth]{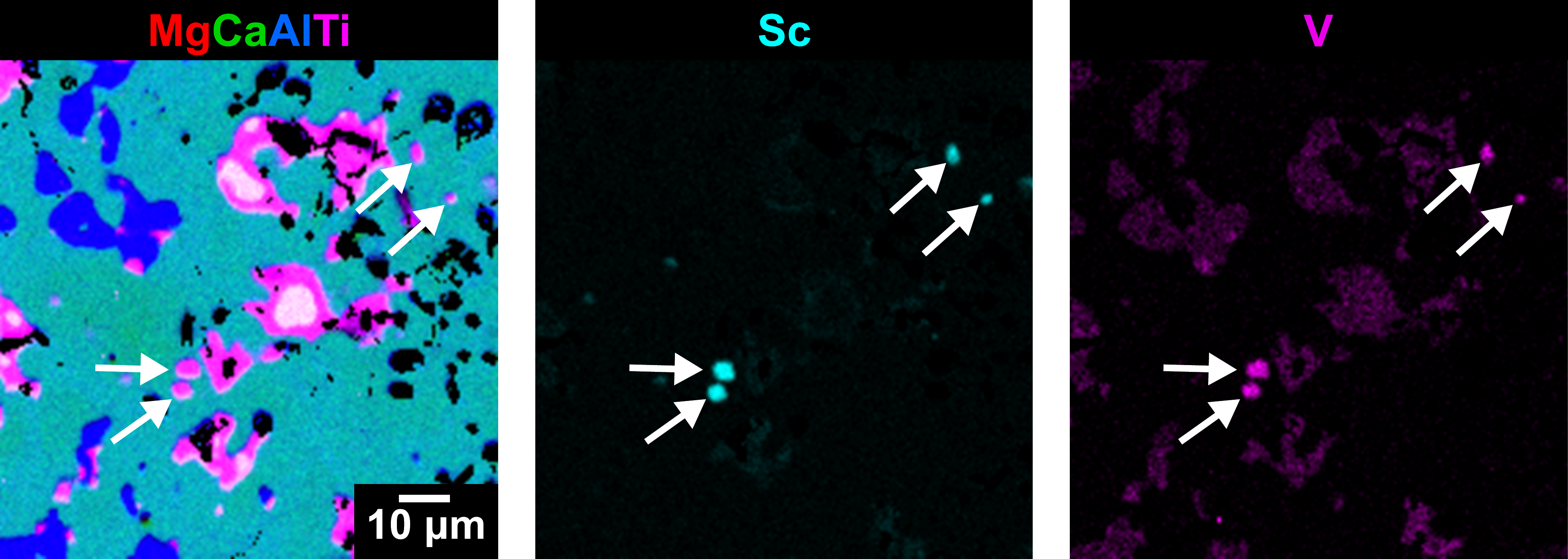}
			\caption[Occurrence of pyroxene in R3C-01-U1]
			{Combined X-ray map in Mg (red), Ca (green), Al (blue) and Ti (pink) and elemental maps in Sc and V $ K\alpha $ X-rays of R3C-01-U1 (area a in \cref{fig:RGBP_R3C-01}). Arrows show occurrences of Ti,V-rich davisite.
			}
			\label{fig:Sc-Px_map}
		\end{figure}
		\clearpage
	\end{landscape}
	
	\begin{figure}[p]
		\centering
		\includegraphics[width=1\linewidth]{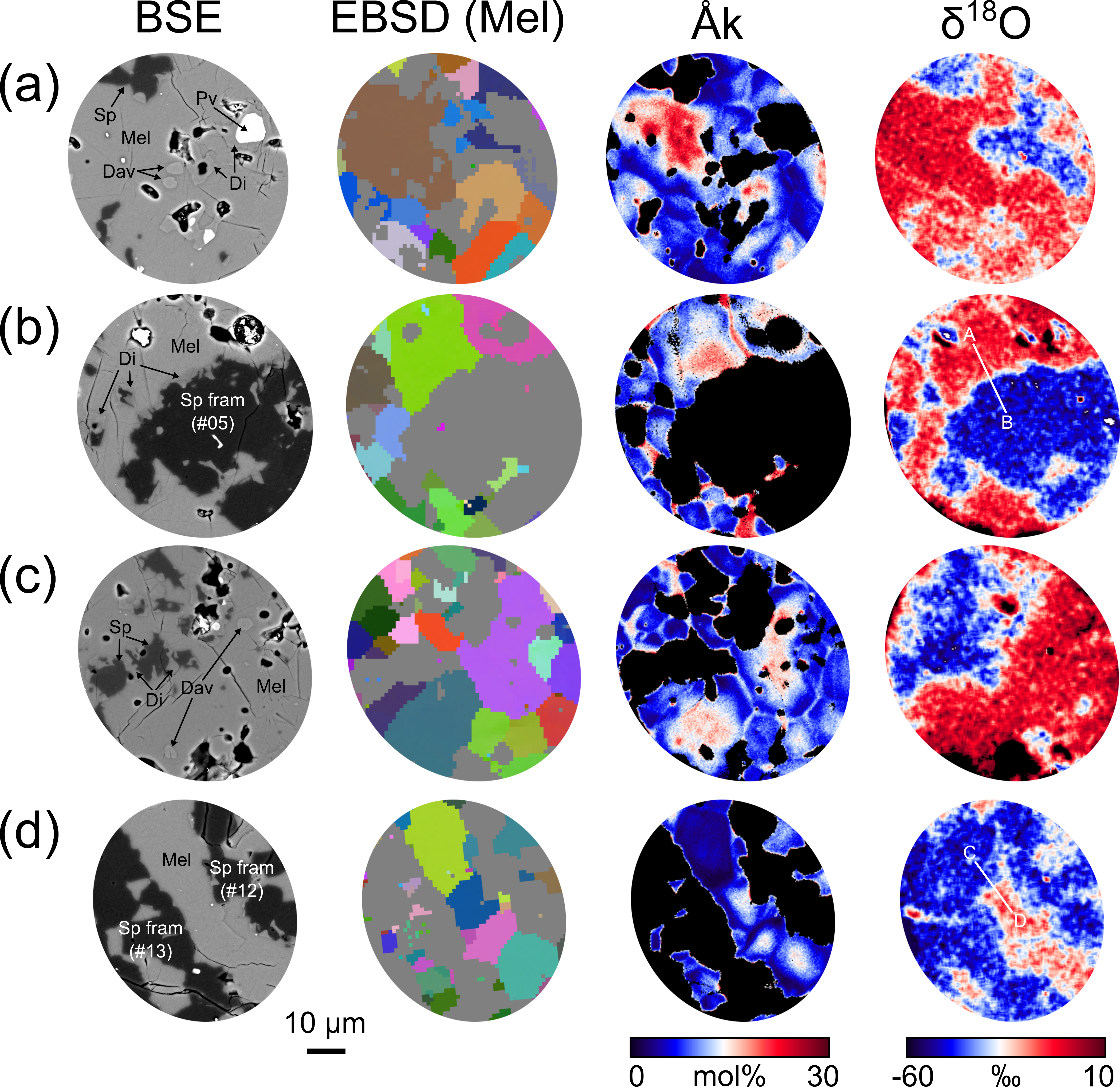}
		\caption[BSE image, EBSD map, \AA{}k map and $\delta$\ce{^{18}O} isotopograph of R3C-01-U1]
		{BSE image, EBSD crystal orientation map of melilite, \aa{}kermanite (\AA{}k) content map of melilite and $\delta$\ce{^{18}O} isotopograph for each analyzed areas. In EBSD maps, similar colors indicate similar crystal orientations. Line profiles along lines A--B and C--D are shown in \cref{fig:SCAPS_quality} and \cref{fig:SCAPS_line}, respectively. The $ \delta $\ce{^{18}O} value around pores and cracks could be affected by differences in topography of the sample surface \citep{,park2012oxygen_SIA}. Abbreviations: Dav--davisite; Di--diopside; Mel--melilite; Pv--perovskite; Sp--spinel; Sp fram--spinel framboid.
		}
		\label{fig:isotopograph}
	\end{figure}
	\clearpage
	
	\begin{figure}[p]
		\centering
		\includegraphics[width=1\linewidth]{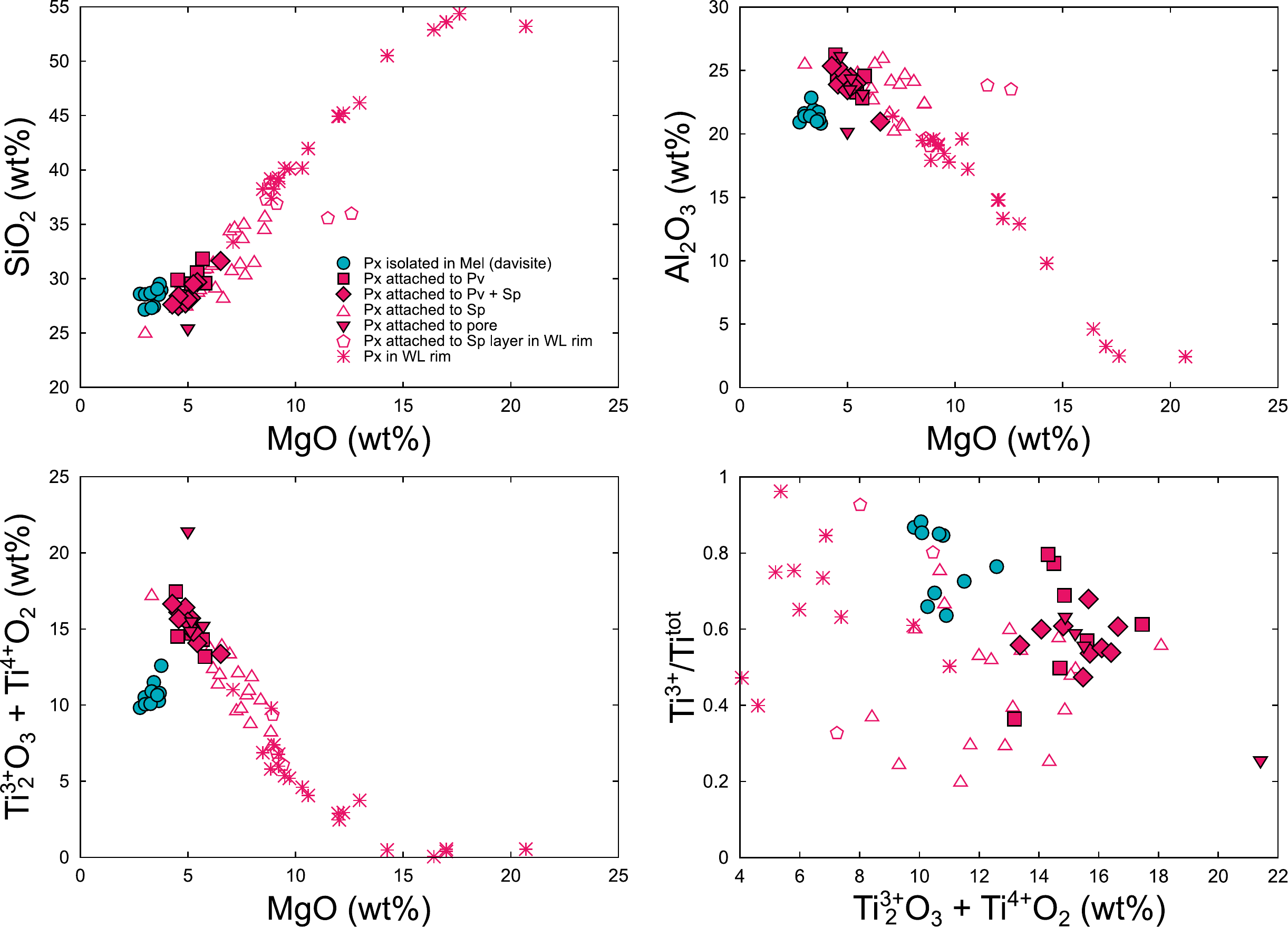}
		\caption[Major elemental composition of pyroxene in R3C-01-U1]
		{Major elemental composition of Al,Ti-rich pyroxene in R3C-01-U1.
			Note that quantitative analyses of Al,Ti-rich pyroxene attached to spinel described above are possibly contaminated by beam overlapping onto spinel due to tiny grain size of pyroxene, as seen in their elevated Al and Mg contents and lower  \ce{Ti^{3+}}/\ce{Ti^{tot}} ratios (open symbols). Importantly, measurements of pyroxene which is not attached to spinel (filled symbols or asterisks) do not show clear contamination by other phases.
			Abbreviations are as in \cref{fig:isotopograph}.
		}
		\label{fig:Px-comp}
	\end{figure}
	
	\begin{figure}[p]
		\centering
		\includegraphics[width=1\linewidth]{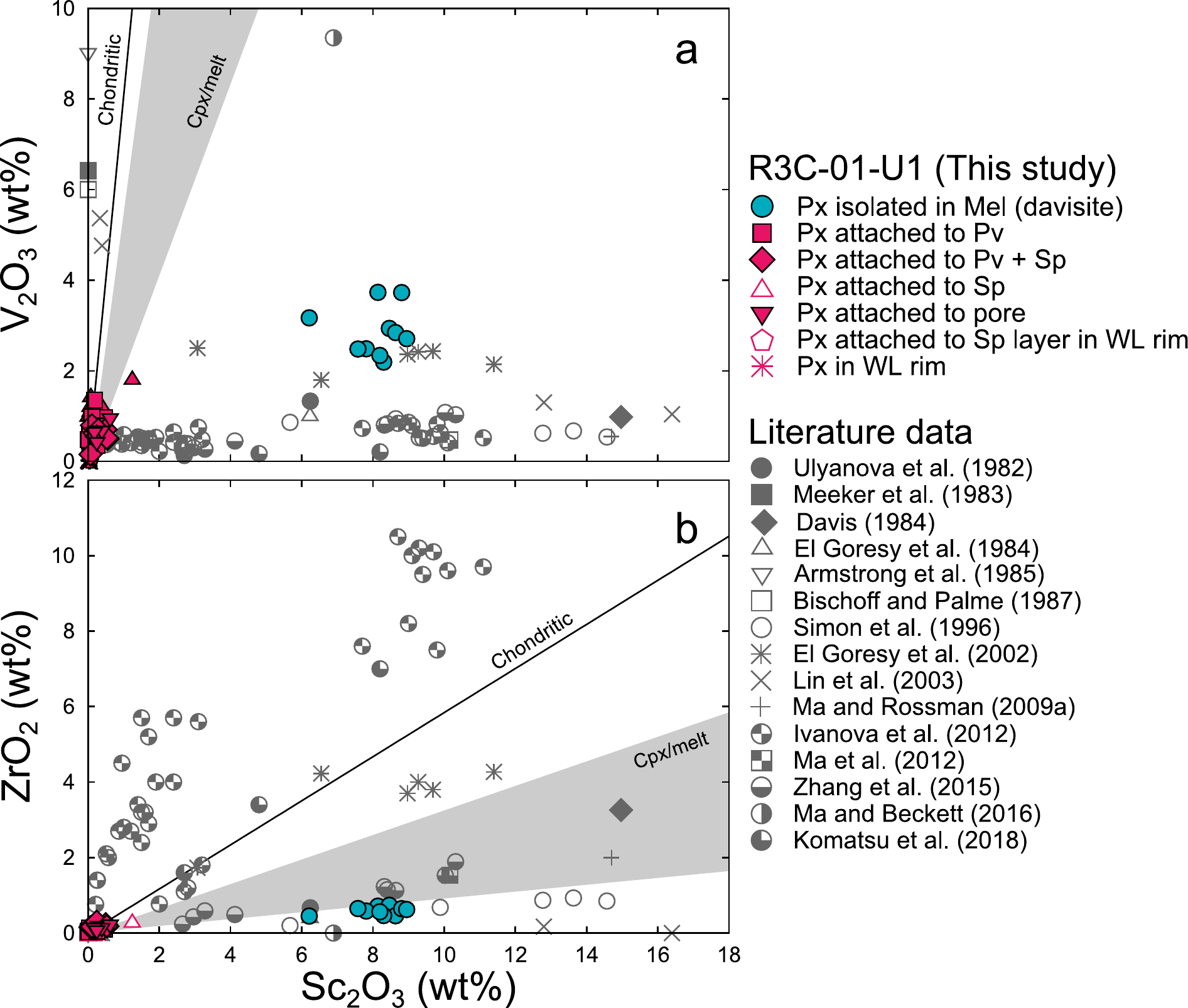}
		\caption[Sc, V and Zr composition of pyroxene in R3C-01-U1]
		{Plots of (a) \ce{V2O3} vs \ce{Sc2O3} and (b) \ce{ZrO2} vs \ce{Sc2O3} of pyroxene in R3C-01-U1. Compositions of Sc-, V-, and/or Zr-rich pyroxene from previous studies \citep{,ulyanov1982efremovka,meeker1983replacement,davis1984scandalously,el1984calcium,armstrong1985willy,bischoff1987composition,simon1996unique,el2002efremovka,lin2003fassaites,ma2009davisite,ivanova2012compound,ma2012panguite,zhang2015mineralogical,ma2016burnettite,komatsu2018first}
			and chondritic ratios \citep{,lodders2003solar} are also shown. Gray areas correspond clinopyroxene/melt fractionation ratio determined by previous studies \citep{,simon1991fassaite,hart1993experimental}.
			Abbreviations are as in \cref{fig:isotopograph}.
		}
		\label{fig:Sc-Px_comp}
	\end{figure}
	\clearpage

	\clearpage

	\clearpage
	
	\begin{figure}[p]
		\centering
		\includegraphics[width=1\linewidth]{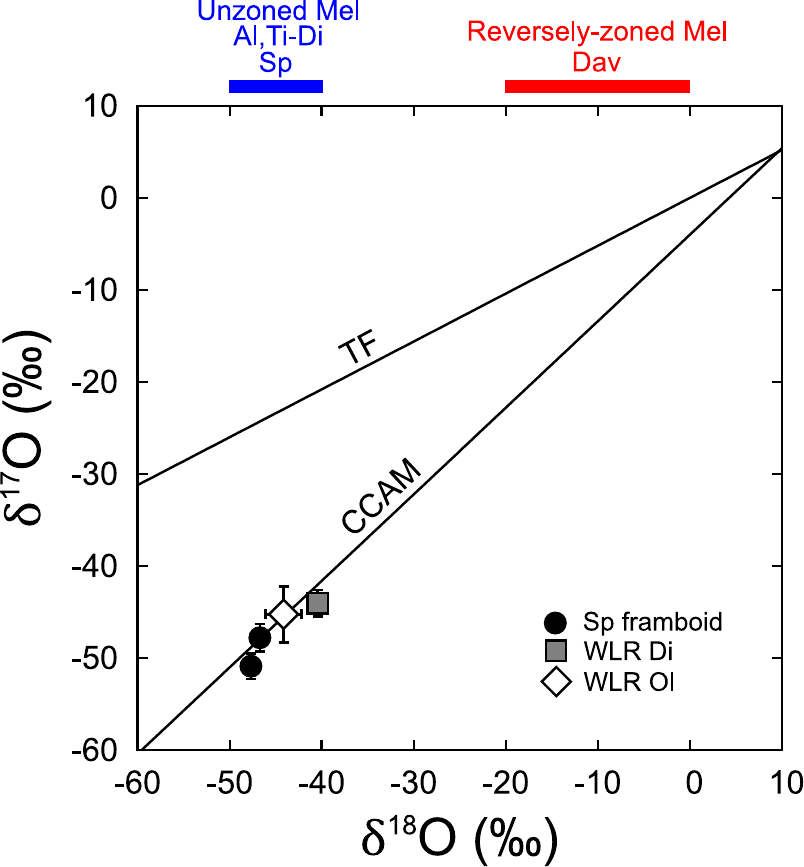}
		\caption[Spot oxygen isotopic composition of minerals in R3C-01-U1]
		{Oxygen three-isotope plots of minerals in R3C-01-U1 measured by spot analysis. Error bars are shown as 1$ \sigma $. The terrestrial fractionation (TF) line and carbonaceous chondrite anhydrous mineral (CCAM) line  \citep{,clayton1977distribution} are shown for references. $ \delta ^{18}$O values of CAI minerals determined by isotope imaging (\cref{fig:isotopograph}) are also shown. Abbreviations: Dav\textendash davisite; Di\textendash diopside; Mel\textendash melilite; Ol\textendash olivine; Sp\textendash spinel; WLR\textendash Wark-Lovering rim. 
		}
		\label{fig:SIMS_spot}
	\end{figure}
	\clearpage
	
	\begin{figure}[p]
		\centering
		\includegraphics[width=1\linewidth]{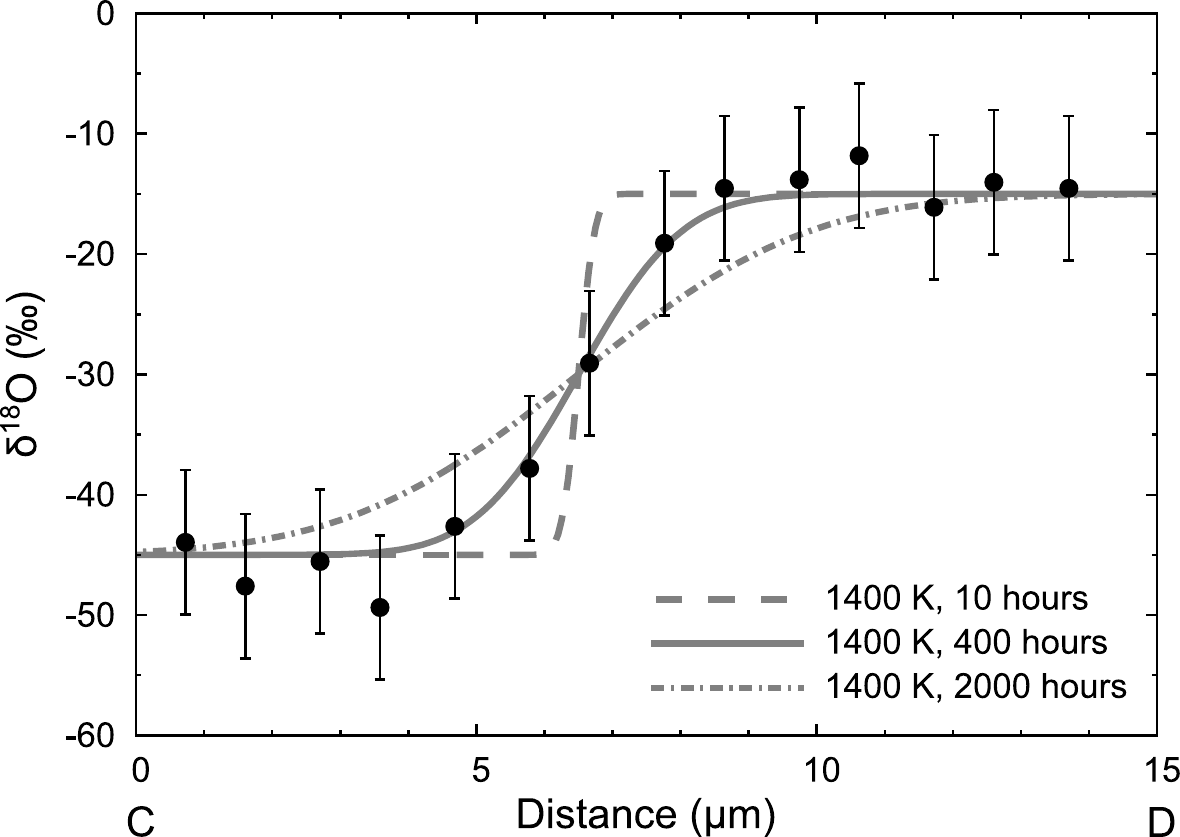}
		\caption{Line $\delta  $\ce{^{18}O} profile for the line C--D in \cref{fig:isotopograph}d. Error bars are shown as 1$ \sigma $.  Gray curves shows the modeled oxygen isotopic zoning in melilite grains heated at 1400 K for 10, 400 and 2000 hours, respectively.
		}
		\label{fig:SCAPS_line}
	\end{figure}
	\clearpage

	\begin{landscape}
		\begin{table}[h]
			\centering
			\begin{threeparttable}
				\caption{Summary of petrological and mineralogical characters used to distinguish five lithological units in R3C-01.}
				\label{tab:R3C-01_units}
				\begin{tabular}{cccccc}
					\toprule
					Unit  & 1 (R3C-01-U1) & 2     & 3     & 4     & 5 \\
					\midrule
					Size (mm) & 1.5 $ \times $ 0.8  & 0.7 $ \times $ 0.4 & 0.3 $ \times $ 0.3 & 0.3 $ \times $ 0.2 & 0.3 $ \times $ 0.2 \\
					Shape & Irregular & Irregular & Subrounded & Subrounded & Nodular \\
					Core-mantle structure & $ - $     & $ - $     & +     & $ - $     & $ - $ \\
					Typical grain size of melilite ($ \mu $m) & 20--50 & $ \sim $20   & $ \sim $50 (core); $ \sim $15 (mantle) & $ \sim $10   & $ \sim $50 \\
					Davisite & +     & $ - $     & $ - $     & $ - $     & $ - $ \\
					Spinel framboids & ++    & +     & +     & $ - $     & $ - $ \\
					Irregular-shaped Al,Ti-rich pyroxene & ++    & $ - $     & +     & ++    & + \\
					Reversely-zoned melilite & + & $ - $     & +  & $ - $     & +  \\
					Perovskite & ++    & +     & ++    & $ - $     & $ - $ \\
					&       &       &       &       &  \\
					\AA{}k (mol\%) in melilite & 5--30 & 5--25 & 5--30 & 5--15 & 10--30 \\
					\ce{V2O3} (wt\%) in spinel & 0.1--0.8 & 0.3--0.4 & 0.3--0.4 & 0.1--0.2 & $ \sim $0.2 \\
					\ce{ZrO2} (wt\%) in perovskite & 0.15--0.35 & 0.10  & 0.05--0.15  & $ - $    & $ - $ \\
					\ce{V2O3} (wt\%) in perovskite & 0.1--0.8 & NA    & 0.7--1.3 & $ - $   & $ - $ \\
					\bottomrule
				\end{tabular}
				\begin{tablenotes}[flushleft]
					\item[] NA--not available.
				\end{tablenotes}
			\end{threeparttable}
		\end{table}
	\end{landscape}
	
	\begin{table}[p]
		\centering
		\begin{threeparttable}
			\footnotesize
			\caption[Major element composition of Al,Ti-rich pyroxene in R3C-01-U1]{Major element composition of Al,Ti-rich pyroxene in R3C-01-U1.}
			\label{tab:R3C-01_px_EPMA}
			\begin{tabular}{lrrrrrrrrrrrrr}
				\toprule
				wt\% & 77\tnotex{tab:Px-a} & 163\tnotex{tab:Px-a}   & 86\tnotex{tab:Px-b}  & 84\tnotex{tab:Px-b}   & 103\tnotex{tab:Px-c}   & 178\tnotex{tab:Px-c} & 75\tnotex{tab:Px-d}  & 172\tnotex{tab:Px-d}  & 283\tnotex{tab:Px-e} & 366\tnotex{tab:Px-e} & 13\tnotex{tab:Px-f}   & 14\tnotex{tab:Px-f}  & 23\tnotex{tab:Px-f} \\
				\midrule
				\ce{SiO2} & 28.4  & 29.5  & 27.7  & 27.6  & 24.9  & 29.1  & 28.6  & 29.8  & 38.9  & 37.3  & 37.4  & 42.0  & 52.9 \\
				\ce{TiO^{tot}_2} & 16.6  & 15.5  & 17.4  & 17.7  & 19.0  & 15.2  & 16.4  & 16.2  & 8.4   & 11.0  & 10.4  & 4.3   & 0.1 \\
				\ce{Ti2O3} & 8.5   & 6.9   & 8.4   & 9.7   & 10.1  & 5.8   & 8.2   & 8.6   & 7.4   & 8.4   & 5.7   & 1.8   & n.a. \\
				\ce{TiO2} & 7.1   & 7.8   & 8.0   & 7.0   & 7.7   & 8.8   & 7.3   & 6.6   & 0.2   & 1.7   & 4.1   & 2.3   & 0.1 \\
				\ce{Al2O3} & 24.5  & 24.2  & 24.5  & 25.3  & 25.6  & 25.7  & 24.4  & 23.2  & 19.0  & 19.6  & 17.9  & 17.2  & 4.6 \\
				\ce{Cr2O3} & 0.0   & 0.0   & 0.0   & 0.0   & 0.1   & 0.0   & 0.1   & 0.0   & 0.1   & b.d.  & 0.1   & 0.0   & 0.0 \\
				\ce{FeO} & 0.0   & 0.1   & b.d.  & 0.1   & 0.0   & 0.1   & b.d.  & b.d.  & 0.1   & 0.1   & b.d.  & 0.1   & b.d. \\
				\ce{MnO} & b.d.  & b.d.  & b.d.  & b.d.  & b.d.  & b.d.  & b.d.  & b.d.  & b.d.  & 0.0   & 0.0   & b.d.  & b.d. \\
				\ce{MgO} & 4.8   & 5.2   & 4.9   & 4.3   & 3.0   & 6.3   & 5.2   & 5.7   & 8.8   & 8.6   & 8.9   & 10.6  & 16.4 \\
				\ce{CaO} & 24.9  & 25.9  & 24.8  & 24.8  & 24.6  & 24.5  & 24.7  & 24.5  & 24.1  & 23.9  & 25.4  & 25.7  & 25.9 \\
				\ce{Na2O} & b.d.  & b.d.  & 0.0   & b.d.  & 0.0   & 0.0   & b.d.  & 0.1   & b.d.  & 0.0   & b.d.  & 0.1   & 0.1 \\
				\ce{V2O3} & 1.0   & 0.6   & 0.5   & 0.5   & 1.8   & 0.5   & 1.0   & 0.7   & 0.4   & 0.2   & 0.6   & 0.3   & 0.0 \\
				\ce{ZrO2} & 0.1   & 0.1   & 0.2   & 0.2   & 0.3   & 0.1   & 0.2   & 0.1   & 0.0  & 0.1   & 0.0   & b.d.  & 0.0 \\
				\ce{Sc2O3} & 0.5   & 0.0   & 0.5   & 0.5   & 1.2   & 0.0   & 0.6   & 0.2   & 0.1   & 0.0   & 0.2   & 0.1   & 0.0 \\
				\ce{Y2O3} & b.d.  & b.d.  & 0.0   & b.d.  & b.d.  & b.d.  & b.d.  & b.d.  & b.d.  & b.d.  & 0.1   & b.d.  & b.d. \\
				Total & 99.8  & 100.5 & 99.6  & 100.0 & 99.4  & 101.0 & 100.3 & 99.5  & 99.1  & 100.1 & 100.5 & 100.2 & 100.2 \\
				\multicolumn{14}{l}{\textit{Cations per 24 oxygen}}                                                                              \\
				Si    & 1.1   & 1.1   & 1.1   & 1.1   & 1.0   & 1.1   & 1.1   & 1.1   & 1.4   & 1.5   & 1.4   & 1.5   & 1.9 \\
				\ce{Ti^{3+}} & 0.3   & 0.2   & 0.3   & 0.3   & 0.3   & 0.2   & 0.3   & 0.3   & 0.2   & 0.3   & 0.2   & 0.1   & n.a. \\
				\ce{Ti^{4+}} & 0.2   & 0.2   & 0.2   & 0.2   & 0.2   & 0.2   & 0.2   & 0.2   & 0.0   & 0.0   & 0.1   & 0.1   & 0.0 \\
				Al    & 1.1   & 1.1   & 1.1   & 1.1   & 1.2   & 1.1   & 1.1   & 1.0   & 0.8   & 0.9   & 0.8   & 0.7   & 0.2 \\
				Cr    & 0.0   & 0.0   & 0.0   & 0.0   & 0.0   & 0.0   & 0.0   & 0.0   & 0.0   & n.a.  & 0.0   & 0.0   & 0.0 \\
				Fe    & 0.0   & 0.0   & n.a.  & 0.0   & 0.0   & 0.0   & n.a.  & n.a.  & 0.0   & 0.0   & n.a.  & 0.0   & n.a. \\
				Mn    & n.a.  & n.a.  & n.a.  & n.a.  & n.a.  & n.a.  & n.a.  & n.a.  & n.a.  & 0.0   & 0.0   & n.a.  & n.a. \\
				Mg    & 0.3   & 0.3   & 0.3   & 0.2   & 0.2   & 0.3   & 0.3   & 0.3   & 0.5   & 0.5   & 0.5   & 0.6   & 0.9 \\
				Ca    & 1.0   & 1.0   & 1.0   & 1.0   & 1.0   & 1.0   & 1.0   & 1.0   & 1.0   & 1.0   & 1.0   & 1.0   & 1.0 \\
				Na    & n.a.  & n.a.  & 0.0   & n.a.  & 0.0   & 0.0   & n.a.  & 0.0   & n.a.   & 0.0   & n.a.  & 0.0   & 0.0 \\
				V     & 0.0   & 0.0   & 0.0   & 0.0   & 0.1   & 0.0   & 0.0   & 0.0   & 0.0   & 0.0   & 0.0   & 0.0   & 0.0 \\
				Zr    & 0.0   & 0.0   & 0.0   & 0.0   & 0.0   & 0.0   & 0.0   & 0.0   & 0.0   & 0.0   & 0.0   & n.a.  & 0.0 \\
				Sc    & 0.0   & 0.0   & 0.0   & 0.0   & 0.0   & 0.0   & 0.0   & 0.0   & 0.0   & 0.0   & 0.0   & 0.0   & 0.0 \\
				Y     & n.a.  & n.a.  & 0.0   & n.a.  & n.a.  & n.a.  & n.a.  & n.a.  & n.a.  & n.a.  & 0.0   & n.a.  & n.a. \\
				\ce{Ti^{3+}}/\ce{Ti^{tot}} & 0.57  & 0.50  & 0.54  & 0.61  & 0.59  & 0.42  & 0.56  & 0.59  & 0.97  & 0.84  & 0.61  & 0.47  & n.a. \\
				\bottomrule
			\end{tabular}%
			\begin{tablenotes}[para,flushleft]
				\scriptsize
				\item[a] Px attached to Pv; \label{tab:Px-a} \item[b] Px attached to Pv + Sp; \label{tab:Px-b} \item[c] Px attached to Sp; \label{tab:Px-c} \item[d] Px attached to pore; \label{tab:Px-d} \item[e] Px WL Sp layer in WL rim; \label{tab:Px-e} \item[f] Px in WL rim.;\label{tab:Px-f} \item b.d.-- below detection limit; \item n.a.-- not available.
			\end{tablenotes}
		\end{threeparttable}
	\end{table}%
	
	\begin{table}[p]
		\centering
		\begin{threeparttable}
			\footnotesize
			\caption[Major element composition of davisite in R3C-01-U1]{Major element composition of davisite in R3C-01-U1.}
			\label{tab:R3C-01_dav_EPMA}
			\begin{tabular}{lrrrrrrrrrr}
				\toprule
				wt\% & 226   & 227   & 228   & 229   & 230   & 293   & 294   & 295   & 296   & 297 \\
				\midrule
				\ce{SiO2} & 27.2  & 28.9  & 27.4  & 27.3  & 28.5  & 29.5  & 28.6  & 28.6  & 28.7  & 29.0 \\
				\ce{TiO^{tot}_2} & 11.3  & 13.6  & 12.4  & 11.6  & 11.0  & 11.8  & 10.8  & 11.0  & 11.0  & 11.7 \\
				\ce{Ti2O3} & 7.1   & 9.4   & 8.1   & 6.7   & 6.5   & 9.0   & 8.4   & 8.8   & 8.5   & 8.9 \\
				\ce{TiO2} & 3.4   & 3.2   & 3.4   & 4.2   & 3.7   & 1.8   & 1.4   & 1.3   & 1.6   & 1.7 \\
				\ce{Al2O3} & 21.6  & 20.8  & 21.9  & 22.8  & 21.7  & 21.1  & 20.9  & 21.4  & 21.4  & 21.0 \\
				\ce{Cr2O3} & 0.1   & 0.1   & 0.2   & b.d.  & 0.1   & 0.1   & 0.1   & 0.2   & b.d.  & 0.1 \\
				\ce{FeO} & 0.1   & b.d.  & 0.1   & 0.1   & 0.1   & 0.1   & b.d.  & b.d.  & 0.1   & b.d. \\
				\ce{MnO} & b.d.  & b.d.  & b.d.  & b.d.  & b.d.  & b.d.  & b.d.  & 0.1   & b.d.  & b.d. \\
				\ce{MgO} & 3.0   & 3.8   & 3.4   & 3.3   & 3.6   & 3.7   & 2.8   & 3.0   & 3.3   & 3.6 \\
				\ce{CaO} & 23.8  & 24.2  & 23.4  & 24.1  & 24.3  & 23.9  & 24.1  & 23.6  & 23.5  & 23.6 \\
				\ce{Na2O} & b.d.  & b.d.  & b.d.  & b.d.  & b.d.  & b.d.  & b.d.  & b.d.  & b.d.  & b.d. \\
				\ce{V2O3} & 3.7   & 3.2   & 2.5   & 2.9   & 2.8   & 2.5   & 3.7   & 2.7   & 2.2   & 2.3 \\
				\ce{ZrO2} & 0.7   & 0.5   & 0.6   & 0.7   & 0.5   & 0.6   & 0.6   & 0.6   & 0.5   & 0.6 \\
				\ce{Sc2O3} & 8.1   & 6.2   & 7.8   & 8.5   & 8.6   & 7.6   & 8.8   & 8.9   & 8.3   & 8.2 \\
				\ce{Y2O3} & b.d.  & b.d.  & b.d.  & b.d.  & b.d.  & b.d.  & b.d.  & b.d.  & b.d.  & b.d. \\
				Total & 99.0  & 100.3 & 98.8  & 100.8 & 100.6 & 99.9  & 99.5  & 99.2  & 98.0  & 99.1 \\
				\multicolumn{11}{l}{\textit{Cations per 24 oxygen}}    \\
				Si    & 1.1   & 1.1   & 1.1   & 1.1   & 1.1   & 1.1   & 1.1   & 1.1   & 1.1   & 1.1 \\
				\ce{Ti^{3+}} & 0.2   & 0.3   & 0.3   & 0.2   & 0.2   & 0.3   & 0.3   & 0.3   & 0.3   & 0.3 \\
				\ce{Ti^{4+}} & 0.1   & 0.1   & 0.1   & 0.1   & 0.1   & 0.1   & 0.0   & 0.0   & 0.0   & 0.1 \\
				Al    & 1.0   & 0.9   & 1.0   & 1.0   & 1.0   & 1.0   & 1.0   & 1.0   & 1.0   & 1.0 \\
				Cr    & 0.0   & 0.0   & 0.0   & n.a.  & 0.0   & 0.0   & 0.0   & 0.0   & n.a.  & 0.0 \\
				Fe    & 0.0   & n.a.  & 0.0   & 0.0   & 0.0   & 0.0   & n.a.  & n.a.  & 0.0   & n.a. \\
				Mn    & n.a.  & n.a.  & n.a.  & n.a.  & n.a.  & n.a.  & n.a.  & 0.0   & n.a.  & n.a. \\
				Mg    & 0.2   & 0.2   & 0.2   & 0.2   & 0.2   & 0.2   & 0.2   & 0.2   & 0.2   & 0.2 \\
				Ca    & 1.0   & 1.0   & 1.0   & 1.0   & 1.0   & 1.0   & 1.0   & 1.0   & 1.0   & 1.0 \\
				Na    & n.a.  & n.a.  & n.a.  & n.a.  & n.a.  & n.a.  & n.a.  & n.a.  & n.a.  & n.a. \\
				V     & 0.1   & 0.1   & 0.1   & 0.1   & 0.1   & 0.1   & 0.1   & 0.1   & 0.1   & 0.1 \\
				Zr    & 0.0   & 0.0   & 0.0   & 0.0   & 0.0   & 0.0   & 0.0   & 0.0   & 0.0   & 0.0 \\
				Sc    & 0.3   & 0.2   & 0.3   & 0.3   & 0.3   & 0.3   & 0.3   & 0.3   & 0.3   & 0.3 \\
				Y     & n.a.  & n.a.  & n.a.  & n.a.  & n.a.  & n.a.  & n.a.  & n.a.  & n.a.  & n.a. \\
				\ce{Ti^{3+}}/\ce{Ti^{tot}} & 0.70  & 0.76  & 0.73  & 0.64  & 0.66  & 0.85  & 0.87  & 0.88  & 0.85  & 0.85 \\
				\bottomrule
			\end{tabular}
			\begin{tablenotes}[flushleft]
				\item[] b.d.-- below detection limit; n.a.-- not available. \label{note}
			\end{tablenotes}
		\end{threeparttable}
	\end{table}
	
	\clearpage
	
	\bibliography{myrefs}

\begin{thebibliography}{152}
\expandafter\ifx\csname natexlab\endcsname\relax\def\natexlab#1{#1}\fi
\providecommand{\url}[1]{\texttt{#1}}
\providecommand{\href}[2]{#2}
\providecommand{\path}[1]{#1}
\providecommand{\DOIprefix}{doi:}
\providecommand{\ArXivprefix}{arXiv:}
\providecommand{\URLprefix}{URL: }
\providecommand{\Pubmedprefix}{pmid:}
\providecommand{\doi}[1]{\href{http://dx.doi.org/#1}{\path{#1}}}
\providecommand{\Pubmed}[1]{\href{pmid:#1}{\path{#1}}}
\providecommand{\bibinfo}[2]{#2}
\ifx\xfnm\relax \def\xfnm[#1]{\unskip,\space#1}\fi
\bibitem[{Al{\'e}on(2018)}]{aleon2018closed}
\bibinfo{author}{Al{\'e}on, J.}, \bibinfo{year}{2018}.
\newblock \bibinfo{title}{{Closed system oxygen isotope redistribution in
  igneous CAIs upon spinel dissolution}}.
\newblock \bibinfo{journal}{Earth and Planetary Science Letters}
  \bibinfo{volume}{482}, \bibinfo{pages}{324--333}.
\newblock \DOIprefix\doi{10.1016/j.epsl.2017.11.027}.
\bibitem[{Al{\'e}on et~al.(2002)Al{\'e}on, Krot and
  McKeegan}]{aleon2002calcium}
\bibinfo{author}{Al{\'e}on, J.}, \bibinfo{author}{Krot, A.N.},
  \bibinfo{author}{McKeegan, K.D.}, \bibinfo{year}{2002}.
\newblock \bibinfo{title}{{Calcium-aluminum-rich inclusions and amoeboid
  olivine aggregates from the CR carbonaceous chondrites}}.
\newblock \bibinfo{journal}{Meteoritics \& Planetary Science}
  \bibinfo{volume}{37}, \bibinfo{pages}{1729--1755}.
\newblock \DOIprefix\doi{10.1111/j.1945-5100.2002.tb01160.x}.
\bibitem[{Al{\'e}on et~al.(2005)Al{\'e}on, Krot, McKeegan, MacPherson and
  Ulyanov}]{aleon2005fine}
\bibinfo{author}{Al{\'e}on, J.}, \bibinfo{author}{Krot, A.N.},
  \bibinfo{author}{McKeegan, K.D.}, \bibinfo{author}{MacPherson, G.J.},
  \bibinfo{author}{Ulyanov, A.A.}, \bibinfo{year}{2005}.
\newblock \bibinfo{title}{{Fine-grained, spinel-rich inclusions from the
  reduced CV chondrite Efremovka: II. Oxygen isotopic compositions}}.
\newblock \bibinfo{journal}{Meteoritics \& Planetary Science}
  \bibinfo{volume}{40}, \bibinfo{pages}{1043--1058}.
\newblock \DOIprefix\doi{10.1111/j.1945-5100.2005.tb00172.x}.
\bibitem[{Al{\'e}on et~al.(2018)Al{\'e}on, Marin-Carbonne, McKeegan and
  El~Goresy}]{aleon2018o}
\bibinfo{author}{Al{\'e}on, J.}, \bibinfo{author}{Marin-Carbonne, J.},
  \bibinfo{author}{McKeegan, K.D.}, \bibinfo{author}{El~Goresy, A.},
  \bibinfo{year}{2018}.
\newblock \bibinfo{title}{{O, Mg, and Si isotope distributions in the complex
  ultrarefractory CAI Efremovka 101.1: Assimilation of ultrarefractory, FUN,
  and regular CAI precursors}}.
\newblock \bibinfo{journal}{Geochimica et Cosmochimica Acta}
  \bibinfo{volume}{232}, \bibinfo{pages}{48--81}.
\newblock \DOIprefix\doi{10.1016/j.gca.2018.04.001}.
\bibitem[{Armstrong et~al.(1985)Armstrong, El~Goresy and
  Wasserburg}]{armstrong1985willy}
\bibinfo{author}{Armstrong, J.T.}, \bibinfo{author}{El~Goresy, A.},
  \bibinfo{author}{Wasserburg, G.J.}, \bibinfo{year}{1985}.
\newblock \bibinfo{title}{{Willy: A prize noble Ur-Fremdling--Its history and
  implications for the formation of Fremdlinge and CAI}}.
\newblock \bibinfo{journal}{Geochimica et Cosmochimica Acta}
  \bibinfo{volume}{49}, \bibinfo{pages}{1001--1022}.
\newblock \DOIprefix\doi{10.1016/0016-7037(85)90315-1}.
\bibitem[{Beckett(1986)}]{beckett1986origin}
\bibinfo{author}{Beckett, J.R.}, \bibinfo{year}{1986}.
\newblock \bibinfo{title}{{The origin of calcium-, aluminum-rich inclusions
  from carbonaceous chondrites: An experimental study}}.
\newblock Ph.D. thesis. University of Chicago.
\bibitem[{Bischoff and Palme(1987)}]{bischoff1987composition}
\bibinfo{author}{Bischoff, A.}, \bibinfo{author}{Palme, H.},
  \bibinfo{year}{1987}.
\newblock \bibinfo{title}{{Composition and mineralogy of refractory-metal-rich
  assemblages from a Ca, Al-rich inclusion in the Allende meteorite}}.
\newblock \bibinfo{journal}{Geochimica et Cosmochimica Acta}
  \bibinfo{volume}{51}, \bibinfo{pages}{2733--2748}.
\newblock \DOIprefix\doi{0.1016/0016-7037(87)90153-0}.
\bibitem[{Bod{\'e}nan et~al.(2014)Bod{\'e}nan, Starkey, Russell, Wright and
  Franchi}]{bodenan2014oxygen}
\bibinfo{author}{Bod{\'e}nan, J.D.}, \bibinfo{author}{Starkey, N.A.},
  \bibinfo{author}{Russell, S.S.}, \bibinfo{author}{Wright, I.P.},
  \bibinfo{author}{Franchi, I.A.}, \bibinfo{year}{2014}.
\newblock \bibinfo{title}{{An oxygen isotope study of Wark-Lovering rims on
  type A CAIs in primitive carbonaceous chondrites}}.
\newblock \bibinfo{journal}{Earth and Planetary Science Letters}
  \bibinfo{volume}{401}, \bibinfo{pages}{327--336}.
\newblock \DOIprefix\doi{10.1016/j.epsl.2014.05.035}.
\bibitem[{Bolser et~al.(2016)Bolser, Zega, Asaduzzaman, Bringuier, Simon,
  Grossman, Thompson and Domanik}]{bolser2016microstructural}
\bibinfo{author}{Bolser, D.}, \bibinfo{author}{Zega, T.J.},
  \bibinfo{author}{Asaduzzaman, A.}, \bibinfo{author}{Bringuier, S.},
  \bibinfo{author}{Simon, S.B.}, \bibinfo{author}{Grossman, L.},
  \bibinfo{author}{Thompson, M.S.}, \bibinfo{author}{Domanik, K.J.},
  \bibinfo{year}{2016}.
\newblock \bibinfo{title}{{Microstructural analysis of Wark-Lovering rims in
  the Allende and Axtell CV3 chondrites: Implications for high-temperature
  nebular processes}}.
\newblock \bibinfo{journal}{Meteoritics \& Planetary Science}
  \bibinfo{volume}{51}, \bibinfo{pages}{743--756}.
\newblock \DOIprefix\doi{10.1111/maps.12620}.
\bibitem[{Bowen(1914)}]{bowen1914ternary}
\bibinfo{author}{Bowen, N.L.}, \bibinfo{year}{1914}.
\newblock \bibinfo{title}{{The ternary system: diopside--forsterite--silica}}.
\newblock \bibinfo{journal}{American Journal of Science} \bibinfo{volume}{38},
  \bibinfo{pages}{207--264}.
\newblock \DOIprefix\doi{10.2475/ajs.s4-38.225.207}.
\bibitem[{Boynton(1975)}]{boynton1975fractionation}
\bibinfo{author}{Boynton, W.V.}, \bibinfo{year}{1975}.
\newblock \bibinfo{title}{{Fractionation in the solar nebula: Condensation of
  yttrium and the rare earth elements}}.
\newblock \bibinfo{journal}{Geochimica et Cosmochimica Acta}
  \bibinfo{volume}{39}, \bibinfo{pages}{569--584}.
\newblock \DOIprefix\doi{10.1016/0016-7037(75)90003-4}.
\bibitem[{Brearley and Jones(1998)}]{brearley1998chondritic}
\bibinfo{author}{Brearley, A.J.}, \bibinfo{author}{Jones, R.H.},
  \bibinfo{year}{1998}.
\newblock \bibinfo{title}{{Chondritic meteorites}}, in:
  \bibinfo{editor}{Papike, J.J.} (Ed.), \bibinfo{booktitle}{Reviews in
  Mineralogy}. \bibinfo{publisher}{Mineralogical Society of America}.
  volume~\bibinfo{volume}{36}, pp. \bibinfo{pages}{3--1--3--398}.
\newblock \DOIprefix\doi{10.2138/rmg.1999.36.3}.
\bibitem[{Brearley and Krot(2013)}]{brearley2013metasomatism}
\bibinfo{author}{Brearley, A.J.}, \bibinfo{author}{Krot, A.N.},
  \bibinfo{year}{2013}.
\newblock \bibinfo{title}{{Metasomatism in the early solar system: The record
  from chondritic meteorites}}, in: \bibinfo{editor}{Harlov, D.},
  \bibinfo{editor}{Austrheim, H.} (Eds.), \bibinfo{booktitle}{Metasomatism and
  the Chemical Transformation of Rock: The Role of Fluids in Terrestrial and
  Extraterrestrial Processes}. \bibinfo{publisher}{Springer}, pp.
  \bibinfo{pages}{659--789}.
\newblock \DOIprefix\doi{10.1007/978-3-642-28394-9}.
\bibitem[{Brownlee et~al.(2006)Brownlee, Tsou, Al{\'e}on, Alexander, Araki,
  Bajt, Baratta, Bastien, Bland, Bleuet, Borg, Bradley, Brearley, Brenker,
  Brennan, Bridges, Browning, Brucato, Bullock, Burchell, Busemann,
  Butterworth, Chaussidon, Cheuvront, Chi, Cintala, Clark, Clemett, Cody,
  Colangeli, Cooper, Cordier, Daghlian, Dai, D{\textquoteright}Hendecourt,
  Djouadi, Dominguez, Duxbury, Dworkin, Ebel, Economou, Fakra, Fairey, Fallon,
  Ferrini, Ferroir, Fleckenstein, Floss, Flynn, Franchi, Fries, Gainsforth,
  Gallien, Genge, Gilles, Gillet, Gilmour, Glavin, Gounelle, Grady, Graham,
  Grant, Green, Grossemy, Grossman, Grossman, Guan, Hagiya, Harvey, Heck,
  Herzog, Hoppe, H{\"o}rz, Huth, Hutcheon, Ignatyev, Ishii, Ito, Jacob,
  Jacobsen, Jacobsen, Jones, Joswiak, Jurewicz, Kearsley, Keller, Khodja,
  Kilcoyne, Kissel, Krot, Langenhorst, Lanzirotti, Le, Leshin, Leitner,
  Lemelle, Leroux, Liu, Luening, Lyon, MacPherson, Marcus, Marhas, Marty,
  Matrajt, McKeegan, Meibom, Mennella, Messenger, Messenger, Mikouchi,
  Mostefaoui, Nakamura, Nakano, Newville, Nittler, Ohnishi, Ohsumi, Okudaira,
  Papanastassiou, Palma, Palumbo, Pepin, Perkins, Perronnet, Pianetta, Rao,
  Rietmeijer, Robert, Rost, Rotundi, Ryan, Sandford, Schwandt, See, Schlutter,
  Sheffield-Parker, Simionovici, Simon, Sitnitsky, Snead, Spencer, Stadermann,
  Steele, Stephan, Stroud, Susini, Sutton, Suzuki, Taheri, Taylor, Teslich,
  Tomeoka, Tomioka, Toppani, Trigo-Rodr{\'\i}guez, Troadec, Tsuchiyama,
  Tuzzolino, Tyliszczak, Uesugi, Velbel, Vellenga, Vicenzi, Vincze, Warren,
  Weber, Weisberg, Westphal, Wirick, Wooden, Wopenka, Wozniakiewicz, Wright,
  Yabuta, Yano, Young, Zare, Zega, Ziegler, Zimmerman, Zinner and
  Zolensky}]{brownlee2006comet}
\bibinfo{author}{Brownlee, D.}, \bibinfo{author}{Tsou, P.},
  \bibinfo{author}{Al{\'e}on, J.}, \bibinfo{author}{Alexander, C.M.O.},
  \bibinfo{author}{Araki, T.}, \bibinfo{author}{Bajt, S.},
  \bibinfo{author}{Baratta, G.A.}, \bibinfo{author}{Bastien, R.},
  \bibinfo{author}{Bland, P.}, \bibinfo{author}{Bleuet, P.},
  \bibinfo{author}{Borg, J.}, \bibinfo{author}{Bradley, J.P.},
  \bibinfo{author}{Brearley, A.}, \bibinfo{author}{Brenker, F.},
  \bibinfo{author}{Brennan, S.}, \bibinfo{author}{Bridges, J.C.},
  \bibinfo{author}{Browning, N.D.}, \bibinfo{author}{Brucato, J.R.},
  \bibinfo{author}{Bullock, E.}, \bibinfo{author}{Burchell, M.J.},
  \bibinfo{author}{Busemann, H.}, \bibinfo{author}{Butterworth, A.},
  \bibinfo{author}{Chaussidon, M.}, \bibinfo{author}{Cheuvront, A.},
  \bibinfo{author}{Chi, M.}, \bibinfo{author}{Cintala, M.J.},
  \bibinfo{author}{Clark, B.C.}, \bibinfo{author}{Clemett, S.J.},
  \bibinfo{author}{Cody, G.}, \bibinfo{author}{Colangeli, L.},
  \bibinfo{author}{Cooper, G.}, \bibinfo{author}{Cordier, P.},
  \bibinfo{author}{Daghlian, C.}, \bibinfo{author}{Dai, Z.},
  \bibinfo{author}{D{\textquoteright}Hendecourt, L.}, \bibinfo{author}{Djouadi,
  Z.}, \bibinfo{author}{Dominguez, G.}, \bibinfo{author}{Duxbury, T.},
  \bibinfo{author}{Dworkin, J.P.}, \bibinfo{author}{Ebel, D.S.},
  \bibinfo{author}{Economou, T.E.}, \bibinfo{author}{Fakra, S.},
  \bibinfo{author}{Fairey, S.A.J.}, \bibinfo{author}{Fallon, S.},
  \bibinfo{author}{Ferrini, G.}, \bibinfo{author}{Ferroir, T.},
  \bibinfo{author}{Fleckenstein, H.}, \bibinfo{author}{Floss, C.},
  \bibinfo{author}{Flynn, G.}, \bibinfo{author}{Franchi, I.A.},
  \bibinfo{author}{Fries, M.}, \bibinfo{author}{Gainsforth, Z.},
  \bibinfo{author}{Gallien, J.P.}, \bibinfo{author}{Genge, M.},
  \bibinfo{author}{Gilles, M.K.}, \bibinfo{author}{Gillet, P.},
  \bibinfo{author}{Gilmour, J.}, \bibinfo{author}{Glavin, D.P.},
  \bibinfo{author}{Gounelle, M.}, \bibinfo{author}{Grady, M.M.},
  \bibinfo{author}{Graham, G.A.}, \bibinfo{author}{Grant, P.G.},
  \bibinfo{author}{Green, S.F.}, \bibinfo{author}{Grossemy, F.},
  \bibinfo{author}{Grossman, L.}, \bibinfo{author}{Grossman, J.N.},
  \bibinfo{author}{Guan, Y.}, \bibinfo{author}{Hagiya, K.},
  \bibinfo{author}{Harvey, R.}, \bibinfo{author}{Heck, P.},
  \bibinfo{author}{Herzog, G.F.}, \bibinfo{author}{Hoppe, P.},
  \bibinfo{author}{H{\"o}rz, F.}, \bibinfo{author}{Huth, J.},
  \bibinfo{author}{Hutcheon, I.D.}, \bibinfo{author}{Ignatyev, K.},
  \bibinfo{author}{Ishii, H.}, \bibinfo{author}{Ito, M.},
  \bibinfo{author}{Jacob, D.}, \bibinfo{author}{Jacobsen, C.},
  \bibinfo{author}{Jacobsen, S.}, \bibinfo{author}{Jones, S.},
  \bibinfo{author}{Joswiak, D.}, \bibinfo{author}{Jurewicz, A.},
  \bibinfo{author}{Kearsley, A.T.}, \bibinfo{author}{Keller, L.P.},
  \bibinfo{author}{Khodja, H.}, \bibinfo{author}{Kilcoyne, A.D.},
  \bibinfo{author}{Kissel, J.}, \bibinfo{author}{Krot, A.},
  \bibinfo{author}{Langenhorst, F.}, \bibinfo{author}{Lanzirotti, A.},
  \bibinfo{author}{Le, L.}, \bibinfo{author}{Leshin, L.A.},
  \bibinfo{author}{Leitner, J.}, \bibinfo{author}{Lemelle, L.},
  \bibinfo{author}{Leroux, H.}, \bibinfo{author}{Liu, M.C.},
  \bibinfo{author}{Luening, K.}, \bibinfo{author}{Lyon, I.},
  \bibinfo{author}{MacPherson, G.J.}, \bibinfo{author}{Marcus, M.A.},
  \bibinfo{author}{Marhas, K.}, \bibinfo{author}{Marty, B.},
  \bibinfo{author}{Matrajt, G.}, \bibinfo{author}{McKeegan, K.},
  \bibinfo{author}{Meibom, A.}, \bibinfo{author}{Mennella, V.},
  \bibinfo{author}{Messenger, K.}, \bibinfo{author}{Messenger, S.},
  \bibinfo{author}{Mikouchi, T.}, \bibinfo{author}{Mostefaoui, S.},
  \bibinfo{author}{Nakamura, T.}, \bibinfo{author}{Nakano, T.},
  \bibinfo{author}{Newville, M.}, \bibinfo{author}{Nittler, L.R.},
  \bibinfo{author}{Ohnishi, I.}, \bibinfo{author}{Ohsumi, K.},
  \bibinfo{author}{Okudaira, K.}, \bibinfo{author}{Papanastassiou, D.A.},
  \bibinfo{author}{Palma, R.}, \bibinfo{author}{Palumbo, M.E.},
  \bibinfo{author}{Pepin, R.O.}, \bibinfo{author}{Perkins, D.},
  \bibinfo{author}{Perronnet, M.}, \bibinfo{author}{Pianetta, P.},
  \bibinfo{author}{Rao, W.}, \bibinfo{author}{Rietmeijer, F.J.M.},
  \bibinfo{author}{Robert, F.}, \bibinfo{author}{Rost, D.},
  \bibinfo{author}{Rotundi, A.}, \bibinfo{author}{Ryan, R.},
  \bibinfo{author}{Sandford, S.A.}, \bibinfo{author}{Schwandt, C.S.},
  \bibinfo{author}{See, T.H.}, \bibinfo{author}{Schlutter, D.},
  \bibinfo{author}{Sheffield-Parker, J.}, \bibinfo{author}{Simionovici, A.},
  \bibinfo{author}{Simon, S.}, \bibinfo{author}{Sitnitsky, I.},
  \bibinfo{author}{Snead, C.J.}, \bibinfo{author}{Spencer, M.K.},
  \bibinfo{author}{Stadermann, F.J.}, \bibinfo{author}{Steele, A.},
  \bibinfo{author}{Stephan, T.}, \bibinfo{author}{Stroud, R.},
  \bibinfo{author}{Susini, J.}, \bibinfo{author}{Sutton, S.R.},
  \bibinfo{author}{Suzuki, Y.}, \bibinfo{author}{Taheri, M.},
  \bibinfo{author}{Taylor, S.}, \bibinfo{author}{Teslich, N.},
  \bibinfo{author}{Tomeoka, K.}, \bibinfo{author}{Tomioka, N.},
  \bibinfo{author}{Toppani, A.}, \bibinfo{author}{Trigo-Rodr{\'\i}guez, J.M.},
  \bibinfo{author}{Troadec, D.}, \bibinfo{author}{Tsuchiyama, A.},
  \bibinfo{author}{Tuzzolino, A.J.}, \bibinfo{author}{Tyliszczak, T.},
  \bibinfo{author}{Uesugi, K.}, \bibinfo{author}{Velbel, M.},
  \bibinfo{author}{Vellenga, J.}, \bibinfo{author}{Vicenzi, E.},
  \bibinfo{author}{Vincze, L.}, \bibinfo{author}{Warren, J.},
  \bibinfo{author}{Weber, I.}, \bibinfo{author}{Weisberg, M.},
  \bibinfo{author}{Westphal, A.J.}, \bibinfo{author}{Wirick, S.},
  \bibinfo{author}{Wooden, D.}, \bibinfo{author}{Wopenka, B.},
  \bibinfo{author}{Wozniakiewicz, P.}, \bibinfo{author}{Wright, I.},
  \bibinfo{author}{Yabuta, H.}, \bibinfo{author}{Yano, H.},
  \bibinfo{author}{Young, E.D.}, \bibinfo{author}{Zare, R.N.},
  \bibinfo{author}{Zega, T.}, \bibinfo{author}{Ziegler, K.},
  \bibinfo{author}{Zimmerman, L.}, \bibinfo{author}{Zinner, E.},
  \bibinfo{author}{Zolensky, M.E.}, \bibinfo{year}{2006}.
\newblock \bibinfo{title}{{Comet 81P/Wild 2 under a microscope}}.
\newblock \bibinfo{journal}{Science} \bibinfo{volume}{314},
  \bibinfo{pages}{1711--1716}.
\newblock \DOIprefix\doi{10.1126/science.1135840}.
\bibitem[{Ciesla(2007)}]{ciesla2007outward}
\bibinfo{author}{Ciesla, F.J.}, \bibinfo{year}{2007}.
\newblock \bibinfo{title}{{Outward transport of high-temperature materials
  around the midplane of the solar nebula}}.
\newblock \bibinfo{journal}{Science} \bibinfo{volume}{318},
  \bibinfo{pages}{613--615}.
\newblock \DOIprefix\doi{10.1126/science.1147273}.
\bibitem[{Clauser and Huenges(1995)}]{clauser1995thermal}
\bibinfo{author}{Clauser, C.}, \bibinfo{author}{Huenges, E.},
  \bibinfo{year}{1995}.
\newblock \bibinfo{title}{{Thermal conductivity of rocks and minerals}}, in:
  \bibinfo{editor}{Ahrens, T.J.} (Ed.), \bibinfo{booktitle}{Rock physics \&
  phase relations: A handbook of physical constants}.
  \bibinfo{publisher}{American Geophysical Union}, pp.
  \bibinfo{pages}{105--126}.
\newblock \DOIprefix\doi{10.1029/RF003p0105}.
\bibitem[{Clayton(1993)}]{clayton1993oxygen}
\bibinfo{author}{Clayton, R.N.}, \bibinfo{year}{1993}.
\newblock \bibinfo{title}{{Oxygen isotopes in meteorites}}.
\newblock \bibinfo{journal}{Annual Review of Earth and Planetary Sciences}
  \bibinfo{volume}{21}, \bibinfo{pages}{115--149}.
\newblock \DOIprefix\doi{10.1146/annurev.ea.21.050193.000555}.
\bibitem[{Clayton(2002)}]{clayton2002solar}
\bibinfo{author}{Clayton, R.N.}, \bibinfo{year}{2002}.
\newblock \bibinfo{title}{{Solar system: Self-shielding in the solar nebula}}.
\newblock \bibinfo{journal}{Nature} \bibinfo{volume}{415},
  \bibinfo{pages}{860--861}.
\newblock \DOIprefix\doi{10.1038/415860b}.
\bibitem[{Clayton et~al.(1977)Clayton, Onuma, Grossman and
  Mayeda}]{clayton1977distribution}
\bibinfo{author}{Clayton, R.N.}, \bibinfo{author}{Onuma, N.},
  \bibinfo{author}{Grossman, L.}, \bibinfo{author}{Mayeda, T.K.},
  \bibinfo{year}{1977}.
\newblock \bibinfo{title}{{Distribution of the pre-solar component in Allende
  and other carbonaceous chondrites}}.
\newblock \bibinfo{journal}{Earth and Planetary Science Letters}
  \bibinfo{volume}{34}, \bibinfo{pages}{209--224}.
\newblock \DOIprefix\doi{10.1016/0012-821X(77)90005-X}.
\bibitem[{Connelly et~al.(2012)Connelly, Bizzarro, Krot, Nordlund, Wielandt and
  Ivanova}]{connelly2012absolute}
\bibinfo{author}{Connelly, J.N.}, \bibinfo{author}{Bizzarro, M.},
  \bibinfo{author}{Krot, A.N.}, \bibinfo{author}{Nordlund, {\AA}.},
  \bibinfo{author}{Wielandt, D.}, \bibinfo{author}{Ivanova, M.A.},
  \bibinfo{year}{2012}.
\newblock \bibinfo{title}{{The absolute chronology and thermal processing of
  solids in the solar protoplanetary disk}}.
\newblock \bibinfo{journal}{Science} \bibinfo{volume}{338},
  \bibinfo{pages}{651--655}.
\newblock \DOIprefix\doi{10.1126/science.1226919}.
\bibitem[{Connolly~Jr and Burnett(2003)}]{connolly2003type}
\bibinfo{author}{Connolly~Jr, H.C.}, \bibinfo{author}{Burnett, D.S.},
  \bibinfo{year}{2003}.
\newblock \bibinfo{title}{{On type B CAI formation: experimental constraints on
  fO2 variations in spinel minor element partitioning and reequilibration
  effects}}.
\newblock \bibinfo{journal}{Geochimica et Cosmochimica Acta}
  \bibinfo{volume}{67}, \bibinfo{pages}{4429--4434}.
\newblock \DOIprefix\doi{10.1016/S0016-7037(03)00271-0}.
\bibitem[{Connolly~Jr et~al.(2003)Connolly~Jr, Burnett and
  McKeegan}]{connolly2003petrogenesis}
\bibinfo{author}{Connolly~Jr, H.C.}, \bibinfo{author}{Burnett, D.S.},
  \bibinfo{author}{McKeegan, K.D.}, \bibinfo{year}{2003}.
\newblock \bibinfo{title}{{The petrogenesis of type B1 Ca-Al-rich inclusions:
  The spinel perspective}}.
\newblock \bibinfo{journal}{Meteoritics \& Planetary Science}
  \bibinfo{volume}{38}, \bibinfo{pages}{197--224}.
\newblock \DOIprefix\doi{10.1111/j.1945-5100.2003.tb00260.x}.
\bibitem[{Cosarinsky et~al.(2005)Cosarinsky, McKeegan, Hutcheon, Weber and
  Fallon}]{cosarinsky2005magnesium}
\bibinfo{author}{Cosarinsky, M.}, \bibinfo{author}{McKeegan, K.D.},
  \bibinfo{author}{Hutcheon, I.D.}, \bibinfo{author}{Weber, P.},
  \bibinfo{author}{Fallon, S.}, \bibinfo{year}{2005}.
\newblock \bibinfo{title}{{Magnesium and oxygen isotopic study of the
  Wark-Lovering rim around a fluffy type-A inclusion from Allende}}, in:
  \bibinfo{booktitle}{Lunar and Planetary Science Conference}, p.
  \bibinfo{pages}{2105}.
\bibitem[{Daly et~al.(2017)Daly, Bland, Saxey, Reddy, Fougerouse, Rickard and
  Forman}]{daly2017nebula}
\bibinfo{author}{Daly, L.}, \bibinfo{author}{Bland, P.A.},
  \bibinfo{author}{Saxey, D.W.}, \bibinfo{author}{Reddy, S.M.},
  \bibinfo{author}{Fougerouse, D.}, \bibinfo{author}{Rickard, W.D.A.},
  \bibinfo{author}{Forman, L.V.}, \bibinfo{year}{2017}.
\newblock \bibinfo{title}{{Nebula sulfidation and evidence for migration of
  "free-floating" refractory metal nuggets revealed by atom probe microscopy}}.
\newblock \bibinfo{journal}{Geology} \bibinfo{volume}{45},
  \bibinfo{pages}{847--850}.
\newblock \DOIprefix\doi{10.1130/G39075.1}.
\bibitem[{Davis(1984)}]{davis1984scandalously}
\bibinfo{author}{Davis, A.M.}, \bibinfo{year}{1984}.
\newblock \bibinfo{title}{{A scandalously refractory inclusion in Ornans}}.
\newblock \bibinfo{journal}{Meteoritics} \bibinfo{volume}{19},
  \bibinfo{pages}{214}.
\bibitem[{Davis and Grossman(1979)}]{davis1979condensation}
\bibinfo{author}{Davis, A.M.}, \bibinfo{author}{Grossman, L.},
  \bibinfo{year}{1979}.
\newblock \bibinfo{title}{{Condensation and fractionation of rare earths in the
  solar nebula}}.
\newblock \bibinfo{journal}{Geochimica et Cosmochimica Acta}
  \bibinfo{volume}{43}, \bibinfo{pages}{1611--1632}.
\newblock \DOIprefix\doi{10.1016/0016-7037(79)90181-9}.
\bibitem[{Davis and Hashimoto(1995)}]{davis1995volatility}
\bibinfo{author}{Davis, A.M.}, \bibinfo{author}{Hashimoto, A.},
  \bibinfo{year}{1995}.
\newblock \bibinfo{title}{{Volatility fractionation of REE and other trace
  elements during vacuum evaporation}}.
\newblock \bibinfo{journal}{Meteoritics} \bibinfo{volume}{30},
  \bibinfo{pages}{500--501}.
\bibitem[{Davis et~al.(1995)Davis, Hashimoto, Clayton and
  Mayeda}]{davis1995isotopic}
\bibinfo{author}{Davis, A.M.}, \bibinfo{author}{Hashimoto, A.},
  \bibinfo{author}{Clayton, R.N.}, \bibinfo{author}{Mayeda, T.K.},
  \bibinfo{year}{1995}.
\newblock \bibinfo{title}{{Isotopic and chemical fractionation during
  evaporation of \ce{CaTiO3}}}, in: \bibinfo{booktitle}{Lunar and Planetary
  Science Conference}, p. \bibinfo{pages}{317}.
\bibitem[{Davis and Hinton(1985)}]{davis1985trace}
\bibinfo{author}{Davis, A.M.}, \bibinfo{author}{Hinton, R.W.},
  \bibinfo{year}{1985}.
\newblock \bibinfo{title}{{Trace element abundances in OSCAR, a scandium-rich
  refractory inclusion from the Ornans meteorite}}.
\newblock \bibinfo{journal}{Meteoritics} \bibinfo{volume}{20},
  \bibinfo{pages}{633--634}.
\bibitem[{Davis et~al.(2018)Davis, Zhang, Greber, Hu, Tissot and
  Dauphas}]{davis2018titanium}
\bibinfo{author}{Davis, A.M.}, \bibinfo{author}{Zhang, J.},
  \bibinfo{author}{Greber, N.D.}, \bibinfo{author}{Hu, J.},
  \bibinfo{author}{Tissot, F.L.H.}, \bibinfo{author}{Dauphas, N.},
  \bibinfo{year}{2018}.
\newblock \bibinfo{title}{{Titanium isotopes and rare earth patterns in CAIs:
  Evidence for thermal processing and gas-dust decoupling in the protoplanetary
  disk}}.
\newblock \bibinfo{journal}{Geochimica et Cosmochimica Acta}
  \bibinfo{volume}{221}, \bibinfo{pages}{275--295}.
\newblock \DOIprefix\doi{10.1016/j.gca.2017.07.032}.
\bibitem[{Desch et~al.(2018)Desch, Kalyaan and Alexander}]{desch2017effect}
\bibinfo{author}{Desch, S.J.}, \bibinfo{author}{Kalyaan, A.},
  \bibinfo{author}{Alexander, C.M.O.}, \bibinfo{year}{2018}.
\newblock \bibinfo{title}{{The effect of Jupiter's formation on the
  distribution of refractory elements and inclusions in meteorites}}.
\newblock \bibinfo{journal}{The Astrophysical Journal Supplement Series}
  \bibinfo{volume}{238}, \bibinfo{pages}{11}.
\newblock \DOIprefix\doi{10.3847/1538-4365/aad95f}.
\bibitem[{Desch et~al.(2010)Desch, Morris, Connolly~Jr and
  Boss}]{desch2010critical}
\bibinfo{author}{Desch, S.J.}, \bibinfo{author}{Morris, M.A.},
  \bibinfo{author}{Connolly~Jr, H.C.}, \bibinfo{author}{Boss, A.P.},
  \bibinfo{year}{2010}.
\newblock \bibinfo{title}{{A critical examination of the X-wind model for
  chondrule and calcium-rich, aluminum-rich inclusion formation and
  radionuclide production}}.
\newblock \bibinfo{journal}{The Astrophysical Journal} \bibinfo{volume}{725},
  \bibinfo{pages}{692}.
\newblock \DOIprefix\doi{10.1088/0004-637X/725/1/692}.
\bibitem[{Dyl et~al.(2011)Dyl, Simon and Young}]{dyl2011valence}
\bibinfo{author}{Dyl, K.A.}, \bibinfo{author}{Simon, J.I.},
  \bibinfo{author}{Young, E.D.}, \bibinfo{year}{2011}.
\newblock \bibinfo{title}{{Valence state of titanium in the Wark-Lovering rim
  of a Leoville CAI as a record of progressive oxidation in the early Solar
  Nebula}}.
\newblock \bibinfo{journal}{Geochimica et Cosmochimica Acta}
  \bibinfo{volume}{75}, \bibinfo{pages}{937--949}.
\newblock \DOIprefix\doi{10.1016/j.gca.2010.09.042}.
\bibitem[{Ebel(2006)}]{ebel2006condensation}
\bibinfo{author}{Ebel, D.S.}, \bibinfo{year}{2006}.
\newblock \bibinfo{title}{{Condensation of rocky material in astrophysical
  environments}}, in: \bibinfo{editor}{Lauretta, D.S.},
  \bibinfo{editor}{McSween~Jr., H.Y.} (Eds.), \bibinfo{booktitle}{Meteorites
  and the Early Solar System II}. \bibinfo{publisher}{The University of Arizona
  Press}, \bibinfo{address}{Tucson}, pp. \bibinfo{pages}{253--277}.
\bibitem[{Ebel et~al.(2012)Ebel, Weisberg and Beckett}]{ebel2012thermochemical}
\bibinfo{author}{Ebel, D.S.}, \bibinfo{author}{Weisberg, M.K.},
  \bibinfo{author}{Beckett, J.R.}, \bibinfo{year}{2012}.
\newblock \bibinfo{title}{{Thermochemical stability of low-iron,
  manganese-enriched olivine in astrophysical environments}}.
\newblock \bibinfo{journal}{Meteoritics \& Planetary Science}
  \bibinfo{volume}{47}, \bibinfo{pages}{585--593}.
\newblock \DOIprefix\doi{10.1111/j.1945-5100.2012.01347.x}.
\bibitem[{El~Goresy et~al.(1984)El~Goresy, Palme, Yabuki, Nagel, Herrwerth and
  Ramdohr}]{el1984calcium}
\bibinfo{author}{El~Goresy, A.}, \bibinfo{author}{Palme, H.},
  \bibinfo{author}{Yabuki, H.}, \bibinfo{author}{Nagel, K.},
  \bibinfo{author}{Herrwerth, I.}, \bibinfo{author}{Ramdohr, P.},
  \bibinfo{year}{1984}.
\newblock \bibinfo{title}{{A calcium-aluminum-rich inclusion from the Essebi
  (CM2) chondrite: Evidence for captured spinel-hibonite spherules and for an
  ultra-refractory rimming sequence}}.
\newblock \bibinfo{journal}{Geochimica et Cosmochimica Acta}
  \bibinfo{volume}{48}, \bibinfo{pages}{2283--2298}.
\newblock \DOIprefix\doi{10.1016/0016-7037(84)90224-2}.
\bibitem[{El~Goresy et~al.(1979)El~Goresy, Ramdohr and Nagel}]{el1979spinel}
\bibinfo{author}{El~Goresy, A.}, \bibinfo{author}{Ramdohr, P.},
  \bibinfo{author}{Nagel, K.}, \bibinfo{year}{1979}.
\newblock \bibinfo{title}{{Spinel framboids and fremdlinge in Allende
  inclusions: Possible sequential markers in the early history of the solar
  system}}, in: \bibinfo{booktitle}{Proceedings of the Lunar and Planetary
  Science Conference}, pp. \bibinfo{pages}{833--850}.
\bibitem[{El~Goresy et~al.(2002)El~Goresy, Zinner, Matsunami, Palme, Spettel,
  Lin and Nazarov}]{el2002efremovka}
\bibinfo{author}{El~Goresy, A.}, \bibinfo{author}{Zinner, E.},
  \bibinfo{author}{Matsunami, S.}, \bibinfo{author}{Palme, H.},
  \bibinfo{author}{Spettel, B.}, \bibinfo{author}{Lin, Y.},
  \bibinfo{author}{Nazarov, M.}, \bibinfo{year}{2002}.
\newblock \bibinfo{title}{{Efremovka 101.1: A CAI with ultrarefractory REE
  patterns and enormous enrichments of Sc, Zr, and Y in fassaite and
  perovskite}}.
\newblock \bibinfo{journal}{Geochimica et Cosmochimica Acta}
  \bibinfo{volume}{66}, \bibinfo{pages}{1459--1491}.
\newblock \DOIprefix\doi{10.1016/S0016-7037(01)00854-7}.
\bibitem[{Fagan et~al.(2004a)Fagan, Krot, Keil and
  Yurimoto}]{fagan2004oxygenalt}
\bibinfo{author}{Fagan, T.J.}, \bibinfo{author}{Krot, A.N.},
  \bibinfo{author}{Keil, K.}, \bibinfo{author}{Yurimoto, H.},
  \bibinfo{year}{2004}a.
\newblock \bibinfo{title}{{Oxygen isotopic alteration in Ca-Al-rich inclusions
  from Efremovka: Nebular or parent body setting?}}
\newblock \bibinfo{journal}{Meteoritics \& Planetary Science}
  \bibinfo{volume}{39}, \bibinfo{pages}{1257--1272}.
\newblock \DOIprefix\doi{10.1111/j.1945-5100.2004.tb00945.x}.
\bibitem[{Fagan et~al.(2004b)Fagan, Krot, Keil and Yurimoto}]{fagan2004oxygen}
\bibinfo{author}{Fagan, T.J.}, \bibinfo{author}{Krot, A.N.},
  \bibinfo{author}{Keil, K.}, \bibinfo{author}{Yurimoto, H.},
  \bibinfo{year}{2004}b.
\newblock \bibinfo{title}{{Oxygen isotopic evolution of amoeboid olivine
  aggregates in the reduced CV3 chondrites Efremovka, Vigarano, and Leoville}}.
\newblock \bibinfo{journal}{Geochimica et Cosmochimica Acta}
  \bibinfo{volume}{68}, \bibinfo{pages}{2591--2611}.
\newblock \DOIprefix\doi{10.1016/j.gca.2003.10.033}.
\bibitem[{Faul and Scott(2006)}]{faul2006grain}
\bibinfo{author}{Faul, U.H.}, \bibinfo{author}{Scott, D.},
  \bibinfo{year}{2006}.
\newblock \bibinfo{title}{{Grain growth in partially molten olivine
  aggregates}}.
\newblock \bibinfo{journal}{Contributions to Mineralogy and Petrology}
  \bibinfo{volume}{151}, \bibinfo{pages}{101--111}.
\newblock \DOIprefix\doi{10.1007/s00410-005-0048-1}.
\bibitem[{Flynn(1972)}]{flynn1972point}
\bibinfo{author}{Flynn, C.P.}, \bibinfo{year}{1972}.
\newblock \bibinfo{title}{{Point Defects and Diffusion}}.
\newblock \bibinfo{publisher}{Clarendon Press}, \bibinfo{address}{Oxford}.
\bibitem[{Grossman(1972)}]{grossman1972condensation}
\bibinfo{author}{Grossman, L.}, \bibinfo{year}{1972}.
\newblock \bibinfo{title}{{Condensation in the primitive solar nebula}}.
\newblock \bibinfo{journal}{Geochimica et Cosmochimica Acta}
  \bibinfo{volume}{36}, \bibinfo{pages}{597--619}.
\newblock \DOIprefix\doi{10.1016/0016-7037(72)90078-6}.
\bibitem[{Grossman et~al.(2008)Grossman, Beckett, Fedkin, Simon and
  Ciesla}]{grossman2008redox}
\bibinfo{author}{Grossman, L.}, \bibinfo{author}{Beckett, J.R.},
  \bibinfo{author}{Fedkin, A.V.}, \bibinfo{author}{Simon, S.B.},
  \bibinfo{author}{Ciesla, F.J.}, \bibinfo{year}{2008}.
\newblock \bibinfo{title}{{Redox conditions in the solar nebula: Observational,
  experimental, and theoretical constraints}}, in: \bibinfo{editor}{MacPherson,
  G.J.} (Ed.), \bibinfo{booktitle}{Reviews in Mineralogy and Geochemistry}.
  \bibinfo{publisher}{Mineralogical Society of America}.
  volume~\bibinfo{volume}{68}, pp. \bibinfo{pages}{93--140}.
\newblock \DOIprefix\doi{10.2138/rmg.2008.68.7}.
\bibitem[{Grossman et~al.(2002)Grossman, Ebel and
  Simon}]{grossman2002formation}
\bibinfo{author}{Grossman, L.}, \bibinfo{author}{Ebel, D.S.},
  \bibinfo{author}{Simon, S.B.}, \bibinfo{year}{2002}.
\newblock \bibinfo{title}{{Formation of refractory inclusions by evaporation of
  condensate precursors}}.
\newblock \bibinfo{journal}{Geochimica et Cosmochimica Acta}
  \bibinfo{volume}{66}, \bibinfo{pages}{145--161}.
\newblock \DOIprefix\doi{10.1016/S0016-7037(01)00731-1}.
\bibitem[{Hart and Dunn(1993)}]{hart1993experimental}
\bibinfo{author}{Hart, S.R.}, \bibinfo{author}{Dunn, T.}, \bibinfo{year}{1993}.
\newblock \bibinfo{title}{Experimental cpx/melt partitioning of 24 trace
  elements}.
\newblock \bibinfo{journal}{Contributions to Mineralogy and Petrology}
  \bibinfo{volume}{113}, \bibinfo{pages}{1--8}.
\newblock \DOIprefix\doi{10.1007/BF00320827}.
\bibitem[{Hashimoto and Grossman(1985)}]{hashimoto1985sem}
\bibinfo{author}{Hashimoto, A.}, \bibinfo{author}{Grossman, L.},
  \bibinfo{year}{1985}.
\newblock \bibinfo{title}{{SEM-petrography of Allende fine-grained
  inclusions}}, in: \bibinfo{booktitle}{Lunar and Planetary Science
  Conference}, pp. \bibinfo{pages}{323--324}.
\bibitem[{Hezel et~al.(2008)Hezel, Russell, Ross and Kearsley}]{hezel2008modal}
\bibinfo{author}{Hezel, D.C.}, \bibinfo{author}{Russell, S.S.},
  \bibinfo{author}{Ross, A.J.}, \bibinfo{author}{Kearsley, A.T.},
  \bibinfo{year}{2008}.
\newblock \bibinfo{title}{{Modal abundances of CAIs: Implications for bulk
  chondrite element abundances and fractionations}}.
\newblock \bibinfo{journal}{Meteoritics \& Planetary Science}
  \bibinfo{volume}{43}, \bibinfo{pages}{1879--1894}.
\newblock \DOIprefix\doi{10.1111/j.1945-5100.2008.tb00649.x}.
\bibitem[{Ishida et~al.(2012)Ishida, Nakamura, Miura and
  Kakazu}]{ishida2012diverse}
\bibinfo{author}{Ishida, H.}, \bibinfo{author}{Nakamura, T.},
  \bibinfo{author}{Miura, H.}, \bibinfo{author}{Kakazu, Y.},
  \bibinfo{year}{2012}.
\newblock \bibinfo{title}{{Diverse mineralogical and oxygen isotopic signatures
  recorded in CV3 carbonaceous chondrites}}.
\newblock \bibinfo{journal}{Polar Science} \bibinfo{volume}{6},
  \bibinfo{pages}{252--262}.
\newblock \DOIprefix\doi{10.1016/j.polar.2012.06.002}.
\bibitem[{Itoh et~al.(2004)Itoh, Kojima and Yurimoto}]{itoh2004petrography}
\bibinfo{author}{Itoh, S.}, \bibinfo{author}{Kojima, H.},
  \bibinfo{author}{Yurimoto, H.}, \bibinfo{year}{2004}.
\newblock \bibinfo{title}{{Petrography and oxygen isotopic compositions in
  refractory inclusions from CO chondrites}}.
\newblock \bibinfo{journal}{Geochimica et Cosmochimica Acta}
  \bibinfo{volume}{68}, \bibinfo{pages}{183--194}.
\newblock \DOIprefix\doi{10.1016/S0016-7037(03)00368-5}.
\bibitem[{Itoh and Yurimoto(2003)}]{itoh2003contemporaneous}
\bibinfo{author}{Itoh, S.}, \bibinfo{author}{Yurimoto, H.},
  \bibinfo{year}{2003}.
\newblock \bibinfo{title}{{Contemporaneous formation of chondrules and
  refractory inclusions in the early Solar System}}.
\newblock \bibinfo{journal}{Nature} \bibinfo{volume}{423},
  \bibinfo{pages}{728--731}.
\newblock \DOIprefix\doi{10.1038/nature01699}.
\bibitem[{Ivanova et~al.(2017)Ivanova, Krot, Nagashima, Ma and
  MacPherson}]{ivanova2017oxygen}
\bibinfo{author}{Ivanova, M.A.}, \bibinfo{author}{Krot, A.N.},
  \bibinfo{author}{Nagashima, K.}, \bibinfo{author}{Ma, C.},
  \bibinfo{author}{MacPherson, G.J.}, \bibinfo{year}{2017}.
\newblock \bibinfo{title}{{Oxygen-isotope composition of ultrarefractory CAI
  from CV chondrite Efremovka}}.
\newblock \bibinfo{journal}{Meteoritics \& Planetary Science}
  \bibinfo{volume}{52}, \bibinfo{pages}{A152}.
\newblock \DOIprefix\doi{10.1111/maps.12934}.
\bibitem[{Ivanova et~al.(2012)Ivanova, Krot, Nagashima and
  MacPherson}]{ivanova2012compound}
\bibinfo{author}{Ivanova, M.A.}, \bibinfo{author}{Krot, A.N.},
  \bibinfo{author}{Nagashima, K.}, \bibinfo{author}{MacPherson, G.J.},
  \bibinfo{year}{2012}.
\newblock \bibinfo{title}{{Compound ultrarefractory CAI-bearing inclusions from
  CV3 carbonaceous chondrites}}.
\newblock \bibinfo{journal}{Meteoritics \& Planetary Science}
  \bibinfo{volume}{47}, \bibinfo{pages}{2107--2127}.
\newblock \DOIprefix\doi{10.1111/maps.12031}.
\bibitem[{Karato(1989)}]{karato1989grain}
\bibinfo{author}{Karato, S.i.}, \bibinfo{year}{1989}.
\newblock \bibinfo{title}{{Grain growth kinetics in olivine aggregates}}.
\newblock \bibinfo{journal}{Tectonophysics} \bibinfo{volume}{168},
  \bibinfo{pages}{255--273}.
\newblock \DOIprefix\doi{10.1016/0040-1951(89)90221-7}.
\bibitem[{Katayama et~al.(2012)Katayama, Itoh and
  Yurimoto}]{katayama2012oxygen}
\bibinfo{author}{Katayama, J.}, \bibinfo{author}{Itoh, S.},
  \bibinfo{author}{Yurimoto, H.}, \bibinfo{year}{2012}.
\newblock \bibinfo{title}{{Oxygen isotopic zoning of reversely zoned melilite
  crystals in a fluffy type A Ca-Al-rich inclusions from the Vigarano
  meteorite}}.
\newblock \bibinfo{journal}{Meteoritics \& Planetary Science}
  \bibinfo{volume}{47}, \bibinfo{pages}{2094--2106}.
\newblock \DOIprefix\doi{10.1111/maps.12034}.
\bibitem[{Kawasaki et~al.(2017a)Kawasaki, Itoh, Sakamoto and
  Yurimoto}]{kawasaki2016chronological}
\bibinfo{author}{Kawasaki, N.}, \bibinfo{author}{Itoh, S.},
  \bibinfo{author}{Sakamoto, N.}, \bibinfo{author}{Yurimoto, H.},
  \bibinfo{year}{2017}a.
\newblock \bibinfo{title}{{Chronological study of oxygen isotope composition
  for the solar protoplanetary disk recorded in a fluffy Type A CAI from
  Vigarano}}.
\newblock \bibinfo{journal}{Geochimica et Cosmochimica Acta}
  \bibinfo{volume}{201}, \bibinfo{pages}{83--102}.
\newblock \DOIprefix\doi{10.1016/j.gca.2015.12.031}.
\bibitem[{Kawasaki et~al.(2012)Kawasaki, Sakamoto and
  Yurimoto}]{kawasaki2012oxygen}
\bibinfo{author}{Kawasaki, N.}, \bibinfo{author}{Sakamoto, N.},
  \bibinfo{author}{Yurimoto, H.}, \bibinfo{year}{2012}.
\newblock \bibinfo{title}{{Oxygen isotopic and chemical zoning of melilite
  crystals in a type A Ca-Al-rich inclusion of Efremovka CV3 chondrite}}.
\newblock \bibinfo{journal}{Meteoritics \& Planetary Science}
  \bibinfo{volume}{47}, \bibinfo{pages}{2084--2093}.
\newblock \DOIprefix\doi{10.1111/maps.12033}.
\bibitem[{Kawasaki et~al.(2017b)Kawasaki, Simon, Grossman, Sakamoto and
  Yurimoto}]{kawasaki2017crystal}
\bibinfo{author}{Kawasaki, N.}, \bibinfo{author}{Simon, S.B.},
  \bibinfo{author}{Grossman, L.}, \bibinfo{author}{Sakamoto, N.},
  \bibinfo{author}{Yurimoto, H.}, \bibinfo{year}{2017}b.
\newblock \bibinfo{title}{{Crystal growth and disequilibrium distribution of
  oxygen isotopes in an igneous Ca-Al-rich inclusion from the Allende
  carbonaceous chondrite}}.
\newblock \bibinfo{journal}{Geochimica et Cosmochimica Acta}
  \bibinfo{volume}{221}, \bibinfo{pages}{318--341}.
\newblock \DOIprefix\doi{10.1016/j.gca.2017.05.035}.
\bibitem[{Kim et~al.(2002)Kim, Yurimoto and Sueno}]{kim2002oxygen}
\bibinfo{author}{Kim, G.L.}, \bibinfo{author}{Yurimoto, H.},
  \bibinfo{author}{Sueno, S.}, \bibinfo{year}{2002}.
\newblock \bibinfo{title}{{Oxygen isotopic composition of a compound Ca-Al-rich
  inclusion from Allende meteorite: Implications for origin of palisade bodies
  and O-isotopic environment in the CAI forming region}}.
\newblock \bibinfo{journal}{Journal of Mineralogical and Petrological Sciences}
  \bibinfo{volume}{97}, \bibinfo{pages}{161--167}.
\newblock \DOIprefix\doi{10.2465/jmps.97.161}.
\bibitem[{Komatsu et~al.(2018)Komatsu, Fagan, Krot, Nagashima, Petaev, Kimura
  and Yamaguchi}]{komatsu2018first}
\bibinfo{author}{Komatsu, M.}, \bibinfo{author}{Fagan, T.J.},
  \bibinfo{author}{Krot, A.N.}, \bibinfo{author}{Nagashima, K.},
  \bibinfo{author}{Petaev, M.I.}, \bibinfo{author}{Kimura, M.},
  \bibinfo{author}{Yamaguchi, A.}, \bibinfo{year}{2018}.
\newblock \bibinfo{title}{{First evidence for silica condensation within the
  solar protoplanetary disk}}.
\newblock \bibinfo{journal}{Proceedings of the National Academy of Sciences}
  \bibinfo{volume}{115}, \bibinfo{pages}{7497--7502}.
\newblock \DOIprefix\doi{10.1073/pnas.1722265115}.
\bibitem[{Komatsu et~al.(2015)Komatsu, Fagan, Mikouchi, Petaev and
  Zolensky}]{komatsu2015lime}
\bibinfo{author}{Komatsu, M.}, \bibinfo{author}{Fagan, T.J.},
  \bibinfo{author}{Mikouchi, T.}, \bibinfo{author}{Petaev, M.I.},
  \bibinfo{author}{Zolensky, M.E.}, \bibinfo{year}{2015}.
\newblock \bibinfo{title}{{LIME silicates in amoeboid olivine aggregates in
  carbonaceous chondrites: Indicator of nebular and asteroidal processes}}.
\newblock \bibinfo{journal}{Meteoritics \& Planetary Science}
  \bibinfo{volume}{50}, \bibinfo{pages}{1271--1294}.
\newblock \DOIprefix\doi{10.1111/maps.12460}.
\bibitem[{Komatsu et~al.(2009)Komatsu, Mikouchi and Miyamoto}]{komatsu2009high}
\bibinfo{author}{Komatsu, M.}, \bibinfo{author}{Mikouchi, T.},
  \bibinfo{author}{Miyamoto, M.}, \bibinfo{year}{2009}.
\newblock \bibinfo{title}{{High-temperature annealing of amoeboid olivine
  aggregates: Heating experiments on olivine--anorthite mixtures}}.
\newblock \bibinfo{journal}{Polar Science} \bibinfo{volume}{3},
  \bibinfo{pages}{31--55}.
\newblock \DOIprefix\doi{10.1016/j.polar.2009.03.002}.
\bibitem[{Kornacki and Fegley~Jr(1986)}]{kornacki1986abundance}
\bibinfo{author}{Kornacki, A.S.}, \bibinfo{author}{Fegley~Jr, B.},
  \bibinfo{year}{1986}.
\newblock \bibinfo{title}{{The abundance and relative volatility of refractory
  trace elements in Allende Ca, Al-rich inclusions: Implications for chemical
  and physical processes in the solar nebula}}.
\newblock \bibinfo{journal}{Earth and Planetary Science Letters}
  \bibinfo{volume}{79}, \bibinfo{pages}{217--234}.
\newblock \DOIprefix\doi{10.1016/0012-821X(86)90180-9}.
\bibitem[{Krot et~al.(2009)Krot, Amelin, Bland, Ciesla, Connelly, Davis, Huss,
  Hutcheon, Makide and Nagashima}]{krot2009origin}
\bibinfo{author}{Krot, A.N.}, \bibinfo{author}{Amelin, Y.},
  \bibinfo{author}{Bland, P.}, \bibinfo{author}{Ciesla, F.J.},
  \bibinfo{author}{Connelly, J.}, \bibinfo{author}{Davis, A.M.},
  \bibinfo{author}{Huss, G.R.}, \bibinfo{author}{Hutcheon, I.D.},
  \bibinfo{author}{Makide, K.}, \bibinfo{author}{Nagashima, K.},
  \bibinfo{year}{2009}.
\newblock \bibinfo{title}{{Origin and chronology of chondritic components: A
  review}}.
\newblock \bibinfo{journal}{Geochimica et Cosmochimica Acta}
  \bibinfo{volume}{73}, \bibinfo{pages}{4963--4997}.
\newblock \DOIprefix\doi{10.1016/j.gca.2008.09.039}.
\bibitem[{Krot et~al.(2018a)Krot, Ma, Nagashima, Davis, Beckett, Simon,
  Komatsu, Fagan, Brenker, Ivanova and Bischoff}]{krot2018mineralogy}
\bibinfo{author}{Krot, A.N.}, \bibinfo{author}{Ma, C.},
  \bibinfo{author}{Nagashima, K.}, \bibinfo{author}{Davis, A.M.},
  \bibinfo{author}{Beckett, J.R.}, \bibinfo{author}{Simon, S.B.},
  \bibinfo{author}{Komatsu, M.}, \bibinfo{author}{Fagan, T.J.},
  \bibinfo{author}{Brenker, F.}, \bibinfo{author}{Ivanova, M.A.},
  \bibinfo{author}{Bischoff, A.}, \bibinfo{year}{2018}a.
\newblock \bibinfo{title}{{Mineralogy, petrography, and oxygen isotopic
  compositions of ultrarefractory inclusions from carbonaceous chondrites}},
  in: \bibinfo{booktitle}{Lunar and Planetary Science Conference}, p.
  \bibinfo{pages}{2416}.
\bibitem[{Krot et~al.(2002)Krot, McKeegan, Leshin, MacPherson and
  Scott}]{krot2002existence}
\bibinfo{author}{Krot, A.N.}, \bibinfo{author}{McKeegan, K.D.},
  \bibinfo{author}{Leshin, L.A.}, \bibinfo{author}{MacPherson, G.J.},
  \bibinfo{author}{Scott, E.R.D.}, \bibinfo{year}{2002}.
\newblock \bibinfo{title}{{Existence of an \ce{^{16}O}-rich gaseous reservoir
  in the solar nebula}}.
\newblock \bibinfo{journal}{Science} \bibinfo{volume}{295},
  \bibinfo{pages}{1051--1054}.
\newblock \DOIprefix\doi{10.1126/science.1068200}.
\bibitem[{Krot et~al.(2008)Krot, Nagashima, Bizzarro, Huss, Davis, Meyer and
  Ulyanov}]{krot2008multiple}
\bibinfo{author}{Krot, A.N.}, \bibinfo{author}{Nagashima, K.},
  \bibinfo{author}{Bizzarro, M.}, \bibinfo{author}{Huss, G.R.},
  \bibinfo{author}{Davis, A.M.}, \bibinfo{author}{Meyer, B.S.},
  \bibinfo{author}{Ulyanov, A.A.}, \bibinfo{year}{2008}.
\newblock \bibinfo{title}{{Multiple generations of refractory inclusions in the
  metal-rich carbonaceous chondrites Acfer 182/214 and Isheyevo}}.
\newblock \bibinfo{journal}{Astrophysical Journal} \bibinfo{volume}{672},
  \bibinfo{pages}{713--721}.
\newblock \DOIprefix\doi{10.1086/521973}.
\bibitem[{Krot et~al.(2018b)Krot, Nagashima, Fintor and
  P\'{a}l-Moln\'{a}r}]{krot2018evidence}
\bibinfo{author}{Krot, A.N.}, \bibinfo{author}{Nagashima, K.},
  \bibinfo{author}{Fintor, K.}, \bibinfo{author}{P\'{a}l-Moln\'{a}r, E.},
  \bibinfo{year}{2018}b.
\newblock \bibinfo{title}{{Evidence for oxygen-isotope exchange in refractory
  inclusions from Kaba (CV3.1) carbonaceous chondrite during fluid-rock
  interaction on the CV parent asteroid}}.
\newblock \bibinfo{journal}{Geochimica et Cosmochimica Acta}
  \DOIprefix\doi{10.1016/j.gca.2018.11.002}.
\bibitem[{Krot et~al.(2015)Krot, Nagashima, Ma and
  Wasserburg}]{krot2015forsterite}
\bibinfo{author}{Krot, A.N.}, \bibinfo{author}{Nagashima, K.},
  \bibinfo{author}{Ma, C.}, \bibinfo{author}{Wasserburg, G.J.},
  \bibinfo{year}{2015}.
\newblock \bibinfo{title}{{Forsterite-bearing Type B CAI with a relict
  eringaite-bearing ultra-refractory CAI}}.
\newblock \bibinfo{journal}{Meteoritics \& Planetary Science}
  \bibinfo{volume}{50}, \bibinfo{pages}{5308}.
\newblock \DOIprefix\doi{10.1111/maps.12501}.
\bibitem[{Krot et~al.(2017a)Krot, Nagashima, Van~Kooten and
  Bizzarro}]{krot2017calcium}
\bibinfo{author}{Krot, A.N.}, \bibinfo{author}{Nagashima, K.},
  \bibinfo{author}{Van~Kooten, E.M.M.E.}, \bibinfo{author}{Bizzarro, M.},
  \bibinfo{year}{2017}a.
\newblock \bibinfo{title}{{Calcium-aluminum-rich inclusions recycled during
  formation of porphyritic chondrules from CH carbonaceous chondrites}}.
\newblock \bibinfo{journal}{Geochimica et Cosmochimica Acta}
  \bibinfo{volume}{201}, \bibinfo{pages}{185--223}.
\newblock \DOIprefix\doi{10.1016/j.gca.2016.09.001}.
\bibitem[{Krot et~al.(2017b)Krot, Nagashima, Van~Kooten and
  Bizzarro}]{krot2017high}
\bibinfo{author}{Krot, A.N.}, \bibinfo{author}{Nagashima, K.},
  \bibinfo{author}{Van~Kooten, E.M.M.E.}, \bibinfo{author}{Bizzarro, M.},
  \bibinfo{year}{2017}b.
\newblock \bibinfo{title}{{High-temperature rims around calcium-aluminum-rich
  inclusions from the CR, CB and CH carbonaceous chondrites}}.
\newblock \bibinfo{journal}{Geochimica et Cosmochimica Acta}
  \bibinfo{volume}{201}, \bibinfo{pages}{155--184}.
\newblock \DOIprefix\doi{10.1016/j.gca.2016.09.031}.
\bibitem[{Krot et~al.(2014)Krot, Nagashima, Wasserburg, Huss, Papanastassiou,
  Davis, Hutcheon and Bizzarro}]{krot2014calcium}
\bibinfo{author}{Krot, A.N.}, \bibinfo{author}{Nagashima, K.},
  \bibinfo{author}{Wasserburg, G.J.}, \bibinfo{author}{Huss, G.R.},
  \bibinfo{author}{Papanastassiou, D.}, \bibinfo{author}{Davis, A.M.},
  \bibinfo{author}{Hutcheon, I.D.}, \bibinfo{author}{Bizzarro, M.},
  \bibinfo{year}{2014}.
\newblock \bibinfo{title}{{Calcium-aluminum-rich inclusions with fractionation
  and unknown nuclear effects (FUN CAIs): I. Mineralogy, petrology, and oxygen
  isotopic compositions}}.
\newblock \bibinfo{journal}{Geochimica et Cosmochimica Acta}
  \bibinfo{volume}{145}, \bibinfo{pages}{206--247}.
\newblock \DOIprefix\doi{10.1016/j.gca.2014.09.027}.
\bibitem[{Krot et~al.(2004)Krot, Petaev, Russell, Itoh, Fagan, Yurimoto,
  Chizmadia, Weisberg, Komatsu and Ulyanov}]{krot2004amoeboid}
\bibinfo{author}{Krot, A.N.}, \bibinfo{author}{Petaev, M.I.},
  \bibinfo{author}{Russell, S.S.}, \bibinfo{author}{Itoh, S.},
  \bibinfo{author}{Fagan, T.J.}, \bibinfo{author}{Yurimoto, H.},
  \bibinfo{author}{Chizmadia, L.}, \bibinfo{author}{Weisberg, M.K.},
  \bibinfo{author}{Komatsu, M.}, \bibinfo{author}{Ulyanov, A.A.},
  \bibinfo{year}{2004}.
\newblock \bibinfo{title}{{Amoeboid olivine aggregates and related objects in
  carbonaceous chondrites: Records of nebular and asteroid processes}}.
\newblock \bibinfo{journal}{Chemie der Erde-Geochemistry} \bibinfo{volume}{64},
  \bibinfo{pages}{185--239}.
\newblock \DOIprefix\doi{10.1016/j.chemer.2004.05.001}.
\bibitem[{Krot et~al.(1995)Krot, Scott and Zolensky}]{krot1995mineralogical}
\bibinfo{author}{Krot, A.N.}, \bibinfo{author}{Scott, E.R.D.},
  \bibinfo{author}{Zolensky, M.E.}, \bibinfo{year}{1995}.
\newblock \bibinfo{title}{{Mineralogical and chemical modification of
  components in CV3 chondrites: Nebular or asteroidal processing?}}
\newblock \bibinfo{journal}{Meteoritics} \bibinfo{volume}{30},
  \bibinfo{pages}{748--775}.
\newblock \DOIprefix\doi{10.1111/j.1945-5100.1995.tb01173.x}.
\bibitem[{Krot et~al.(2001)Krot, Ulyanov, Meibom and Keil}]{krot2001forsterite}
\bibinfo{author}{Krot, A.N.}, \bibinfo{author}{Ulyanov, A.A.},
  \bibinfo{author}{Meibom, A.}, \bibinfo{author}{Keil, K.},
  \bibinfo{year}{2001}.
\newblock \bibinfo{title}{{Forsterite-rich accretionary rims around
  calcium-aluminum-rich inclusions from the reduced CV3 chondrite Efremovka}}.
\newblock \bibinfo{journal}{Meteoritics \& Planetary Science}
  \bibinfo{volume}{36}, \bibinfo{pages}{611--628}.
\newblock \DOIprefix\doi{10.1111/j.1945-5100.2001.tb01904.x}.
\bibitem[{Kruijer et~al.(2017)Kruijer, Burkhardt, Budde and
  Kleine}]{kruijer2017age}
\bibinfo{author}{Kruijer, T.S.}, \bibinfo{author}{Burkhardt, C.},
  \bibinfo{author}{Budde, G.}, \bibinfo{author}{Kleine, T.},
  \bibinfo{year}{2017}.
\newblock \bibinfo{title}{{Age of Jupiter inferred from the distinct genetics
  and formation times of meteorites}}.
\newblock \bibinfo{journal}{Proceedings of the National Academy of Sciences}
  \bibinfo{volume}{114}, \bibinfo{pages}{6712--6716}.
\newblock \DOIprefix\doi{10.1073/pnas.1704461114}.
\bibitem[{Lee et~al.(2008)Lee, Bergin and Lyons}]{lee2008oxygen}
\bibinfo{author}{Lee, J.E.}, \bibinfo{author}{Bergin, E.A.},
  \bibinfo{author}{Lyons, J.R.}, \bibinfo{year}{2008}.
\newblock \bibinfo{title}{{Oxygen isotope anomalies of the Sun and the original
  environment of the solar system}}.
\newblock \bibinfo{journal}{Meteoritics \& Planetary Science}
  \bibinfo{volume}{43}, \bibinfo{pages}{1351--1362}.
\newblock \DOIprefix\doi{10.1111/j.1945-5100.2008.tb00702.x}.
\bibitem[{Lin et~al.(2003)Lin, Kimura and Wang}]{lin2003fassaites}
\bibinfo{author}{Lin, Y.}, \bibinfo{author}{Kimura, M.}, \bibinfo{author}{Wang,
  D.}, \bibinfo{year}{2003}.
\newblock \bibinfo{title}{{Fassaites in compact type A Ca-Al-rich inclusions in
  the Ningqiang carbonaceous chondrite: Evidence for partial melting in the
  nebula}}.
\newblock \bibinfo{journal}{Meteoritics \& Planetary Science}
  \bibinfo{volume}{38}, \bibinfo{pages}{407--417}.
\newblock \DOIprefix\doi{10.1111/j.1945-5100.2003.tb00276.x}.
\bibitem[{Lodders(2003)}]{lodders2003solar}
\bibinfo{author}{Lodders, K.}, \bibinfo{year}{2003}.
\newblock \bibinfo{title}{{Solar system abundances and condensation
  temperatures of the elements}}.
\newblock \bibinfo{journal}{Astrophysical Journal} \bibinfo{volume}{591},
  \bibinfo{pages}{1220--1247}.
\newblock \DOIprefix\doi{10.1086/375492}.
\bibitem[{Lunning et~al.(2016)Lunning, Corrigan, McSween, Tenner, Kita and
  Bodnar}]{lunning2016cv}
\bibinfo{author}{Lunning, N.G.}, \bibinfo{author}{Corrigan, C.M.},
  \bibinfo{author}{McSween, H.Y.}, \bibinfo{author}{Tenner, T.J.},
  \bibinfo{author}{Kita, N.T.}, \bibinfo{author}{Bodnar, R.J.},
  \bibinfo{year}{2016}.
\newblock \bibinfo{title}{{CV and CM chondrite impact melts}}.
\newblock \bibinfo{journal}{Geochimica et Cosmochimica Acta}
  \bibinfo{volume}{189}, \bibinfo{pages}{338--358}.
\newblock \DOIprefix\doi{10.1016/j.gca.2016.05.038}.
\bibitem[{Lyons and Young(2005)}]{lyons2005co}
\bibinfo{author}{Lyons, J.R.}, \bibinfo{author}{Young, E.D.},
  \bibinfo{year}{2005}.
\newblock \bibinfo{title}{{CO} self-shielding as the origin of oxygen isotope
  anomalies in the early solar nebula}.
\newblock \bibinfo{journal}{Nature} \bibinfo{volume}{435},
  \bibinfo{pages}{317--320}.
\newblock \DOIprefix\doi{10.1038/nature03557}.
\bibitem[{Ma(2012)}]{ma2012discovery}
\bibinfo{author}{Ma, C.}, \bibinfo{year}{2012}.
\newblock \bibinfo{title}{{Discovery of meteoritic eringaite, $\rm
  Ca_3(Sc,Y,Ti)_2Si_3O_{12} $, the first solar garnet?}}
\newblock \bibinfo{journal}{Meteoritics \& Planetary Science}
  \bibinfo{volume}{47}, \bibinfo{pages}{A256}.
\newblock \DOIprefix\doi{10.1111/j.1945-5100.2012.01401_2.x}.
\bibitem[{Ma and Beckett(2016)}]{ma2016burnettite}
\bibinfo{author}{Ma, C.}, \bibinfo{author}{Beckett, J.R.},
  \bibinfo{year}{2016}.
\newblock \bibinfo{title}{{Burnettite, \ce{CaVAlSiO6}, and paqueite,
  \ce{Ca3TiSi2(Al2Ti)O14}, two new minerals from Allende: Clues to the
  evolution of a V-rich Ca-Al-rich inclusion}}, in: \bibinfo{booktitle}{Lunar
  and Planetary Science Conference}, p. \bibinfo{pages}{1595}.
\bibitem[{Ma et~al.(2011)Ma, Beckett, Tschauner and
  Rossman}]{ma2011thortveitite}
\bibinfo{author}{Ma, C.}, \bibinfo{author}{Beckett, J.R.},
  \bibinfo{author}{Tschauner, O.}, \bibinfo{author}{Rossman, G.R.},
  \bibinfo{year}{2011}.
\newblock \bibinfo{title}{{Thortveitite (\ce{Sc2Si2O7}), the first solar
  silicate?}}
\newblock \bibinfo{journal}{Meteoritics \& Planetary Science}
  \bibinfo{volume}{46}, \bibinfo{pages}{A144}.
\newblock \DOIprefix\doi{10.1111/j.1945-5100.2011.01221.x}.
\bibitem[{Ma and Rossman(2009a)}]{ma2009davisite}
\bibinfo{author}{Ma, C.}, \bibinfo{author}{Rossman, G.R.},
  \bibinfo{year}{2009}a.
\newblock \bibinfo{title}{{Davisite, \ce{CaScAlSiO6}, a new pyroxene from the
  Allende meteorite}}.
\newblock \bibinfo{journal}{American Mineralogist} \bibinfo{volume}{94},
  \bibinfo{pages}{845--848}.
\newblock \DOIprefix\doi{10.2138/am.2009.3310}.
\bibitem[{Ma and Rossman(2009b)}]{ma2009grossmanite}
\bibinfo{author}{Ma, C.}, \bibinfo{author}{Rossman, G.R.},
  \bibinfo{year}{2009}b.
\newblock \bibinfo{title}{{Grossmanite, \ce{CaTi^{3+}AlSiO6}, a new pyroxene
  from the Allende meteorite}}.
\newblock \bibinfo{journal}{American Mineralogist} \bibinfo{volume}{94},
  \bibinfo{pages}{1491--1494}.
\newblock \DOIprefix\doi{10.2138/am.2009.3310}.
\bibitem[{Ma et~al.(2012)Ma, Tschauner, Beckett, Rossman and
  Liu}]{ma2012panguite}
\bibinfo{author}{Ma, C.}, \bibinfo{author}{Tschauner, O.},
  \bibinfo{author}{Beckett, J.R.}, \bibinfo{author}{Rossman, G.R.},
  \bibinfo{author}{Liu, W.}, \bibinfo{year}{2012}.
\newblock \bibinfo{title}{{Panguite, ($\rm \ce{Ti^{4+}}, \ce{Sc}, \ce{Al},
  \ce{Mg}, \ce{Zr}, \ce{Ca})_{1.8} O_3$, a new ultra-refractory titania mineral
  from the Allende meteorite: Synchrotron micro-diffraction and EBSD}}.
\newblock \bibinfo{journal}{American Mineralogist} \bibinfo{volume}{97},
  \bibinfo{pages}{1219--1225}.
\newblock \DOIprefix\doi{10.2138/am.2012.4027}.
\bibitem[{Ma et~al.(2013)Ma, Tschauner, Beckett, Rossman and
  Liu}]{ma2013kangite}
\bibinfo{author}{Ma, C.}, \bibinfo{author}{Tschauner, O.},
  \bibinfo{author}{Beckett, J.R.}, \bibinfo{author}{Rossman, G.R.},
  \bibinfo{author}{Liu, W.}, \bibinfo{year}{2013}.
\newblock \bibinfo{title}{{Kangite, ($\rm \ce{Sc}, \ce{Ti}, \ce{Al}, \ce{Zr},
  \ce{Mg}, \ce{Ca}, \square)_2O_3$}, a new ultra-refractory scandia mineral
  from the allende meteorite: Synchrotron micro-laue diffraction and electron
  backscatter diffraction}.
\newblock \bibinfo{journal}{American Mineralogist} \bibinfo{volume}{98},
  \bibinfo{pages}{870--878}.
\newblock \DOIprefix\doi{10.2138/am.2013.4290}.
\bibitem[{Ma et~al.(2017)Ma, Yoshizaki, Krot, Beckett, Nakamura, Nagashima,
  Muto and Ivanova}]{ma2017discovery}
\bibinfo{author}{Ma, C.}, \bibinfo{author}{Yoshizaki, T.},
  \bibinfo{author}{Krot, A.N.}, \bibinfo{author}{Beckett, J.R.},
  \bibinfo{author}{Nakamura, T.}, \bibinfo{author}{Nagashima, K.},
  \bibinfo{author}{Muto, J.}, \bibinfo{author}{Ivanova, M.A.},
  \bibinfo{year}{2017}.
\newblock \bibinfo{title}{{Discovery of rubinite, \ce{Ca3Ti^{3+}_2Si3O_{12}}, a
  new garnet mineral in refractory inclusions from carbonaceous chondrites}}.
\newblock \bibinfo{journal}{Meteoritics \& Planetary Science}
  \bibinfo{volume}{52}, \bibinfo{pages}{A211}.
\newblock \DOIprefix\doi{10.1111/maps.12934}.
\bibitem[{MacPherson(2014)}]{macpherson2014calcium}
\bibinfo{author}{MacPherson, G.J.}, \bibinfo{year}{2014}.
\newblock \bibinfo{title}{{Calcium-aluminum-rich inclusions in chondritic
  meteorites}}, in: \bibinfo{editor}{Holland, H.D.}, \bibinfo{editor}{Turekian,
  K.K.} (Eds.), \bibinfo{booktitle}{Meteorites and Cosmochemical Processes}.
  \bibinfo{publisher}{Elsevier}, \bibinfo{address}{Oxford}.
  volume~\bibinfo{volume}{1} of \textit{\bibinfo{series}{Treatise on
  Geochemistry (Second Edition)}}, pp. \bibinfo{pages}{139--179}.
\newblock \DOIprefix\doi{10.1016/B978-0-08-095975-7.00105-4}.
\bibitem[{MacPherson and Grossman(1984)}]{macpherson1984fluffy}
\bibinfo{author}{MacPherson, G.J.}, \bibinfo{author}{Grossman, L.},
  \bibinfo{year}{1984}.
\newblock \bibinfo{title}{{"Fluffy" Type A Ca-, Al-rich inclusions in the
  Allende meteorite}}.
\newblock \bibinfo{journal}{Geochimica et Cosmochimica Acta}
  \bibinfo{volume}{48}, \bibinfo{pages}{29--46}.
\newblock \DOIprefix\doi{10.1016/0016-7037(84)90347-8}.
\bibitem[{MacPherson and Huss(2005)}]{macpherson2005petrogenesis}
\bibinfo{author}{MacPherson, G.J.}, \bibinfo{author}{Huss, G.R.},
  \bibinfo{year}{2005}.
\newblock \bibinfo{title}{{Petrogenesis of Al-rich chondrules: Evidence from
  bulk compositions and phase equilibria}}.
\newblock \bibinfo{journal}{Geochimica et Cosmochimica Acta}
  \bibinfo{volume}{69}, \bibinfo{pages}{3099--3127}.
\newblock \DOIprefix\doi{10.1016/j.gca.2004.12.022}.
\bibitem[{Makide et~al.(2009)Makide, Nagashima, Krot, Huss, Hutcheon and
  Bischoff}]{makide2009oxygen}
\bibinfo{author}{Makide, K.}, \bibinfo{author}{Nagashima, K.},
  \bibinfo{author}{Krot, A.N.}, \bibinfo{author}{Huss, G.R.},
  \bibinfo{author}{Hutcheon, I.D.}, \bibinfo{author}{Bischoff, A.},
  \bibinfo{year}{2009}.
\newblock \bibinfo{title}{{Oxygen-and magnesium-isotope compositions of
  calcium-aluminum-rich inclusions from CR2 carbonaceous chondrites}}.
\newblock \bibinfo{journal}{Geochimica et Cosmochimica Acta}
  \bibinfo{volume}{73}, \bibinfo{pages}{5018--5050}.
\newblock \DOIprefix\doi{10.1016/j.gca.2009.01.042}.
\bibitem[{Mane et~al.(2016)Mane, Bose, Defouilloy, Kita, MacPherson and
  Wadhwa}]{mane2016formation}
\bibinfo{author}{Mane, P.}, \bibinfo{author}{Bose, M.},
  \bibinfo{author}{Defouilloy, C.}, \bibinfo{author}{Kita, N.T.},
  \bibinfo{author}{MacPherson, G.J.}, \bibinfo{author}{Wadhwa, M.},
  \bibinfo{year}{2016}.
\newblock \bibinfo{title}{{Formation timescales of Wark-Lovering rims around
  calcium-aluminum rich inclusions}}, in: \bibinfo{booktitle}{Lunar and
  Planetary Science Conference}, p. \bibinfo{pages}{2560}.
\bibitem[{Mane et~al.(2015)Mane, Bose and Wadhwa}]{mane2015resolved}
\bibinfo{author}{Mane, P.}, \bibinfo{author}{Bose, M.},
  \bibinfo{author}{Wadhwa, M.}, \bibinfo{year}{2015}.
\newblock \bibinfo{title}{{Resolved time difference between calcium aluminum
  rich inclusions and their Wark Lovering rims inferred from Al-Mg chronology
  of two inclusions from a CV3 carbonaceous chondrite}}, in:
  \bibinfo{booktitle}{Lunar and Planetary Science Conference}, p.
  \bibinfo{pages}{2898}.
\bibitem[{Matzel et~al.(2015)Matzel, Jacobsen and Simon}]{matzel2015aluminum}
\bibinfo{author}{Matzel, J.}, \bibinfo{author}{Jacobsen, B.},
  \bibinfo{author}{Simon, J.I.}, \bibinfo{year}{2015}.
\newblock \bibinfo{title}{{Aluminum-magnesium chronology of the rim of a
  Murchison type A CAI}}, in: \bibinfo{booktitle}{Meteoritics \& Planetary
  Science}, p. \bibinfo{pages}{A241}.
\newblock \DOIprefix\doi{10.1111/maps.12501}.
\bibitem[{Matzel et~al.(2013)Matzel, Simon, Hutcheon, Jacobsen, Simon and
  Grossman}]{matzel2013oxygen}
\bibinfo{author}{Matzel, J.E.P.}, \bibinfo{author}{Simon, J.I.},
  \bibinfo{author}{Hutcheon, I.D.}, \bibinfo{author}{Jacobsen, B.},
  \bibinfo{author}{Simon, S.B.}, \bibinfo{author}{Grossman, L.},
  \bibinfo{year}{2013}.
\newblock \bibinfo{title}{{Oxygen isotope measurements of a rare Murchison Type
  A CAI and its rim}}, in: \bibinfo{booktitle}{Lunar and Planetary Science
  Conference}, p. \bibinfo{pages}{2632}.
\bibitem[{McKeegan et~al.(2006)McKeegan, Al{\'e}on, Bradley, Brownlee,
  Busemann, Butterworth, Chaussidon, Fallon, Floss, Gilmour, Gounelle, Graham,
  Guan, Heck, Hoppe, Hutcheon, Huth, Ishii, Ito, Jacobsen, Kearsley, Leshin,
  Liu, Lyon, Marhas, Marty, Matrajt, Meibom, Messenger, Mostefaoui,
  Mukhopadhyay, Nakamura-Messenger, Nittler, Palma, Pepin, Papanastassiou,
  Robert, Schlutter, Snead, Stadermann, Stroud, Tsou, Westphal, Young, Ziegler,
  Zimmermann and Zinner}]{mckeegan2006isotopic}
\bibinfo{author}{McKeegan, K.D.}, \bibinfo{author}{Al{\'e}on, J.},
  \bibinfo{author}{Bradley, J.}, \bibinfo{author}{Brownlee, D.},
  \bibinfo{author}{Busemann, H.}, \bibinfo{author}{Butterworth, A.},
  \bibinfo{author}{Chaussidon, M.}, \bibinfo{author}{Fallon, S.},
  \bibinfo{author}{Floss, C.}, \bibinfo{author}{Gilmour, J.},
  \bibinfo{author}{Gounelle, M.}, \bibinfo{author}{Graham, G.},
  \bibinfo{author}{Guan, Y.}, \bibinfo{author}{Heck, P.R.},
  \bibinfo{author}{Hoppe, P.}, \bibinfo{author}{Hutcheon, I.D.},
  \bibinfo{author}{Huth, J.}, \bibinfo{author}{Ishii, H.},
  \bibinfo{author}{Ito, M.}, \bibinfo{author}{Jacobsen, S.B.},
  \bibinfo{author}{Kearsley, A.}, \bibinfo{author}{Leshin, L.A.},
  \bibinfo{author}{Liu, M.C.}, \bibinfo{author}{Lyon, I.},
  \bibinfo{author}{Marhas, K.}, \bibinfo{author}{Marty, B.},
  \bibinfo{author}{Matrajt, G.}, \bibinfo{author}{Meibom, A.},
  \bibinfo{author}{Messenger, S.}, \bibinfo{author}{Mostefaoui, S.},
  \bibinfo{author}{Mukhopadhyay, S.}, \bibinfo{author}{Nakamura-Messenger, K.},
  \bibinfo{author}{Nittler, L.}, \bibinfo{author}{Palma, R.},
  \bibinfo{author}{Pepin, R.O.}, \bibinfo{author}{Papanastassiou, D.A.},
  \bibinfo{author}{Robert, F.}, \bibinfo{author}{Schlutter, D.},
  \bibinfo{author}{Snead, C.J.}, \bibinfo{author}{Stadermann, F.J.},
  \bibinfo{author}{Stroud, R.}, \bibinfo{author}{Tsou, P.},
  \bibinfo{author}{Westphal, A.}, \bibinfo{author}{Young, E.D.},
  \bibinfo{author}{Ziegler, K.}, \bibinfo{author}{Zimmermann, L.},
  \bibinfo{author}{Zinner, E.}, \bibinfo{year}{2006}.
\newblock \bibinfo{title}{{Isotopic compositions of cometary matter returned by
  Stardust}}.
\newblock \bibinfo{journal}{Science} \bibinfo{volume}{314},
  \bibinfo{pages}{1724--1728}.
\newblock \DOIprefix\doi{10.1126/science.1135992}.
\bibitem[{McKeegan et~al.(2000)McKeegan, Chaussidon and
  Robert}]{mckeegan2000incorporation}
\bibinfo{author}{McKeegan, K.D.}, \bibinfo{author}{Chaussidon, M.},
  \bibinfo{author}{Robert, F.}, \bibinfo{year}{2000}.
\newblock \bibinfo{title}{{Incorporation of short-lived \ce{^{10}Be} in a
  calcium-aluminum-rich inclusion from the Allende meteorite}}.
\newblock \bibinfo{journal}{Science} \bibinfo{volume}{289},
  \bibinfo{pages}{1334--1337}.
\newblock \DOIprefix\doi{10.1126/science.289.5483.1334}.
\bibitem[{McKeegan et~al.(2011)McKeegan, Kallio, Heber, Jarzebinski, Mao,
  Coath, Kunihiro, Wiens, Nordholt, Moses~Jr., Reisenfeld, Jurewicz and
  Burnett}]{mckeegan2011oxygen}
\bibinfo{author}{McKeegan, K.D.}, \bibinfo{author}{Kallio, A.P.A.},
  \bibinfo{author}{Heber, V.S.}, \bibinfo{author}{Jarzebinski, G.},
  \bibinfo{author}{Mao, P.H.}, \bibinfo{author}{Coath, C.D.},
  \bibinfo{author}{Kunihiro, T.}, \bibinfo{author}{Wiens, R.C.},
  \bibinfo{author}{Nordholt, J.E.}, \bibinfo{author}{Moses~Jr., R.W.},
  \bibinfo{author}{Reisenfeld, D.B.}, \bibinfo{author}{Jurewicz, A.J.G.},
  \bibinfo{author}{Burnett, D.S.}, \bibinfo{year}{2011}.
\newblock \bibinfo{title}{{The oxygen isotopic composition of the Sun inferred
  from captured solar wind}}.
\newblock \bibinfo{journal}{Science} \bibinfo{volume}{332},
  \bibinfo{pages}{1528--1532}.
\newblock \DOIprefix\doi{10.1126/science.1204636}.
\bibitem[{Meeker et~al.(1983)Meeker, Wasserburg and
  Armstrong}]{meeker1983replacement}
\bibinfo{author}{Meeker, G.P.}, \bibinfo{author}{Wasserburg, G.J.},
  \bibinfo{author}{Armstrong, J.T.}, \bibinfo{year}{1983}.
\newblock \bibinfo{title}{{Replacement textures in CAI and implications
  regarding planetary metamorphism}}.
\newblock \bibinfo{journal}{Geochimica et Cosmochimica Acta}
  \bibinfo{volume}{47}, \bibinfo{pages}{707--721}.
\newblock \DOIprefix\doi{10.1016/0016-7037(83)90105-9}.
\bibitem[{Nagasawa et~al.(2001)Nagasawa, Suzuki, Ito and
  Morioka}]{nagasawa2001diffusion}
\bibinfo{author}{Nagasawa, H.}, \bibinfo{author}{Suzuki, T.},
  \bibinfo{author}{Ito, M.}, \bibinfo{author}{Morioka, M.},
  \bibinfo{year}{2001}.
\newblock \bibinfo{title}{{Diffusion in single crystal of melilite:
  Interdiffusion of Al + Al vs. Mg + Si}}.
\newblock \bibinfo{journal}{Physics and Chemistry of Minerals}
  \bibinfo{volume}{28}, \bibinfo{pages}{706--710}.
\newblock \DOIprefix\doi{10.1007/s002690100212}.
\bibitem[{Nichols and Mackwell(1991)}]{nichols1991grain}
\bibinfo{author}{Nichols, S.J.}, \bibinfo{author}{Mackwell, S.J.},
  \bibinfo{year}{1991}.
\newblock \bibinfo{title}{{Grain growth in porous olivine aggregates}}.
\newblock \bibinfo{journal}{Physics and Chemistry of Minerals}
  \bibinfo{volume}{18}, \bibinfo{pages}{269--278}.
\newblock \DOIprefix\doi{10.1007/BF00202580}.
\bibitem[{Noonan et~al.(1977)Noonan, Nelen, Fredriksson and
  Newbury}]{noonan1977zr}
\bibinfo{author}{Noonan, A.F.}, \bibinfo{author}{Nelen, J.},
  \bibinfo{author}{Fredriksson, K.}, \bibinfo{author}{Newbury, D.},
  \bibinfo{year}{1977}.
\newblock \bibinfo{title}{{Zr-Y oxides and high-alkali glass in an ameboid
  inclusion from Ornans}}.
\newblock \bibinfo{journal}{Meteoritics} \bibinfo{volume}{12},
  \bibinfo{pages}{332}.
\bibitem[{Ohashi(1978)}]{ohashi1978studies}
\bibinfo{author}{Ohashi, H.}, \bibinfo{year}{1978}.
\newblock \bibinfo{title}{{Studies on \ce{CaScAlSiO6}-pyroxene}}.
\newblock \bibinfo{journal}{The Journal of the Japanese Association of
  Mineralogists, Petrologists and Economic Geologists} \bibinfo{volume}{73},
  \bibinfo{pages}{58--61}.
\newblock \DOIprefix\doi{10.2465/ganko1941.73.58}.
\bibitem[{Osborn and Schairer(1941)}]{osborn1941ternary}
\bibinfo{author}{Osborn, E.F.}, \bibinfo{author}{Schairer, J.F.},
  \bibinfo{year}{1941}.
\newblock \bibinfo{title}{{The ternary system
  pseudowollastonite--\aa{}kermanite--gehlenite}}.
\newblock \bibinfo{journal}{American Journal of Science} \bibinfo{volume}{239},
  \bibinfo{pages}{715--763}.
\newblock \DOIprefix\doi{10.2475/ajs.239.10.715}.
\bibitem[{Park et~al.(2012a)Park, Wakaki, Sakamoto, Kobayashi and
  Yurimoto}]{park2012oxygen}
\bibinfo{author}{Park, C.}, \bibinfo{author}{Wakaki, S.},
  \bibinfo{author}{Sakamoto, N.}, \bibinfo{author}{Kobayashi, S.},
  \bibinfo{author}{Yurimoto, H.}, \bibinfo{year}{2012}a.
\newblock \bibinfo{title}{{Oxygen isotopic composition of the solar nebula gas
  inferred from high-precision isotope imaging of melilite crystals in an
  Allende CAI}}.
\newblock \bibinfo{journal}{Meteoritics \& Planetary Science}
  \bibinfo{volume}{47}, \bibinfo{pages}{2070--2083}.
\newblock \DOIprefix\doi{10.1111/maps.12032}.
\bibitem[{Park et~al.(2012b)Park, Wakaki and Yurimoto}]{park2012oxygen_SIA}
\bibinfo{author}{Park, C.}, \bibinfo{author}{Wakaki, S.},
  \bibinfo{author}{Yurimoto, H.}, \bibinfo{year}{2012}b.
\newblock \bibinfo{title}{{Oxygen isotopic variations in a type A Ca--Al-rich
  inclusion revealed by high-precision secondary ion mass spectrometry analysis
  with micrometer resolution}}.
\newblock \bibinfo{journal}{Surface and Interface Analysis}
  \bibinfo{volume}{44}, \bibinfo{pages}{678--681}.
\newblock \DOIprefix\doi{10.1002/sia.3871}.
\bibitem[{Petaev and Wood(1998)}]{petaev1998condensation}
\bibinfo{author}{Petaev, M.I.}, \bibinfo{author}{Wood, J.A.},
  \bibinfo{year}{1998}.
\newblock \bibinfo{title}{{The condensation with partial isolation (CWPI) model
  of condensation in the solar nebula}}.
\newblock \bibinfo{journal}{Meteoritics \& Planetary Science}
  \bibinfo{volume}{33}, \bibinfo{pages}{1123--1137}.
\newblock \DOIprefix\doi{10.1111/j.1945-5100.1998.tb01717.x}.
\bibitem[{Petaev and Wood(2005)}]{petaev2005meteoritic}
\bibinfo{author}{Petaev, M.I.}, \bibinfo{author}{Wood, J.A.},
  \bibinfo{year}{2005}.
\newblock \bibinfo{title}{{Meteoritic constraints on temperatures, pressures,
  cooling rates, chemical compositions and modes of condensation in the solar
  nebula}}, in: \bibinfo{editor}{Krot, A.N.}, \bibinfo{editor}{Scott, E.R.D.},
  \bibinfo{editor}{Reipurth, B.} (Eds.), \bibinfo{booktitle}{Chondrites and the
  Protoplanetary Disk}. \bibinfo{publisher}{Astronomical Society of the
  Pacific}, \bibinfo{address}{San Francisco}. volume \bibinfo{volume}{341} of
  \textit{\bibinfo{series}{ASP Conference Series}}, pp.
  \bibinfo{pages}{373--406}.
\bibitem[{Rubin(2011)}]{rubin2011origin}
\bibinfo{author}{Rubin, A.E.}, \bibinfo{year}{2011}.
\newblock \bibinfo{title}{{Origin of the differences in
  refractory-lithophile-element abundances among chondrite groups}}.
\newblock \bibinfo{journal}{Icarus} \bibinfo{volume}{213},
  \bibinfo{pages}{547--558}.
\newblock \DOIprefix\doi{10.1016/j.icarus.2011.04.003}.
\bibitem[{Rubin(2012)}]{rubin2012new}
\bibinfo{author}{Rubin, A.E.}, \bibinfo{year}{2012}.
\newblock \bibinfo{title}{{A new model for the origin of Type-B and Fluffy
  Type-A CAIs: Analogies to remelted compound chondrules}}.
\newblock \bibinfo{journal}{Meteoritics \& Planetary Science}
  \bibinfo{volume}{47}, \bibinfo{pages}{1062--1074}.
\newblock \DOIprefix\doi{10.1111/j.1945-5100.2012.01374.x}.
\bibitem[{Rubin(2018)}]{rubin2018differences}
\bibinfo{author}{Rubin, A.E.}, \bibinfo{year}{2018}.
\newblock \bibinfo{title}{{Differences in chemical, physical, and collective
  properties between carbonaceous and non-carbonaceous magmatic iron
  meteorites}}, in: \bibinfo{booktitle}{Lunar and Planetary Science
  Conference}, p. \bibinfo{pages}{1034}.
\bibitem[{Ryerson and McKeegan(1994)}]{ryerson1994determination}
\bibinfo{author}{Ryerson, F.J.}, \bibinfo{author}{McKeegan, K.D.},
  \bibinfo{year}{1994}.
\newblock \bibinfo{title}{{Determination of oxygen self-diffusion in
  {\aa}kermanite, anorthite, diopside, and spinel: Implications for oxygen
  isotopic anomalies and the thermal histories of Ca-Al-rich inclusions}}.
\newblock \bibinfo{journal}{Geochimica et Cosmochimica Acta}
  \bibinfo{volume}{58}, \bibinfo{pages}{3713--3734}.
\newblock \DOIprefix\doi{10.1016/0016-7037(94)90161-9}.
\bibitem[{Sakamoto et~al.(2008)Sakamoto, Itoh and
  Yurimoto}]{sakamoto2008discovery}
\bibinfo{author}{Sakamoto, N.}, \bibinfo{author}{Itoh, S.},
  \bibinfo{author}{Yurimoto, H.}, \bibinfo{year}{2008}.
\newblock \bibinfo{title}{{Discovery of \ce{^{17,18}O}-rich material from
  meteorite by direct-imaging method using stigmatic-SIMS and 2D ion
  detector}}.
\newblock \bibinfo{journal}{Applied Surface Science} \bibinfo{volume}{255},
  \bibinfo{pages}{1458--1460}.
\newblock \DOIprefix\doi{10.1016/j.apsusc.2008.05.052}.
\bibitem[{Scott et~al.(2018)Scott, Krot and Sanders}]{scott2018isotopic}
\bibinfo{author}{Scott, E.R.D.}, \bibinfo{author}{Krot, A.N.},
  \bibinfo{author}{Sanders, I.S.}, \bibinfo{year}{2018}.
\newblock \bibinfo{title}{{Isotopic dichotomy among meteorites and its bearing
  on the protoplanetary disk}}.
\newblock \bibinfo{journal}{The Astrophysical Journal} \bibinfo{volume}{854},
  \bibinfo{pages}{164}.
\newblock \DOIprefix\doi{10.3847/1538-4357/aaa5a5}.
\bibitem[{Shu et~al.(1996)Shu, Shang and Lee}]{shu1996toward}
\bibinfo{author}{Shu, F.H.}, \bibinfo{author}{Shang, H.}, \bibinfo{author}{Lee,
  T.}, \bibinfo{year}{1996}.
\newblock \bibinfo{title}{{Toward an astrophysical theory of chondrites}}.
\newblock \bibinfo{journal}{Science} \bibinfo{volume}{271},
  \bibinfo{pages}{1545--1552}.
\newblock \DOIprefix\doi{10.1126/science.271.5255.1545}.
\bibitem[{Simon et~al.(2011)Simon, Hutcheon, Simon, Matzel, Ramon, Weber,
  Grossman and DePaolo}]{simon2011oxygen}
\bibinfo{author}{Simon, J.I.}, \bibinfo{author}{Hutcheon, I.D.},
  \bibinfo{author}{Simon, S.B.}, \bibinfo{author}{Matzel, J.E.P.},
  \bibinfo{author}{Ramon, E.C.}, \bibinfo{author}{Weber, P.K.},
  \bibinfo{author}{Grossman, L.}, \bibinfo{author}{DePaolo, D.J.},
  \bibinfo{year}{2011}.
\newblock \bibinfo{title}{{Oxygen isotope variations at the margin of a CAI
  records circulation within the solar nebula}}.
\newblock \bibinfo{journal}{Science} \bibinfo{volume}{331},
  \bibinfo{pages}{1175--1178}.
\newblock \DOIprefix\doi{10.1126/science.1197970}.
\bibitem[{Simon et~al.(2016)Simon, Matzel, Simon, Hutcheon, Ross, Weber and
  Grossman}]{simon2016oxygen}
\bibinfo{author}{Simon, J.I.}, \bibinfo{author}{Matzel, J.E.P.},
  \bibinfo{author}{Simon, S.B.}, \bibinfo{author}{Hutcheon, I.D.},
  \bibinfo{author}{Ross, D.K.}, \bibinfo{author}{Weber, P.K.},
  \bibinfo{author}{Grossman, L.}, \bibinfo{year}{2016}.
\newblock \bibinfo{title}{{Oxygen isotopic variations in the outer margins and
  Wark-Lovering rims of refractory inclusions}}.
\newblock \bibinfo{journal}{Geochimica et Cosmochimica Acta}
  \bibinfo{volume}{186}, \bibinfo{pages}{242--276}.
\newblock \DOIprefix\doi{10.1016/j.gca.2016.04.025}.
\bibitem[{Simon et~al.(2005)Simon, Young, Russell, Tonui, Dyl and
  Manning}]{simon2005short}
\bibinfo{author}{Simon, J.I.}, \bibinfo{author}{Young, E.D.},
  \bibinfo{author}{Russell, S.S.}, \bibinfo{author}{Tonui, E.K.},
  \bibinfo{author}{Dyl, K.A.}, \bibinfo{author}{Manning, C.E.},
  \bibinfo{year}{2005}.
\newblock \bibinfo{title}{{A short timescale for changing oxygen fugacity in
  the solar nebula revealed by high-resolution $\rm ^{26}Al-^{26}Mg$ dating of
  CAI rims}}.
\newblock \bibinfo{journal}{Earth and Planetary Science Letters}
  \bibinfo{volume}{238}, \bibinfo{pages}{272--283}.
\newblock \DOIprefix\doi{10.1016/j.epsl.2005.08.004}.
\bibitem[{Simon et~al.(1996)Simon, Davis and Grossman}]{simon1996unique}
\bibinfo{author}{Simon, S.B.}, \bibinfo{author}{Davis, A.M.},
  \bibinfo{author}{Grossman, L.}, \bibinfo{year}{1996}.
\newblock \bibinfo{title}{{A unique ultrarefractory inclusion from the
  Murchison meteorite}}.
\newblock \bibinfo{journal}{Meteoritics \& Planetary Science}
  \bibinfo{volume}{31}, \bibinfo{pages}{106--115}.
\newblock \DOIprefix\doi{10.1111/j.1945-5100.1996.tb02060.x}.
\bibitem[{Simon and Grossman(1997)}]{simon1997situ}
\bibinfo{author}{Simon, S.B.}, \bibinfo{author}{Grossman, L.},
  \bibinfo{year}{1997}.
\newblock \bibinfo{title}{{\textit{In situ} formation of palisade bodies in
  calcium, aluminum-rich refractory inclusions}}.
\newblock \bibinfo{journal}{Meteoritics \& Planetary Science}
  \bibinfo{volume}{32}, \bibinfo{pages}{61--70}.
\newblock \DOIprefix\doi{10.1111/j.1945-5100.1997.tb01241.x}.
\bibitem[{Simon and Grossman(2004)}]{simon2004preferred}
\bibinfo{author}{Simon, S.B.}, \bibinfo{author}{Grossman, L.},
  \bibinfo{year}{2004}.
\newblock \bibinfo{title}{{A preferred method for the determination of bulk
  compositions of coarse-grained refractory inclusions and some implications of
  the results}}.
\newblock \bibinfo{journal}{Geochimica et Cosmochimica Acta}
  \bibinfo{volume}{68}, \bibinfo{pages}{4237--4248}.
\newblock \DOIprefix\doi{10.1016/j.gca.2004.04.008}.
\bibitem[{Simon and Grossman(2006)}]{simon2006comparative}
\bibinfo{author}{Simon, S.B.}, \bibinfo{author}{Grossman, L.},
  \bibinfo{year}{2006}.
\newblock \bibinfo{title}{{A comparative study of melilite and fassaite in
  Types B1 and B2 refractory inclusions}}.
\newblock \bibinfo{journal}{Geochimica et Cosmochimica Acta}
  \bibinfo{volume}{70}, \bibinfo{pages}{780--798}.
\newblock \DOIprefix\doi{10.1016/j.gca.2005.09.018}.
\bibitem[{Simon and Grossman(2011)}]{simon2011refractory}
\bibinfo{author}{Simon, S.B.}, \bibinfo{author}{Grossman, L.},
  \bibinfo{year}{2011}.
\newblock \bibinfo{title}{{Refractory inclusions in the unique carbonaceous
  chondrite Acfer 094}}.
\newblock \bibinfo{journal}{Meteoritics \& Planetary Science}
  \bibinfo{volume}{46}, \bibinfo{pages}{1197--1216}.
\newblock \DOIprefix\doi{10.1111/j.1945-5100.2011.01224.x}.
\bibitem[{Simon et~al.(1991)Simon, Grossman and Davis}]{simon1991fassaite}
\bibinfo{author}{Simon, S.B.}, \bibinfo{author}{Grossman, L.},
  \bibinfo{author}{Davis, A.M.}, \bibinfo{year}{1991}.
\newblock \bibinfo{title}{{Fassaite composition trends during crystallization
  of Allende Type B refractory inclusion melts}}.
\newblock \bibinfo{journal}{Geochimica et Cosmochimica Acta}
  \bibinfo{volume}{55}, \bibinfo{pages}{2635--2655}.
\newblock \DOIprefix\doi{10.1016/0016-7037(91)90379-J}.
\bibitem[{Simon et~al.(2008)Simon, Joswiak, Ishii, Bradley, Chi, Grossman,
  Al\'{e}on, Brownlee, Fallon, Hutcheon, Matrajt and
  McKeegan}]{simon2008refractory}
\bibinfo{author}{Simon, S.B.}, \bibinfo{author}{Joswiak, D.J.},
  \bibinfo{author}{Ishii, H.A.}, \bibinfo{author}{Bradley, J.P.},
  \bibinfo{author}{Chi, M.}, \bibinfo{author}{Grossman, L.},
  \bibinfo{author}{Al\'{e}on, J.}, \bibinfo{author}{Brownlee, D.E.},
  \bibinfo{author}{Fallon, S.}, \bibinfo{author}{Hutcheon, I.D.},
  \bibinfo{author}{Matrajt, G.}, \bibinfo{author}{McKeegan, K.D.},
  \bibinfo{year}{2008}.
\newblock \bibinfo{title}{{A refractory inclusion returned by Stardust from
  comet 81P/Wild 2}}.
\newblock \bibinfo{journal}{Meteoritics \& Planetary Science}
  \bibinfo{volume}{43}, \bibinfo{pages}{1861--1877}.
\newblock \DOIprefix\doi{10.1111/j.1945-5100.2008.tb00648.x}.
\bibitem[{Simon et~al.(2018)Simon, Krot, Nagashima, K{\"o}{\"o}p and
  Davis}]{simon2018condensate}
\bibinfo{author}{Simon, S.B.}, \bibinfo{author}{Krot, A.N.},
  \bibinfo{author}{Nagashima, K.}, \bibinfo{author}{K{\"o}{\"o}p, L.},
  \bibinfo{author}{Davis, A.M.}, \bibinfo{year}{2018}.
\newblock \bibinfo{title}{{Condensate refractory inclusions from the CO3.00
  chondrite Dominion Range 08006: Petrography, mineral chemistry, and isotopic
  compositions}}.
\newblock \bibinfo{journal}{Geochimica et Cosmochimica Acta}
  \DOIprefix\doi{10.1016/j.gca.2018.11.029}.
\bibitem[{Simon et~al.(2007)Simon, Sutton and Grossman}]{simon2007valence}
\bibinfo{author}{Simon, S.B.}, \bibinfo{author}{Sutton, S.R.},
  \bibinfo{author}{Grossman, L.}, \bibinfo{year}{2007}.
\newblock \bibinfo{title}{{Valence of titanium and vanadium in pyroxene in
  refractory inclusion interiors and rims}}.
\newblock \bibinfo{journal}{Geochimica et Cosmochimica Acta}
  \bibinfo{volume}{71}, \bibinfo{pages}{3098--3118}.
\newblock \DOIprefix\doi{10.1016/j.gca.2007.03.027}.
\bibitem[{Sossi et~al.(2017)Sossi, Moynier, Chaussidon, Villeneuve, Kato and
  Gounelle}]{sossi2017early}
\bibinfo{author}{Sossi, P.A.}, \bibinfo{author}{Moynier, F.},
  \bibinfo{author}{Chaussidon, M.}, \bibinfo{author}{Villeneuve, J.},
  \bibinfo{author}{Kato, C.}, \bibinfo{author}{Gounelle, M.},
  \bibinfo{year}{2017}.
\newblock \bibinfo{title}{Early solar system irradiation quantified by linked
  vanadium and beryllium isotope variations in meteorites}.
\newblock \bibinfo{journal}{Nature Astronomy} \bibinfo{volume}{1},
  \bibinfo{pages}{0055}.
\newblock \DOIprefix\doi{10.1038/s41550-017-0055}.
\bibitem[{Sugiura et~al.(2009)Sugiura, Petaev, Kimura, Miyazaki and
  Hiyagon}]{sugiura2009nebular}
\bibinfo{author}{Sugiura, N.}, \bibinfo{author}{Petaev, M.I.},
  \bibinfo{author}{Kimura, M.}, \bibinfo{author}{Miyazaki, A.},
  \bibinfo{author}{Hiyagon, H.}, \bibinfo{year}{2009}.
\newblock \bibinfo{title}{{Nebular history of amoeboid olivine aggregates}}.
\newblock \bibinfo{journal}{Meteoritics \& Planetary Science}
  \bibinfo{volume}{44}, \bibinfo{pages}{559--572}.
\newblock \DOIprefix\doi{10.1111/j.1945-5100.2009.tb00751.x}.
\bibitem[{Sylvester et~al.(1993)Sylvester, Simon and
  Grossman}]{sylvester1993refractory}
\bibinfo{author}{Sylvester, P.J.}, \bibinfo{author}{Simon, S.B.},
  \bibinfo{author}{Grossman, L.}, \bibinfo{year}{1993}.
\newblock \bibinfo{title}{{Refractory inclusions from the Leoville, Efremovka,
  and Vigarano C3V chondrites: Major element differences between Types A and B,
  and extraordinary refractory siderophile element compositions}}.
\newblock \bibinfo{journal}{Geochimica et Cosmochimica Acta}
  \bibinfo{volume}{57}, \bibinfo{pages}{3763--3784}.
\newblock \DOIprefix\doi{10.1016/0016-7037(93)90154-O}.
\bibitem[{Takayanagi et~al.(1999)Takayanagi, Nakamura, Yurimoto, Kunihiro,
  Nagashima and Kosaka}]{takayanagi1999stacked}
\bibinfo{author}{Takayanagi, I.}, \bibinfo{author}{Nakamura, J.},
  \bibinfo{author}{Yurimoto, H.}, \bibinfo{author}{Kunihiro, T.},
  \bibinfo{author}{Nagashima, K.}, \bibinfo{author}{Kosaka, K.},
  \bibinfo{year}{1999}.
\newblock \bibinfo{title}{{A Stacked CMOS APS for charge particle detection and
  its noise performance}}, in: \bibinfo{booktitle}{Proceedings of the 1999 IEEE
  Workshop on Charge-Coupled Devices and Advanced Image Sensors}, pp.
  \bibinfo{pages}{159--162}.
\bibitem[{Tanaka and Masuda(1973)}]{tanaka1973rare}
\bibinfo{author}{Tanaka, T.}, \bibinfo{author}{Masuda, A.},
  \bibinfo{year}{1973}.
\newblock \bibinfo{title}{{Rare-earth elements in matrix, inclusions, and
  chondrules of the Allende meteorite}}.
\newblock \bibinfo{journal}{Icarus} \bibinfo{volume}{19},
  \bibinfo{pages}{523--530}.
\newblock \DOIprefix\doi{10.1016/0019-1035(73)90079-1}.
\bibitem[{Taylor et~al.(2004)Taylor, McKeegan and Krot}]{taylor200426al}
\bibinfo{author}{Taylor, D.J.}, \bibinfo{author}{McKeegan, K.D.},
  \bibinfo{author}{Krot, A.N.}, \bibinfo{year}{2004}.
\newblock \bibinfo{title}{{\ce{^{26}Al} in Efremovka CAI E44L--Resolved time
  interval between interior and rim formation in a highly fractionated compact
  type A CAI}}, in: \bibinfo{booktitle}{Workshop on Chondrites and the
  Protoplanetary Disk}, p. \bibinfo{pages}{197}.
\bibitem[{Taylor et~al.(2005)Taylor, McKeegan and Krot}]{taylor2005high}
\bibinfo{author}{Taylor, D.J.}, \bibinfo{author}{McKeegan, K.D.},
  \bibinfo{author}{Krot, A.N.}, \bibinfo{year}{2005}.
\newblock \bibinfo{title}{{High resolution \ce{^{26}Al} chronology: Resolved
  time interval between rim and interior of a highly fractionated compact Type
  A CAI from Efremovka}}, in: \bibinfo{booktitle}{Lunar and Planetary Science
  Conference}, p. \bibinfo{pages}{2121}.
\bibitem[{Ulyanov et~al.(1982)Ulyanov, Korina, Nazarov and
  Sherbovsky}]{ulyanov1982efremovka}
\bibinfo{author}{Ulyanov, A.A.}, \bibinfo{author}{Korina, M.I.},
  \bibinfo{author}{Nazarov, M.A.}, \bibinfo{author}{Sherbovsky, E.Y.},
  \bibinfo{year}{1982}.
\newblock \bibinfo{title}{{Efremovka CAI's: Mineralogical and petrological
  data}}, in: \bibinfo{booktitle}{Lunar and Planetary Science Conference}, p.
  \bibinfo{pages}{813}.
\bibitem[{Ushikubo et~al.(2017)Ushikubo, Tenner, Hiyagon and
  Kita}]{ushikubo2016long}
\bibinfo{author}{Ushikubo, T.}, \bibinfo{author}{Tenner, T.J.},
  \bibinfo{author}{Hiyagon, H.}, \bibinfo{author}{Kita, N.T.},
  \bibinfo{year}{2017}.
\newblock \bibinfo{title}{{A long duration of the \ce{^{16}O}-rich reservoir in
  the solar nebula, as recorded in fine-grained refractory inclusions from the
  least metamorphosed carbonaceous chondrites}}.
\newblock \bibinfo{journal}{Geochimica et Cosmochimica Acta}
  \bibinfo{volume}{201}, \bibinfo{pages}{103--122}.
\newblock \DOIprefix\doi{10.1016/j.gca.2016.08.032}.
\bibitem[{Van~Kooten et~al.(2016)Van~Kooten, Wielandt, Schiller, Nagashima,
  Thomen, Larsen, Olsen, Nordlund, Krot and Bizzarro}]{van2016isotopic}
\bibinfo{author}{Van~Kooten, E.M.M.E.}, \bibinfo{author}{Wielandt, D.},
  \bibinfo{author}{Schiller, M.}, \bibinfo{author}{Nagashima, K.},
  \bibinfo{author}{Thomen, A.}, \bibinfo{author}{Larsen, K.K.},
  \bibinfo{author}{Olsen, M.B.}, \bibinfo{author}{Nordlund, {\AA}.},
  \bibinfo{author}{Krot, A.N.}, \bibinfo{author}{Bizzarro, M.},
  \bibinfo{year}{2016}.
\newblock \bibinfo{title}{{Isotopic evidence for primordial molecular cloud
  material in metal-rich carbonaceous chondrites}}.
\newblock \bibinfo{journal}{Proceedings of the National Academy of Sciences}
  \bibinfo{volume}{113}, \bibinfo{pages}{2011--2016}.
\newblock \DOIprefix\doi{10.1073/pnas.1518183113}.
\bibitem[{Wark and Lovering(1977)}]{wark1977marker}
\bibinfo{author}{Wark, D.A.}, \bibinfo{author}{Lovering, J.F.},
  \bibinfo{year}{1977}.
\newblock \bibinfo{title}{{Marker events in the early evolution of the solar
  system-Evidence from rims on Ca-Al-rich inclusions in carbonaceous
  chondrites}}, in: \bibinfo{booktitle}{Proceedings of the Lunar and Planetary
  Science Conference}, pp. \bibinfo{pages}{95--112}.
\bibitem[{Wark and Lovering(1982)}]{wark1982nature}
\bibinfo{author}{Wark, D.A.}, \bibinfo{author}{Lovering, J.F.},
  \bibinfo{year}{1982}.
\newblock \bibinfo{title}{{The nature and origin of type B1 and B2 Ca$\cdotp
  $Al-rich inclusions in the Allende meteorite}}.
\newblock \bibinfo{journal}{Geochimica et Cosmochimica Acta}
  \bibinfo{volume}{46}, \bibinfo{pages}{2581--2594}.
\newblock \DOIprefix\doi{10.1016/0016-7037(82)90379-9}.
\bibitem[{Wasson et~al.(2001)Wasson, Yurimoto and Russell}]{wasson200116}
\bibinfo{author}{Wasson, J.T.}, \bibinfo{author}{Yurimoto, H.},
  \bibinfo{author}{Russell, S.S.}, \bibinfo{year}{2001}.
\newblock \bibinfo{title}{{\ce{^{16}O}-rich melilite in CO3.0 chondrites:
  Possible formation of common, \ce{^{16}O}-poor melilite by aqueous
  alteration}}.
\newblock \bibinfo{journal}{Geochimica et Cosmochimica Acta}
  \bibinfo{volume}{65}, \bibinfo{pages}{4539--4549}.
\newblock \DOIprefix\doi{10.1016/S0016-7037(01)00738-4}.
\bibitem[{Whattam et~al.(2008)Whattam, Hewins, Cohen, Seaton and
  Prior}]{whattam2008granoblastic}
\bibinfo{author}{Whattam, S.A.}, \bibinfo{author}{Hewins, R.H.},
  \bibinfo{author}{Cohen, B.A.}, \bibinfo{author}{Seaton, N.C.},
  \bibinfo{author}{Prior, D.J.}, \bibinfo{year}{2008}.
\newblock \bibinfo{title}{{Granoblastic olivine aggregates in magnesian
  chondrules: Planetesimal fragments or thermally annealed solar nebula
  condensates?}}
\newblock \bibinfo{journal}{Earth and Planetary Science Letters}
  \bibinfo{volume}{269}, \bibinfo{pages}{200--211}.
\newblock \DOIprefix\doi{10.1016/j.epsl.2008.02.013}.
\bibitem[{Yoneda and Grossman(1995)}]{yoneda1995condensation}
\bibinfo{author}{Yoneda, S.}, \bibinfo{author}{Grossman, L.},
  \bibinfo{year}{1995}.
\newblock \bibinfo{title}{{Condensation of CaO-MgO-\ce{Al2O3}-\ce{SiO2} liquids
  from cosmic gases}}.
\newblock \bibinfo{journal}{Geochimica et Cosmochimica Acta}
  \bibinfo{volume}{59}, \bibinfo{pages}{3413--3444}.
\newblock \DOIprefix\doi{10.1016/0016-7037(95)00214-K}.
\bibitem[{Yurimoto et~al.(1998)Yurimoto, Ito and Nagasawa}]{yurimoto1998oxygen}
\bibinfo{author}{Yurimoto, H.}, \bibinfo{author}{Ito, M.},
  \bibinfo{author}{Nagasawa, H.}, \bibinfo{year}{1998}.
\newblock \bibinfo{title}{{Oxygen isotope exchange between refractory inclusion
  in Allende and solar nebula gas}}.
\newblock \bibinfo{journal}{Science} \bibinfo{volume}{282},
  \bibinfo{pages}{1874--1877}.
\newblock \DOIprefix\doi{10.1126/science.282.5395.1874}.
\bibitem[{Yurimoto et~al.(2008)Yurimoto, Krot, Choi, Aleon, Kunihiro and
  Brearley}]{yurimoto2008oxygen}
\bibinfo{author}{Yurimoto, H.}, \bibinfo{author}{Krot, A.N.},
  \bibinfo{author}{Choi, B.G.}, \bibinfo{author}{Aleon, J.},
  \bibinfo{author}{Kunihiro, T.}, \bibinfo{author}{Brearley, A.J.},
  \bibinfo{year}{2008}.
\newblock \bibinfo{title}{{Oxygen isotopes of chondritic components}}, in:
  \bibinfo{editor}{MacPherson, G.J.} (Ed.), \bibinfo{booktitle}{Reviews in
  Mineralogy and Geochemistry}. volume~\bibinfo{volume}{68}, pp.
  \bibinfo{pages}{141--186}.
\newblock \DOIprefix\doi{10.2138/rmg.2008.68.8}.
\bibitem[{Yurimoto and Kuramoto(2004)}]{yurimoto2004molecular}
\bibinfo{author}{Yurimoto, H.}, \bibinfo{author}{Kuramoto, K.},
  \bibinfo{year}{2004}.
\newblock \bibinfo{title}{{Molecular cloud origin for the oxygen isotope
  heterogeneity in the solar system}}.
\newblock \bibinfo{journal}{Science} \bibinfo{volume}{305},
  \bibinfo{pages}{1763--1766}.
\newblock \DOIprefix\doi{10.1126/science.1100989}.
\bibitem[{Yurimoto et~al.(1989)Yurimoto, Morioka and
  Nagasawa}]{yurimoto1989diffusion}
\bibinfo{author}{Yurimoto, H.}, \bibinfo{author}{Morioka, M.},
  \bibinfo{author}{Nagasawa, H.}, \bibinfo{year}{1989}.
\newblock \bibinfo{title}{{Diffusion in single crystals of melilite: I.
  Oxygen}}.
\newblock \bibinfo{journal}{Geochimica et Cosmochimica Acta}
  \bibinfo{volume}{53}, \bibinfo{pages}{2387--2394}.
\newblock \DOIprefix\doi{10.1016/0016-7037(89)90360-8}.
\bibitem[{Yurimoto et~al.(2003)Yurimoto, Nagashima and
  Kunihiro}]{yurimoto2003high}
\bibinfo{author}{Yurimoto, H.}, \bibinfo{author}{Nagashima, K.},
  \bibinfo{author}{Kunihiro, T.}, \bibinfo{year}{2003}.
\newblock \bibinfo{title}{{High precision isotope micro-imaging of materials}}.
\newblock \bibinfo{journal}{Applied Surface Science} \bibinfo{volume}{203},
  \bibinfo{pages}{793--797}.
\newblock \DOIprefix\doi{10.1016/S0169-4332(02)00825-5}.
\bibitem[{Zhang et~al.(2015)Zhang, Ma, Sakamoto, Wang, Hsu and
  Yurimoto}]{zhang2015mineralogical}
\bibinfo{author}{Zhang, A.C.}, \bibinfo{author}{Ma, C.},
  \bibinfo{author}{Sakamoto, N.}, \bibinfo{author}{Wang, R.C.},
  \bibinfo{author}{Hsu, W.B.}, \bibinfo{author}{Yurimoto, H.},
  \bibinfo{year}{2015}.
\newblock \bibinfo{title}{{Mineralogical anatomy and implications of a
  Ti-Sc-rich ultrarefractory inclusion from Sayh al Uhaymir 290 CH3
  chondrite}}.
\newblock \bibinfo{journal}{Geochimica et Cosmochimica Acta}
  \bibinfo{volume}{163}, \bibinfo{pages}{27--39}.
\newblock \DOIprefix\doi{10.1016/j.gca.2015.04.052}.
\bibitem[{Zhang(2010)}]{zhang2010diffusion}
\bibinfo{author}{Zhang, Y.}, \bibinfo{year}{2010}.
\newblock \bibinfo{title}{Diffusion in minerals and melts: Theoretical
  background}, in: \bibinfo{editor}{Zhang, Y.}, \bibinfo{editor}{Cherniak,
  D.J.} (Eds.), \bibinfo{booktitle}{Reviews in Mineralogy and Geochemistry}.
  \bibinfo{publisher}{Mineralogical Society of America}.
  volume~\bibinfo{volume}{72}, pp. \bibinfo{pages}{5--59}.
\newblock \DOIprefix\doi{10.2138/rmg.2010.72.2}.
\bibitem[{Zolensky et~al.(2006)Zolensky, Zega, Yano, Wirick, Westphal,
  Weisberg, Weber, Warren, Velbel, Tsuchiyama, Tsou, Toppani, Tomioka, Tomeoka,
  Teslich, Taheri, Susini, Stroud, Stephan, Stadermann, Snead, Simon,
  Simionovici, See, Robert, Rietmeijer, Rao, Perronnet, Papanastassiou,
  Okudaira, Ohsumi, Ohnishi, Nakamura-Messenger, Nakamura, Mostefaoui,
  Mikouchi, Meibom, Matrajt, Marcus, Leroux, Lemelle, Le, Lanzirotti,
  Langenhorst, Krot, Keller, Kearsley, Joswiak, Jacob, Ishii, Harvey, Hagiya,
  Grossman, Grossman, Graham, Gounelle, Gillet, Genge, Flynn, Ferroir, Fallon,
  Ebel, Dai, Cordier, Clark, Chi, Butterworth, Brownlee, Bridges, Brennan,
  Brearley, Bradley, Bleuet, Bland and Bastien}]{zolensky2006mineralogy}
\bibinfo{author}{Zolensky, M.E.}, \bibinfo{author}{Zega, T.J.},
  \bibinfo{author}{Yano, H.}, \bibinfo{author}{Wirick, S.},
  \bibinfo{author}{Westphal, A.J.}, \bibinfo{author}{Weisberg, M.K.},
  \bibinfo{author}{Weber, I.}, \bibinfo{author}{Warren, J.L.},
  \bibinfo{author}{Velbel, M.A.}, \bibinfo{author}{Tsuchiyama, A.},
  \bibinfo{author}{Tsou, P.}, \bibinfo{author}{Toppani, A.},
  \bibinfo{author}{Tomioka, N.}, \bibinfo{author}{Tomeoka, K.},
  \bibinfo{author}{Teslich, N.}, \bibinfo{author}{Taheri, M.},
  \bibinfo{author}{Susini, J.}, \bibinfo{author}{Stroud, R.},
  \bibinfo{author}{Stephan, T.}, \bibinfo{author}{Stadermann, F.J.},
  \bibinfo{author}{Snead, C.J.}, \bibinfo{author}{Simon, S.B.},
  \bibinfo{author}{Simionovici, A.}, \bibinfo{author}{See, T.H.},
  \bibinfo{author}{Robert, F.}, \bibinfo{author}{Rietmeijer, F.J.M.},
  \bibinfo{author}{Rao, W.}, \bibinfo{author}{Perronnet, M.C.},
  \bibinfo{author}{Papanastassiou, D.A.}, \bibinfo{author}{Okudaira, K.},
  \bibinfo{author}{Ohsumi, K.}, \bibinfo{author}{Ohnishi, I.},
  \bibinfo{author}{Nakamura-Messenger, K.}, \bibinfo{author}{Nakamura, T.},
  \bibinfo{author}{Mostefaoui, S.}, \bibinfo{author}{Mikouchi, T.},
  \bibinfo{author}{Meibom, A.}, \bibinfo{author}{Matrajt, G.},
  \bibinfo{author}{Marcus, M.A.}, \bibinfo{author}{Leroux, H.},
  \bibinfo{author}{Lemelle, L.}, \bibinfo{author}{Le, L.},
  \bibinfo{author}{Lanzirotti, A.}, \bibinfo{author}{Langenhorst, F.},
  \bibinfo{author}{Krot, A.N.}, \bibinfo{author}{Keller, L.P.},
  \bibinfo{author}{Kearsley, A.T.}, \bibinfo{author}{Joswiak, D.},
  \bibinfo{author}{Jacob, D.}, \bibinfo{author}{Ishii, H.},
  \bibinfo{author}{Harvey, R.}, \bibinfo{author}{Hagiya, K.},
  \bibinfo{author}{Grossman, L.}, \bibinfo{author}{Grossman, J.N.},
  \bibinfo{author}{Graham, G.A.}, \bibinfo{author}{Gounelle, M.},
  \bibinfo{author}{Gillet, P.}, \bibinfo{author}{Genge, M.J.},
  \bibinfo{author}{Flynn, G.}, \bibinfo{author}{Ferroir, T.},
  \bibinfo{author}{Fallon, S.}, \bibinfo{author}{Ebel, D.S.},
  \bibinfo{author}{Dai, Z.R.}, \bibinfo{author}{Cordier, P.},
  \bibinfo{author}{Clark, B.}, \bibinfo{author}{Chi, M.},
  \bibinfo{author}{Butterworth, A.L.}, \bibinfo{author}{Brownlee, D.E.},
  \bibinfo{author}{Bridges, J.C.}, \bibinfo{author}{Brennan, S.},
  \bibinfo{author}{Brearley, A.}, \bibinfo{author}{Bradley, J.P.},
  \bibinfo{author}{Bleuet, P.}, \bibinfo{author}{Bland, P.A.},
  \bibinfo{author}{Bastien, R.}, \bibinfo{year}{2006}.
\newblock \bibinfo{title}{{Mineralogy and petrology of comet 81P/Wild 2 nucleus
  samples}}.
\newblock \bibinfo{journal}{Science} \bibinfo{volume}{314},
  \bibinfo{pages}{1735--1739}.
\newblock \DOIprefix\doi{10.1126/science.1135842}.

\end{thebibliography}
	
	\newpage
	
	\appendix
	\setcounter{figure}{0}
	\setcounter{table}{0}
	\renewcommand{\thefigure}{A\arabic{figure}}
	\renewcommand{\thetable}{A\arabic{table}}
	\renewcommand{\thesection}{A\arabic{section}}
	
	\section*{Appendix A: Supplementary data}
	
		\begin{figure}[h]
		\centering
		\includegraphics[width=0.7\linewidth]{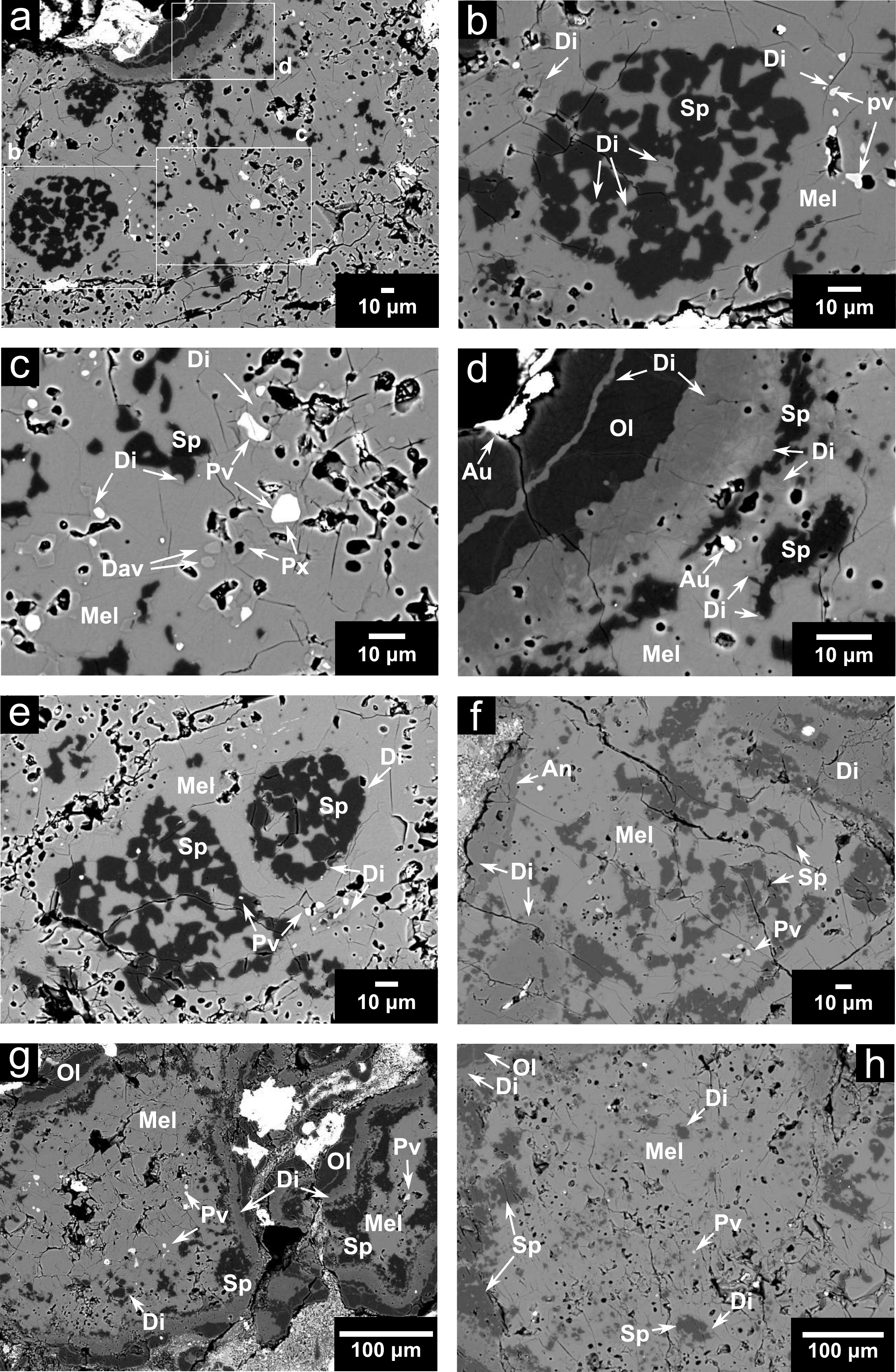}
		\caption{BSE images of Unit 1 (R3C-01-U1) (a--e), Unit 2 (f), Unit 3 and Unit 5 (g) and Unit 4 (h) in R3C-01. Abbreviations are as in the main text.
		}
		\label{fig:BSE_R3C-01_details}
	\end{figure}

\clearpage
	
	\begin{figure}[p]
		\centering
		\includegraphics[width=1\linewidth]{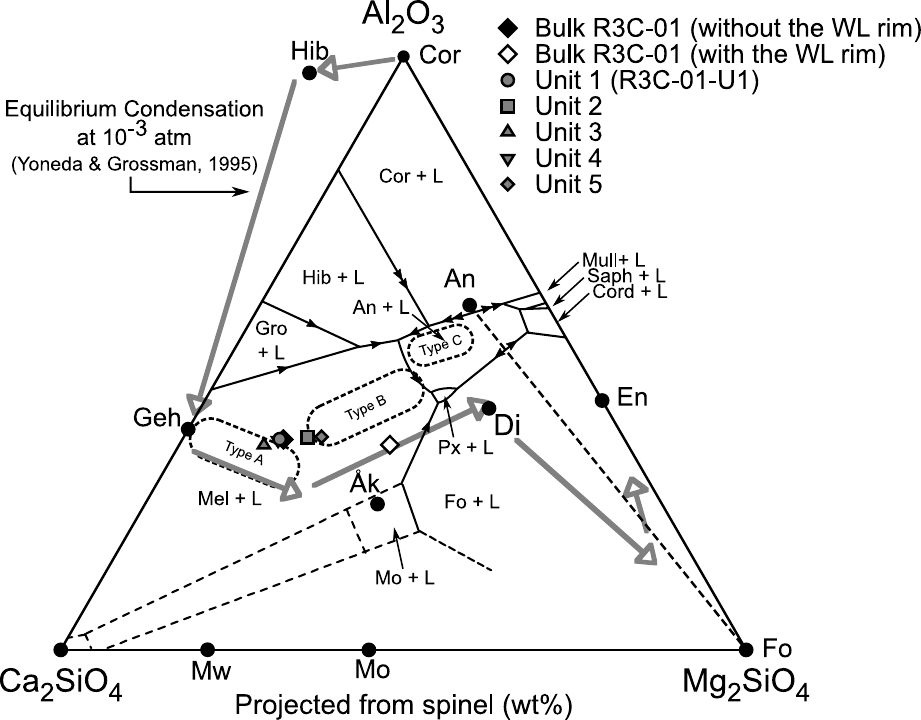}
		\caption{Bulk compositions of R3C-01 and its five lithological units compared to typical types A, B and C CAIs \citep{,macpherson2014calcium}, projected from spinel (\ce{MgAl2O4}) onto the plane \ce{Al2O3}-\ce{Mg2SiO4}-\ce{Ca2SiO4}, modified after \citet{,macpherson2005petrogenesis}. 		Boundary curves are spinel-saturated liquidus equilibria. Gray arrows show the calculated trend for total condensed  solids  formed  from  equilibrium  condensation  of  a  solar  gas  at  $ 10^{-3} $ bar \citep{,yoneda1995condensation}. Abbreviations: \AA{}k\textendash \aa{}kermanite; An\textendash anorthite; Cor\textendash corundum; Cord\textendash cordierite; Di\textendash diopside; En\textendash enstatite; Fo\textendash forsterite; Geh\textendash gehlenite; Gro\textendash grossite; Hib\textendash hibonite; L\textendash liquid; Mel\textendash melilite solid solution; Mo\textendash monticellite; Mull\textendash mullite; Mw\textendash merwinite; Px\textendash pyroxene; Saph\textendash saphirine.
		}
		\label{fig:bulk_tplot}
	\end{figure}
	\clearpage
	
	\begin{figure}[p]
		\centering
		\includegraphics[width=1\linewidth]{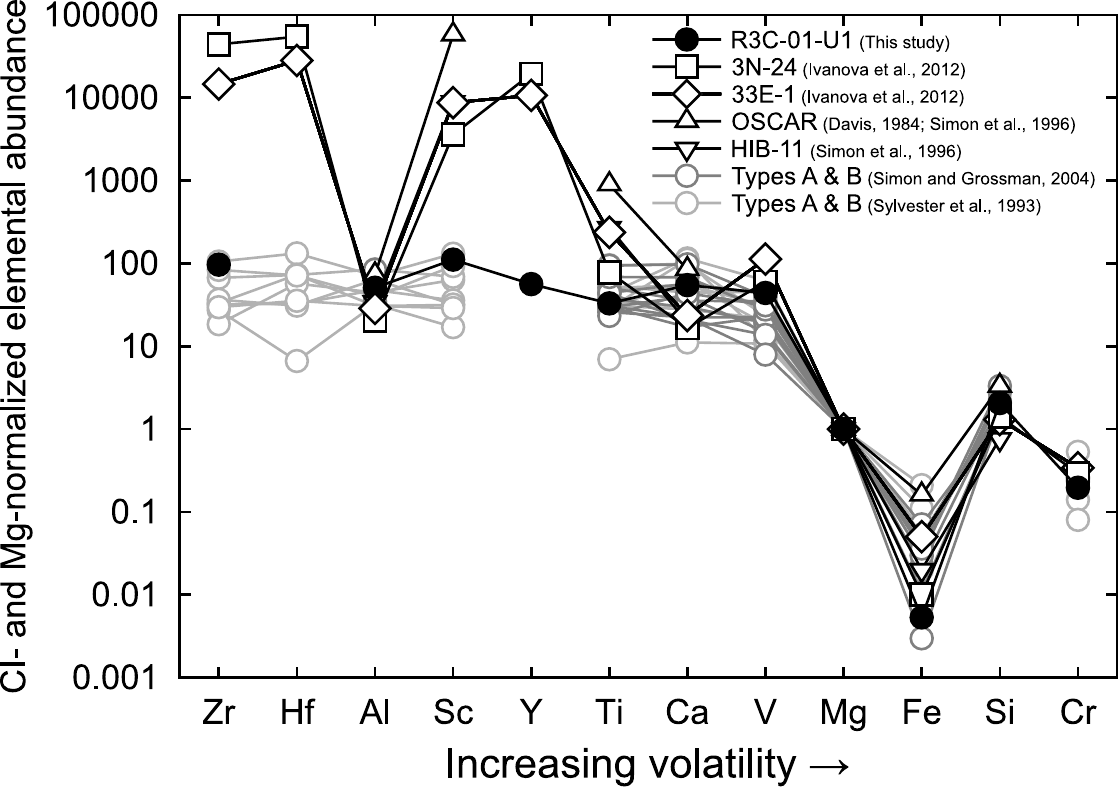}
		\caption[CI-normalized bulk composition of R3C-01-U1]
		{Bulk elemental compositions of R3C-01-U1 (this study), UR CAIs 3N-24, 33E-1 \citep{,ivanova2012compound}, HIB-11 \citep{,simon1996unique}, OSCAR \citep{,davis1984scandalously,simon1996unique}, and types A and B CAIs \citep{,sylvester1993refractory,simon2004preferred} normalized to  CI chondrite \citep{,lodders2003solar} and Mg abundances. Elements are plotted in order of increasing volatility \citep{,lodders2003solar}.
		}
		\label{fig:bulk_sdiagram}
	\end{figure}
	\clearpage

	\begin{table}[h]
		\centering
		\begin{threeparttable}
			\caption[Major element composition of melilite in R3C-01-U1]{Major element composition of melilite in R3C-01-U1.}
			\label{tab:R3C-01_mel_EPMA}
			\begin{tabular}{lrrrrrrrrrr}
				\toprule
				wt\% & 1     & 3     & 4     & 5     & 7     & 8     & 9     & 12    & 324   & 331 \\
				\midrule
				\ce{SiO2} & 22.5  & 22.5  & 22.2  & 24.7  & 24.8  & 25.9  & 27.7  & 24.5  & 28.2  & 28.8 \\
				\ce{TiO^{tot}_2} & 0.0   & 0.0   & 0.1   & b.d.  & 0.0   & b.d.  & b.d.  & 0.3   & 0.0   & 0.1 \\
				\ce{Al2O3} & 34.0  & 34.6  & 35.5  & 30.4  & 31.5  & 30.2  & 28.0  & 31.5  & 28.0  & 27.5 \\
				\ce{Cr2O3} & 0.0   & 0.0   & 0.0   & b.d.  & b.d.  & b.d.  & 0.0   & b.d.  & 0.1   & 0.0 \\
				\ce{FeO} & b.d.  & 0.0   & 0.0   & 0.0   & 0.0   & 0.0   & 0.1   & b.d.  & 0.1   & 0.0 \\
				\ce{MnO} & 0.0   & b.d.  & b.d.  & b.d.  & b.d.  & b.d.  & 0.0   & b.d.  & b.d.  & 0.0 \\
				\ce{MgO} & 0.9   & 0.7   & 0.6   & 2.4   & 2.1   & 2.8   & 3.7   & 2.1   & 3.3   & 3.7 \\
				\ce{CaO} & 40.6  & 40.4  & 40.7  & 40.6  & 41.1  & 40.8  & 41.5  & 40.8  & 39.4  & 39.5 \\
				\ce{Na2O} & b.d.  & b.d.  & b.d.  & b.d.  & 0.0   & b.d.  & b.d.  & 0.0   & b.d.  & 0.0 \\
				\ce{V2O3} & b.d.  & b.d.  & b.d.  & 0.0   & b.d.  & b.d.  & 0.0   & b.d.  & 0.0   & 0.0 \\
				\ce{ZrO2} & b.d.  & b.d.  & b.d.  & 0.1   & b.d.  & b.d.  & b.d.  & 0.1   & 0.1   & b.d. \\
				\ce{HfO2} & b.d.  & b.d.  & b.d.  & b.d.  & b.d.  & 0.1   & 0.1   & b.d.  & b.d.  & b.d. \\
				\ce{Sc2O3} & b.d.  & b.d.  & b.d.  & b.d.  & 0.0   & b.d.  & 0.0   & 0.0   & 0.0   & 0.0 \\
				\ce{Y2O3} & b.d.  & b.d.  & b.d.  & b.d.  & 0.0   & 0.1   & b.d.  & b.d.  & b.d.  & b.d. \\
				Total & 98.1  & 98.2  & 99.1  & 98.3  & 99.6  & 99.9  & 101.0 & 99.2  & 99.2  & 99.6 \\
				\AA{}k\# & 6     & 5     & 4     & 17    & 14    & 19    & 25    & 14    & 26    & 28 \\
				\bottomrule
			\end{tabular}
			\begin{tablenotes}[flushleft]
				\item[] b.d.-- below detection limit.
			\end{tablenotes}
		\end{threeparttable}
	\end{table}

	\begin{landscape}
		\begin{table}[p]
			\centering
			{\footnotesize
			\begin{threeparttable}
				\caption[Major element composition of spinel in R3C-01-U1]{Major element composition of spinel in framboids and the WL rim in  R3C-01-U1.}
				\label{tab:R3C-01_sp_EPMA}
				\begin{tabular}{lrrrrrrrrrrrrrrrrrrrrrrrrrr}
					\toprule
					& \multicolumn{11}{c}{Framboids in the inner CAI ($ n=5 $ for each)}                        &       & \multicolumn{11}{c}{Framboids attached to the WL rim ($ n=5 $ for each)}                  &       & \multicolumn{2}{c}{WL rim} \\
					\cline{2-12} \cline{14-24} 
					& \multicolumn{2}{c}{\#12} &       & \multicolumn{2}{c}{\#15} &       & \multicolumn{2}{c}{\#13} &       & \multicolumn{2}{c}{\#11} &       & \multicolumn{2}{c}{\#10} &       & \multicolumn{2}{c}{\#5} &       & \multicolumn{2}{c}{\#14} &       & \multicolumn{2}{c}{\#2} &       & \multicolumn{2}{c}{($ n=10 $)} \\
					wt\%  & Ave.   & $1\sigma$  &       & Ave.   & $1\sigma$  &       & Ave.   & $1\sigma$  &       & Ave.   & $1\sigma$  &       & Ave.   & $1\sigma$  &       & Ave.   & $1\sigma$  &      & Ave.   & $1\sigma$  &       & Ave.   & $1\sigma$  &       & Ave.   & $1\sigma$  \\
					\midrule
					\ce{SiO2} & 0.0   & 0.0   &       & 0.1   & 0.0   &       & b.d.  & n.a.  &       & 0.0   & n.a.  &       & 0.8   & 1.2   &       & 0.0   & 0.0   &       & 0.0   & 0.0   &       & 0.1   & 0.0   &       & 0.4   & 0.9 \\
					\ce{TiO^{tot}_2} & 0.2   & 0.1   &       & 0.2   & 0.1   &       & 0.2   & 0.1   &       & 0.3   & 0.1   &       & 0.4   & 0.2   &       & 0.2   & 0.1   &       & 0.3   & 0.1   &       & 0.2   & 0.0   &       & 0.2   & 0.2 \\
					\ce{Al2O3} & 69.9  & 0.3   &       & 69.9  & 0.3   &       & 69.9  & 0.2   &       & 69.7  & 0.2   &       & 68.7  & 2.3   &       & 70.2  & 0.2   &       & 70.1  & 0.2   &       & 69.8  & 0.1   &       & 69.5  & 1.4 \\
					\ce{Cr2O3} & 0.2   & 0.0   &       & 0.2   & 0.0   &       & 0.1   & 0.0   &       & 0.1   & 0.0   &       & 0.1   & 0.0   &       & 0.1   & 0.0   &       & 0.1   & 0.0   &       & 0.1   & 0.0   &       & 0.2   & 0.0 \\
					\ce{FeO} & 0.1   & 0.0   &       & 0.1   & 0.0   &       & 0.1   & 0.0   &       & 0.1   & 0.0   &       & 0.1   & 0.0   &       & 0.1   & 0.0   &       & 0.3   & 0.1   &       & 0.2   & 0.0   &       & 0.2   & 0.1 \\
					\ce{MnO} & 0.0   & 0.0   &       & b.d.  & n.a.  &       & 0.0   & 0.0   &       & 0.0   & 0.0   &       & b.d.  & n.a.  &       & 0.0   & n.a.  &       & 0.0   & 0.0   &       & 0.0   & 0.0   &       & 0.0   & 0.0 \\
					\ce{MgO} & 27.9  & 0.1   &       & 27.8  & 0.2   &       & 27.8  & 0.2   &       & 27.6  & 0.1   &       & 26.9  & 1.2   &       & 27.7  & 0.1   &       & 27.7  & 0.2   &       & 27.6  & 0.1   &       & 27.5  & 0.5 \\
					\ce{CaO} & 0.2   & 0.0   &       & 0.3   & 0.1   &       & 0.2   & 0.0   &       & 0.1   & 0.0   &       & 0.3   & 0.1   &       & 0.1   & 0.0   &       & 0.2   & 0.1   &       & 0.1   & 0.0   &       & 0.3   & 0.2 \\
					\ce{Na2O} & 0.0   & n.a.  &       & b.d.  & n.a.  &       & 0.1   & n.a.  &       & 0.0   & n.a.  &       & 0.0   & 0.0   &       & 0.0   & 0.0   &       & 0.0   & 0.0   &       & 0.0   & 0.0   &       & 0.0   & 0.0 \\
					\ce{V2O3} & 0.8   & 0.0   &       & 0.6   & 0.1   &       & 0.6   & 0.0   &       & 0.5   & 0.0   &       & 0.4   & 0.0   &       & 0.4   & 0.1   &       & 0.3   & 0.0   &       & 0.3   & 0.0   &       & 0.2   & 0.0 \\
					\ce{ZrO2} & 0.1   & 0.0   &       & 0.0   & n.a.  &       & 0.1   & n.a.  &       & b.d.  & n.a.  &       & 0.0   & 0.0   &       & 0.0   & n.a.  &       & 0.0   & n.a.  &       & b.d.  & n.a.  &       & 0.0   & n.a. \\
					\ce{Sc2O3} & b.d.  & n.a.  &       & b.d.  & n.a.  &       & b.d.  & n.a.  &       & b.d.  & n.a.  &       & b.d.  & n.a.  &       & b.d.  & n.a.  &       & b.d.  & n.a.  &       & b.d.  & n.a.  &       & b.d.  & n.a. \\
					Total & 99.2  & 0.3   &       & 99.1  & 0.4   &       & 98.9  & 0.3   &       & 98.6  & 0.2   &       & 99.0  & 0.7   &       & 98.8  & 0.2   &       & 99.0  & 0.4   &       & 98.4  & 0.1   &       & 98.7  & 0.4 \\
					\bottomrule
				\end{tabular}
				\begin{tablenotes}[flushleft]
					\item[] b.d.-- below detection limit; n.a.-- not available.
				\end{tablenotes}
			\end{threeparttable}
		}
		\end{table}
	\end{landscape}
	
	\begin{table}[p]
		\centering
		\begin{threeparttable}
			\caption[Major element composition of perovskite in R3C-01-U1]{Major element composition of perovskite in R3C-01-U1.}
			\label{tab:R3C-01_pv_EPMA}
			\begin{tabular}{lrrrrrr}
				\toprule
				wt\%  & 10    & 85    & 89    & 96    & 107   & 246 \\
				\midrule
				\ce{SiO2} & b.d.  & b.d.  & 1.0   & 0.2   & b.d.  & 0.1 \\
				\ce{TiO^{tot}_2} & 56.4  & 57.3  & 55.5  & 56.6  & 53.0  & 56.0 \\
				\ce{Al2O3} & 1.0   & 0.5   & 2.2   & 0.5   & 6.1   & 0.3 \\
				\ce{Cr2O3} & 0.0   & 0.0   & b.d.  & b.d.  & b.d.  & b.d. \\
				\ce{FeO} & 0.0   & 0.0   & 0.2   & 0.2   & 0.0   & b.d. \\
				\ce{MnO} & b.d.  & 0.0   & 0.0   & 0.0   & b.d.  & 0.0 \\
				\ce{MgO} & 0.3   & 0.1   & 0.1   & 0.0   & 3.1   & 0.0 \\
				\ce{CaO} & 40.3  & 40.7  & 40.7  & 41.0  & 38.1  & 40.4 \\
				\ce{Na2O} & b.d.  & 0.0   & b.d.  & 0.1   & b.d.  & b.d. \\
				\ce{V2O3} & 0.2   & 0.2   & 0.1   & 0.8   & 0.7   & 0.4 \\
				\ce{ZrO2} & 0.2   & 0.2   & 0.3   & 0.2   & 0.1   & 0.3 \\
				\ce{HfO2} & b.d.  & b.d.  & b.d.  & b.d.  & b.d.  & b.d. \\
				\ce{Sc2O3} & 0.1   & 0.0   & 0.0   & 0.1   & 0.1   & 0.1 \\
				\ce{Y2O3} & 0.1   & 0.1   & 0.3   & 0.3   & 0.1   & 0.2 \\
				Total & 98.4  & 99.3  & 100.5 & 99.9  & 101.2 & 97.8 \\
				\bottomrule
			\end{tabular}
			\begin{tablenotes}[flushleft]
				\item[] b.d.-- below detection limit.
			\end{tablenotes}
		\end{threeparttable}
	\end{table}
	\clearpage
	
	\begin{table}[p]
		\centering
		\begin{threeparttable}
			\caption[Major element composition of olivine in the WL rim on R3C-01-U1]{Major element composition of olivine in the WL rim on R3C-01-U1.}
			\label{tab:R3C-01_ol_EPMA}
			\begin{tabular}{lrrrrrrrrrr}
				\toprule
				wt\%  & 576   & 579   & 479   & 225   & 454   & 442   & 221   & 389   & 578   & 464 \\
				\midrule
				\ce{SiO2} & 42.6  & 42.7  & 42.6  & 41.5  & 42.3  & 43.0  & 43.2  & 43.8  & 42.3  & 42.6 \\
				\ce{TiO^{tot}_2} & b.d.  & b.d.  & b.d.  & b.d.  & b.d.  & b.d.  & b.d.  & b.d.  & b.d.  & b.d. \\
				\ce{Al2O3} & 0.0   & 0.1   & 0.0   & 0.0   & 0.0   & 0.0   & b.d.  & 0.0   & b.d.  & 0.1 \\
				\ce{Cr2O3} & 0.2   & 0.2   & 0.5   & 0.3   & 0.2   & 0.2   & 0.1   & b.d.  & 0.1   & 0.2 \\
				\ce{FeO} & 0.2   & 0.2   & 0.4   & 0.9   & 0.8   & 1.5   & 0.4   & 0.6   & 0.1   & 1.1 \\
				\ce{MnO} & 0.2   & 0.3   & 0.4   & 0.7   & 0.5   & 0.6   & 0.1   & 0.1   & b.d.  & 0.2 \\
				\ce{MgO} & 56.8  & 55.2  & 57.1  & 56.6  & 57.0  & 56.4  & 57.2  & 57.4  & 56.7  & 56.1 \\
				\ce{CaO} & 0.2   & 0.2   & 0.2   & 0.1   & 0.1   & 0.2   & 0.2   & 0.2   & 0.5   & 0.1 \\
				\ce{Na2O} & b.d.  & b.d.  & b.d.  & b.d.  & b.d.  & b.d.  & b.d.  & b.d.  & b.d.  & b.d. \\
				\ce{K2O} & b.d.  & b.d.  & b.d.  & b.d.  & b.d.  & b.d.  & b.d.  & b.d.  & b.d.  & b.d. \\
				\ce{NiO} & 0.1   & b.d.  & b.d.  & b.d.  & b.d.  & b.d.  & b.d.  & 0.1   & 0.0   & 0.1 \\
				Total & 100.3 & 98.9  & 101.2 & 100.2 & 100.9 & 101.9 & 101.1 & 102.2 & 99.8  & 100.5 \\
				Fa\#  & 0.2   & 0.2   & 0.4   & 0.9   & 0.7   & 1.5   & 0.3   & 0.5   & 0.1   & 1.1 \\
				MnO/FeO & 1.4   & 1.3   & 1.0   & 0.7   & 0.7   & 0.4   & 0.3   & 0.2   & n.a.  & 0.1 \\
				\bottomrule			
			\end{tabular}
			\begin{tablenotes}[flushleft]
				\item[] b.d.-- below detection limit; n.a.-- not available.
			\end{tablenotes}
		\end{threeparttable}
	\end{table}
	\clearpage

	
	\begin{landscape}
	\begin{table}[p]
		\centering
		\caption{Oxygen isotope compositions of individual minerals in R3C-01 and its Wark-Lovering (WL) rim determined by spot SIMS measurements. $ \Delta $\ce{^{17}O} is a deviation from the terrestrial fractionation line: $ \Delta \ce{^{17}O} = \delta \ce{^{17}O} - 0.52 \times \delta \ce{^{18}O} $ (\textperthousand).}
		\begin{tabular}{cccccccc}
			\toprule
			Phase & Spot \#    & $ \delta $\ce{^{18}O} (\textperthousand)  & 1$ \sigma $ (\textperthousand)   & $ \delta $\ce{^{17}O} (\textperthousand)   & 1$ \sigma $  (\textperthousand)  & $ \Delta $\ce{^{17}O}  (\textperthousand)   & 1$ \sigma $  (\textperthousand)   \\
			\midrule
			Spinel framboid (\#01)    & 173   & $ -46.7 $ & 0.7   & $ -47.8 $ & 1.5 & $ -23.5 $ & 1.5 \\
			Spinel framboid (\#05)    & 174   & $ -47.7 $ & 0.8   & $ -50.9  $& 1.4 & $ -26.1 $ & 1.4 \\
			Diopside in the WL rim    & 175   & $ -40.4 $ & 1.0   & $ -44.0 $ & 1.4 & $ -23.0 $ & 1.5 \\
			Olivine in the WL rim    & 172   & $ -44.1 $ & 2.0   & $ -45.2 $ & 3.0 & $ -22.3 $ & 3.2 \\
			\bottomrule
		\end{tabular}%
		\label{tab:SIMS_spot}%
	\end{table}%
\end{landscape}
	\clearpage
	
	\setcounter{figure}{0}
	\setcounter{table}{0}
	\renewcommand{\thefigure}{B\arabic{figure}}
	\renewcommand{\thetable}{B\arabic{table}}
	\renewcommand{\thesection}{B\arabic{section}}
	
	\section*{Appendix B: Quality of isotopography}
	
	Here we evaluate the quality of isotopography following the method used in \citet{,park2012oxygen}. The $ \delta $\ce{^{18}O} value of spinel and melilite are nearly constant in \cref{fig:isotopograph}b ($ \sim -47 $\textperthousand ~and $ \sim -10 $\textperthousand, respectively). Since oxygen diffusivity in spinel is quite small \citep[e.g.,][]{,ryerson1994determination}, we assume that melilite and spinel have never exchanged oxygen isotopes after the CAI formation and therefore there is no oxygen isotopic zoning in these phases in \cref{fig:isotopograph}b. Based on the width between 16\% and 84\% values of a difference between \ce{^{16}O}-poor melilite and \ce{^{16}O}-rich spinel, the spatial resolution of the isotopography is estimated to be $ \sim $ 1 $ \mu $m, which is high enough to obtain oxygen isotopic composition of most of individual mineral grains in the CAI R3C-01. 
	
	\begin{figure}[p]
		\centering
		\includegraphics[width=1\linewidth]{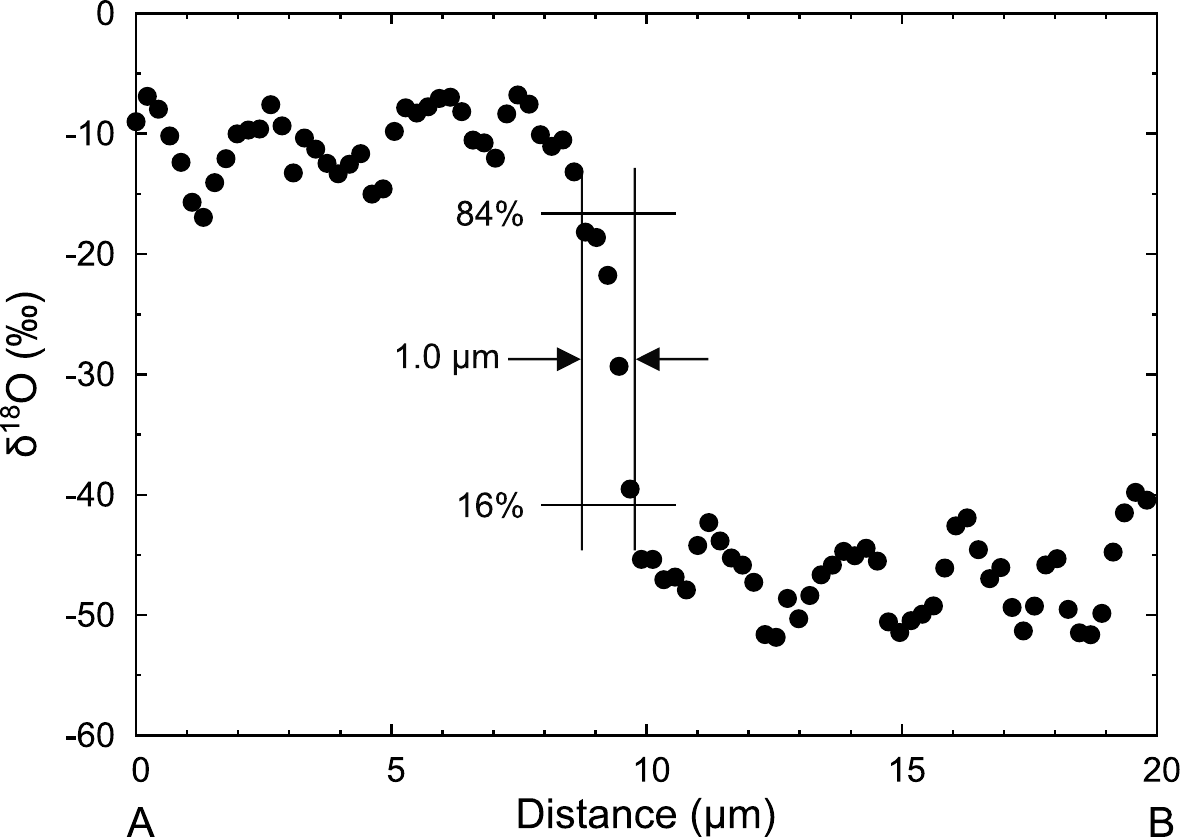}
		\caption{Line $ \delta $\ce{^{18}O} profile for the line A--B in \cref{fig:isotopograph}b.
		}
		\label{fig:SCAPS_quality}
	\end{figure}
	\clearpage

\end{document}